\documentstyle[eqsecnum,rmp,aps,epsf,harvard]{revtex}
\begin{document}
\draft
\citationstyle{dcu}

\title{ Astrometry and geodesy with radio interferometry: experiments,
 models, results}
\author{Ojars J. Sovers}
\address{ Jet Propulsion Laboratory, California Institute of Technology,
 Pasadena, California 91109 }
\author{John L. Fanselow}
\address{ Jet Propulsion Laboratory, California Institute of Technology,
 Pasadena, California 91109 \\ and Remote Sensing Analysis Systems, Inc.,
 Altadena, California 91001 }
\author{Christopher S. Jacobs}
\address{ Jet Propulsion Laboratory, California Institute of Technology,
 Pasadena, California 91109 }
%
%
\date{\today}
\maketitle
\begin{abstract}
Interferometry at radio frequencies between Earth-based receivers separated
by intercontinental distances has made significant contributions to
astrometry and geophysics during the past three decades.  Analyses of
such Very Long Baseline Interferometric (VLBI) experiments now permit
measurements of relative positions of points on the Earth's surface,
and angles between celestial objects, at the levels of better than 1~cm and
1~nanoradian, respectively.  The relative angular positions of extragalactic
radio sources inferred from this technique presently form the best
realization of an inertial reference frame.  This review summarizes the
current status of radio interferometric measurements for astrometric
and geodetic applications.  It emphasizes the theoretical models
that are required to extract results from the VLBI observables at present
accuracy levels.   An unusually broad cross-section of physics contributes
to the required modeling.  Both special and general relativity need to
be considered in properly formulating the geometric part of the propagation
delay.  While high-altitude atmospheric charged particle (ionospheric)
effects are easily calibrated for measurements employing two well separated
frequencies, the contribution of the neutral atmosphere at lower altitudes
is more difficult to remove.  In fact, mismodeling of the troposphere
remains the dominant error source.  Plate tectonic motions of the observing
stations need to be taken into account, as well as the non-point-like
intensity distributions of many sources.  Numerous small periodic and
quasi-periodic tidal effects also make important contributions to space
geodetic observables at the centimeter level, and some of these are just
beginning to be characterized.  Another area of current rapid advances is
the specification of the orientation of the Earth's spin axis in inertial
space: nutation and precession.  Highlights of the achievements of VLBI
are presented in four areas: reference frames, Earth orientation, atmospheric
effects on microwave propagation, and relativity.  The order-of-magnitude
improvement of accuracy that was achieved during the last decade has
provided essential input to geophysical models of the Earth's internal
structure.  Most aspects of VLBI modeling are also directly applicable
to interpretation of other space geodetic measurements, such as active
and passive ranging to Earth-orbiting satellites, interplanetary
spacecraft, and the Moon.
\end{abstract}
\framebox{\makebox
{\tt ~~~~~~~~~~~~~~To be published, Reviews of Modern Physics,
Vol.~70, Oct.~1998~~~~~~~~~~~~~~ } }


\tableofcontents
\subsection*{Symbols and abbreviations}
\begin{tabbing}
$A_{ij}$~~~~~~~~~~~~ \=nutation amplitudes in longitude \\
$\bf{B}$       \>baseline vector \\
$B_{ij}$       \>nutation amplitudes in obliquity \\
BIH            \>Bureau International de l'Heure \\
BIPM           \>Bureau International des Poids et Mesures \\
CDP            \>Crustal Dynamics Project \\
CIO            \>Conventional International Origin \\
$C_{\psi j}$   \>planetary nutation amplitudes in longitude (Eq.~\ref{eqksps})
 \\
$C_{\varepsilon j}$   \>planetary nutation amplitudes in obliquity
  (Eq.~\ref{eqksep}) \\
$c$            \>speed of light \\
$D$            \>mean elongation of Moon from Sun \\
$E$            \>elevation angle \\
$F$            \>latitude argument of Moon \\
$f$            \>flattening factor of the Earth \\
$f$            \>pressure loading factor \\
$G$            \>universal gravitational constant \\
GPS            \>Global Positioning System \\
GSFC           \>Goddard Space Flight Center \\
$g$            \>gravitational acceleration at Earth's surface \\
$g$            \>angle for barycentric dynamic time (Eq.~(\ref{eqmnanom}) \\
$H$            \>hour angle \\
$h$            \>altitude above reference ellipsoid \\
$h_\Upsilon$   \>hour angle of mean equinox of date \\
$h_i$          \>vertical Love number ($i$ = 2,3: quadrupole, octupole) \\
$h_{1,2}$      \>ionosphere or troposphere height limits \\
IAU            \>International Astronomical Union \\
ICRF           \>International Celestial Reference Frame \\
IERS           \>International Earth Rotation Service \\
IRIS           \>International Radio Interferometric Surveying \\
ITRF           \>International Terrestrial Reference Frame \\
IUGG           \>International Union of Geodesy and Geophysics \\
JD             \>Julian date \\
$\widehat{\bf{k}}$ \>unit vector in signal propagation direction \\
LLR            \>Lunar Laser Ranging \\
$l$            \>mean anomaly of Moon \\
$l^\prime$     \>mean anomaly of Sun \\
$l_{\rm E}$          \>mean longitude of Earth \\
$l_{\rm J}$          \>mean longitude of Jupiter \\
$l_{\rm M}$          \>mean longitude of Mars \\
$l_{\rm S}$          \>mean longitude of Saturn \\
$l_{\rm V}$          \>mean longitude of Venus \\
$l_i$          \>horizontal Love number ($i$ = 2,3: quadrupole, octupole) \\
mas            \>milliarcsecond \\
MERIT          \>Monitor Earth Rotation and Intercompare Techniques \\
MIT            \>Massachusetts Institute of Technology \\
$m$            \>speed of precession in right ascension \\
$m_p$          \>mass of body $p$ \\
$M{\rm _w}$    \>first moment of wet troposphere refractivity \\
${\cal M}_{\rm d,w}$  \>Mapping functions for dry, wet troposphere \\
${\bf N}$      \>nutation transformation matrix \\
$N$            \>atmospheric refractivity \\
$N^\prime$     \>lunar node argument \\
NASA           \>National Aeronautics and Space Administration \\
NEOS           \>National Earth Orientation Service \\
NMF            \>Niell mapping function \\
NNR            \>No Net Rotation \\
NOAA           \>National Oceanic and Atmospheric Administration \\
NRAO           \>National Radio Astronomy Observatory \\
Nuvel          \>New velocity model for plate tectonics \\
$n$            \>refractive index \\
$n$            \>speed of precession in declination \\
${\bf P}$      \>precession transformation matrix \\
$\bar p$       \>extended pressure anomaly \\
$p_{\rm A}$          \>general precession \\
$p_{\rm LS}$       \>lunisolar precession \\
$p_{\rm PL}$       \>planetary precession \\
$p_{\rm STD}$  \>reference atmospheric pressure \\
${\bf Q}$      \>rotation matrix for terrestrial to celestial transformation \\
RA             \>Right ascension \\
$R_{\rm E}$    \>Earth equatorial radius \\
${\bf{R}}_c$   \>Earth center coordinates in SSB frame \\
${\bf{R}}_p$   \>position of perturbing body in geocentric celestial system \\
$\widehat{\bf{R}}_p$   \>geocentric unit vector to perturbing body \\
$R_{\rm EG}$    \>distance from Earth to gravitating body $G$ \\
$\widehat{\bf{r}}$ \>unit vector in radial direction \\
$r_0$          \>classical electron radius \\
${\bf{r}}_0$     \>station position in terrestrial system, excluding
                   motions \\
${\bf{r}}_c$     \>station position in geocentric celestial system \\
${\bf{r}}_s$     \>phase shifted station position \\
$\widehat{\bf{r}}_s$     \>geocentric unit vector to station position \\
${\bf{r}}_t$     \>station position in terrestrial frame \\
${\bf{r}}_{1,2}$      \>position of station 1,2 in SSB frame \\
${\bf{r}}_{\rm date}$     \>station position of date \\
$r_{\rm sp}$       \>station cylindrical radius from spin axis \\
$S(E)$         \>slant range factor \\
S/X            \>S-band + X-band \\
SGP            \>Space Geodesy Project \\
SLR            \>Satellite Laser Ranging \\
SSB            \>Solar System Barycenter \\
$S_{\psi j}$   \>planetary nutation amplitudes in longitude \\
$S_{\varepsilon j}$   \>planetary nutation amplitudes in obliquity \\
$\widehat {\bf{s}}$   \>geocentric celestial unit vector to source \\
$\widehat {\bf{s}}_0$ \>SSB celestial unit vector to source \\
TDB            \>Temps Dynamique Barycentrique \\
TDT            \>Terrestrial Dynamic Time \\
$T$            \>time in centuries since J2000.0 \\
$t_0$          \>reference time \\
$t_1$          \>time of arrival of wave front at station 1 \\
$t_1^{\prime}$ \>proper time of arrival of wave front at station 1 \\
$t_2^{\*}$     \>time of arrival of wave front at station 2 \\
$t_e$          \>time of emission by source \\
$t_{\rm tr}$       \>light transit time \\
${\bf U}$            \>UT1 transformation matrix \\
$U$            \>gravitational potential \\
UT1            \>Universal Time 1 \\
UTC            \>Universal Time Coordinated \\
UTPM           \>Universal Time and Polar Motion \\
$u,v$          \>projections of ${\bf{B}}$ on plane of sky \\
${\bf V}{\bf W}$ \>transformation matrix from local to equatorial frame
 (Eqs.~\ref{eqv}) \\
VEN            \>Vertical, East, North (local geodetic coordinates) \\
VLA            \>Very Large Array \\
VLBA           \>Very Long Baseline Array \\
VLBI           \>Very Long Baseline Interferometry \\
$V_i$          \>astronomical argument of tidal constituent $i$ \\
$W$            \>atmospheric temperature lapse rate \\
WVR            \>Water Vapor Radiometer \\
${\bf X}$      \>polar motion transformation matrix, x component \\
$x_i,y_i,z_i$  \>Cartesian coordinates of station $i$ \\
$x_i^0,y_i^0,z_i^0$  \>Cartesian coordinates of station $i$ at reference
 time $t_0$ \\
${\dot x}_i,{\dot y}_i,{\dot z_i}$   \>Cartesian velocities of station $i$ \\
${\bf Y}$      \>polar motion transformation matrix, y component \\
$Z$            \>auxiliary angle for precession (Eq.~\ref{eqzed}) \\
$Z_{\rm d,w}$  \>zenith dry, wet troposphere delay \\
ZMOA           \>Zhu, Mathews, Oceans, Anelasticity (nutation model) \\
$\alpha$       \>right ascension \\
$\alpha_{\rm E}$     \>equation of the equinoxes \\
$\bbox{\beta}_{1,2}$   \>velocity of station 1, 2 \\
$\gamma _{_{\rm PPN}}$ \>general relativity (Parametrized Post-Newtonian)
 gamma factor \\
$\bf{\Delta}$  \>total tidal shift in terrestrial coordinate system \\
${\bf{\Delta}}_{\rm atm}$  \>atmospheric loading station position shift \\
$\Delta_{\rm gd}$  \>ionosphere contribution to group delay \\
$\Delta h$     \>height of antenna reference point above surface \\
${\bf{\Delta}}_{\rm ocn}$  \>ocean loading station position shift \\
$\Delta p$     \>local pressure anomaly \\
$\Delta_{\rm pd}$  \>ionosphere contribution to phase delay \\
${\bf{\Delta}}_{\rm pol}$  \>pole tide station position shift \\
${\bf{\Delta}}_{\rm sol}$  \>solid tide station position shift \\
${\bf \Delta_{x,y,z}}$  \>components of perturbation rotation matrix \\
$\Delta G_p$   \>gravitational contribution to coordinate time delay,
                   body $p$ \\
$\Delta G_p^\prime$   \>gravitational contribution to proper time delay,
                   body $p$ \\
$\Delta U$     \>Earth gravitational potential \\
$\Delta\Theta$ \>gravitational deflection \\
$\Delta\Theta_l$  \>tidal contribution to UTPM \\
$\Delta\Theta_{l^\prime}$  \>companion tidal contribution to UTPM \\
$\Delta\psi$   \>nutation in (celestial) longitude \\
$\Delta\psi^0$  \>out of phase nutation in longitude \\
$\Delta\psi^{\rm f}$  \>free core nutation in longitude \\
$\Delta\varepsilon$  \>nutation in obliquity \\
$\Delta\varepsilon^0$  \>out of phase nutation in obliquity \\
$\Delta\varepsilon^{\rm f}$  \>free core nutation in obliquity \\
$\Delta\tau_s$ \>source structure contribution to delay \\
$\Delta{{\dot \tau}}_s$   \>source structure contribution to delay rate \\
$\delta$       \>declination \\
$\bbox{\delta}$  \>tidal displacement in local (VEN) coordinates \\
$\delta_{1,2,3}^i$  \>Cartesian local solid tidal displacements ($i$ = 2,3:
  quadrupole, octupole) \\
$\delta\psi$   \>empirical correction to longitude \\
$\delta\varepsilon$  \>empirical correction to obliquity \\
$\bbox{\varepsilon}_{1,2}$   \>curved wave front expansion quantities
  (Eqs.~\ref{eqeps}) \\
$\overline\varepsilon$   \>mean obliquity \\
$\overline\varepsilon_0$ \>mean obliquity at J2000.0 \\
$\zeta$        \>auxiliary angle for precession (Eq.~\ref{eqzeta}) \\
$\Theta$       \>auxiliary angle for precession (Eq.~\ref{eqtheta}) \\
$\Theta$       \>gravitational deflection angle \\
$\Theta_{1,2,3}$ \>angular coordinates of Earth rotation axis \\
$\widehat{\bbox{\lambda}}$   \>unit vector in longitude direction \\
$\lambda_s$    \>station longitude \\
$\mu$as        \>microarcsecond \\
$\mu_p$        \>$Gm_p$ \\
$\nu_{\rm S}$  \>S-band VLBI frequency $\approx$2.3 GHz \\
$\nu_{\rm X}$  \>X-band VLBI frequency $\approx$8.4 GHz \\
$\nu_g$        \>electron gyrofrequency \\
$\nu_p$        \>electron plasma frequency \\
$\xi_{ij}$     \>amplitude of component $j$ of tidal constituent $i$ \\
$\rho$         \>number density of electrons \\
${\displaystyle \Sigma}$       \>SSB reference frame \\
${\displaystyle \Sigma^{\prime}}$ \>geocentric celestial reference frame \\
$\sigma$          \>correlation function \\
$\tau$         \>geometric delay \\
$\tau_c$       \>delay due to clock imperfections \\
$\tau_{\rm gd}$    \>group delay \\
$\tau_{\rm pd}$    \>phase delay \\
$\tau_{\rm sr}$    \>delay due to antenna subreflector motion \\
$\tau_{\rm tr}$    \>troposphere delay \\
$\widehat{\bbox{\phi}}$   \>unit vector in latitude direction \\
$\varphi_{ij}$ \>phase of component $j$ of tidal constituent $i$ \\
$\phi_s$       \>structure phase \\
$\phi_s$       \>geocentric station latitude \\
$\phi_{s({\rm gd})}$ \>geodetic station latitude \\
$\psi$         \>phase shift of solid tidal effect \\
$\Omega$       \>mean longitude of ascending lunar node \\
${\bf \Omega}$ \>perturbation transformation matrix \\
$\omega_{\rm E}$     \>rotational speed of Earth \\
$\omega_{\rm f}$     \>free core nutation frequency \\
$\omega_{\rm xyz}^j$  \>Cartesian components of angular velocity of
 tectonic plate $j$ \\
\end{tabbing}

\section{ INTRODUCTION }
\label{intro}

  Astrometry and geodesy have undergone a revolution during the past
three decades.  This revolution was initiated in the mid-1960s by the
advent of interferometry at radio frequencies using antennas separated
by thousands of kilometers.  Developments during the following decades
refined this ``Very Long Baseline Interferometry'' (VLBI) technique
to reach its present capability of point positioning on the Earth's
surface at the centimeter level, and angular positioning for point-like
extragalactic radio sources at the sub-milliarcsecond (nanoradian) level.
These refinements are expected to continue until insurmountable problems
are reached, probably in the sub-millimeter accuracy regime.
Geodetic measurements on the Earth's surface have been enormously
expanded and densified by satellite tracking methods developed during
the past decade, but VLBI remains the prime technique for astrometry
employing natural radio sources.  It is unique in its ability to measure
the Earth's orientation in an inertial frame of reference.

  Detection and processing of the extremely weak signals from distant
radio sources requires long integration times, sophisticated data
acquisition systems, and customized computers.  Analyses of VLBI observables
involve an uncommonly broad cross-section of the disciplines of physical
science, ranging from consideration of the effects of the Earth's internal
structure on its dynamics, to the tectonic plate motions and various
terrestrial tidal effects, to quantification of turbulence in the atmosphere,
to special relativistic description of the radio signals traveling from the
distant sources, to general relativistic retardation.  With the exception of
gravimetric and oceanographic experiments, VLBI is perhaps the most demanding
technique for many aspects of global Earth models.  It therefore provides
results that are used in a number of related fields.  While extension of
experiments to platforms in Earth orbit and elsewhere in the Solar
System is in its infancy \cite{Burke91,Hirab,Adam95},
the present review is limited to Earth-based observations.
A large portion of the modeling we shall discuss here is common to VLBI
and Earth-satellite techniques such as laser ranging to reflectors on
the Moon (lunar laser ranging, LLR) and ranging to passive (Lageos, SLR)
and active satellites (Global Positioning System, GPS).  For these
reasons, it is important to review the complete details of current
models of VLBI observables.  Following the principle that the accuracy
of a theoretical model should exceed the accuracy of the experiments which
it interprets by at least an order of magnitude, the VLBI model should
ideally be complete at the 1 picosecond level.  Unfortunately this requirement
is currently not satisfied by a number of parts of the model.  As experimental
techniques are refined, possibly to improve the accuracy by another order
of magnitude, numerous aspects of the model will need to be re-examined.

  The field of radio astronomy was initiated in the 1930s by the discovery
of celestial radio emission by Jansky \citeyear{Jansky32,Jansky33}.
Development of instrumentation continued, and in the 1940s
\citename{Reber40}~\citeyear{Reber40,Reber44}
used a parabolic antenna to construct the first maps of
radio emission from the Milky Way galaxy at frequencies of 160 and
480 MHz.  Subsequent evolution of radio astronomy through the post-war
years has been traced in the comprehensive collection of memoirs
edited by \citeasnoun{Sullivan}.  Following the discovery of the powerful
non-thermal radio sources called quasars \cite{Schmidt63}, which
are extremely distant and emit
vast quantities of energy, intense interest developed in characterizing
their intrinsic nature, as well as in using them as the basis for a
novel astronomical reference frame.  The 3rd Cambridge survey
\cite{Edge59} and the Parkes survey \cite{Bolton79} identified many
of the radio sources used today for geodetic and astrometric VLBI.

  The basic ideas of the VLBI technique were first demonstrated in
experiments detecting decametric bursts of Jupiter by a group at the
University of Florida in 1965 \cite{Carr65,May67}.  Subsequent
history can be traced from the first extragalactic interferometric
observations by the NRAO-Arecibo group \cite{Bare67} at 0.6 GHz
on a 220-km baseline, the MIT-NRAO group
(\citename{Moran67}~\citeyear{Moran67}; 1.7 GHz, 845 km),
and the Canadian Long Baseline Interferometry group
(\citename{Broten67}~\citeyear{Broten67}; 0.4 GHz, 3074 km).
Early work in geodesy, astrometry, and clock synchronization
was done in 1969 \cite{Hinter72,Cohen72}, yielding
accuracies in distances of 2$-$5 meters and in source positions on the
order of 1 arc second.  After the first decade spent in developing the
technique, numerous applications matured in succeeding years using
networks of fixed antennas.  Extension to transportable antennas was
also implemented during the 1970s \cite{MacDor74,Ong76,DavTrsk85}.
During that period, several basic reviews of VLBI instrumentation,
experiments, and analysis appeared as chapters in the series ``Methods
of Experimental Physics: Astrophysics'' \cite{Meeks76}.  Some Ph.~D.
dissertations of that era are those of \citeasnoun{Whitney74},
\citeasnoun{Robrtsn75},
and \citeasnoun{Ma78}.  After the currently standard Mark III VLBI data
acquisition
system was developed \cite{Rogers83}, large-scale international
cooperation began in 1979 with the establishment of the Crustal Dynamics
Project (CDP) by NASA \cite{Bswrth93}.  In the
1980s, radio interferometry progressed to the point of yielding some
of the most accurately measured physical quantities.  Increasingly larger
and more sensitive antennas were used to extend observations to ever
weaker sources of emission, and the results have made important contributions
to our understanding of a wide variety of astrophysical and geodynamic
phenomena.  Such advances in instrumentation culminated in the construction
of instruments such as the Very Large Array (VLA) in 1980 \cite{Hjellm82},
and the Very Long Baseline Array (VLBA) in 1993 \cite{Napier94}.

  Useful references which provide an introduction to the principles of very
long baseline interferometry are several articles in the above-mentioned
volume 12 of ``Methods of Experimental Physics''
\cite{Rogers76,Pooley76,Moran76,Vessot76,Shapiro76}, the book
by \citeasnoun{Thmpsn86}, and two reports by \citename{Thomas81}
~\citeyear{Thomas81,Thomas87}.  Additional background material can be found
in the review
of geophysical VLBI applications by \citeasnoun{Robrtsn91}, and in more detail
and more recently, in the three-volume compilation that summarizes the
accomplishments of NASA's Crustal Dynamics Project during the past decade
\cite{Smith93}.  Various periodic reviews, including the
quadrennial IUGG geodesy reports (\citeasnoun{Clark79}, \citeasnoun{Carter83};
\citeasnoun{Robrtsn87}; \citeasnoun{Ray91}; \citeasnoun{Herr95}), the
IERS annual reports
\citeaffixed{IERS96a}{$e.g.$}, and the yearly reports on CDP results
[$e.g.$~\citename{Ma92}~\citeyear{Ma92}, more recently at
{\tt{http://lupus.gsfc.nasa.gov/vlbi.html}}], are also good sources for
detailed information concerning current VLBI techniques and results.
The International Earth Rotation Service (IERS) periodically publishes
a compilation of standard models recommended for analyses of space
geodetic data \citeyear{IERS92,IERS96b}; also available at
{\tt{http://maia.usno.navy.mil/conventions.html}}.  In large part, the
model description in the present paper is in agreement with the
specifications of these ``IERS Standards''.

  The intent of this paper is to review astrometric and geodetic
VLBI from a description of the experiments through analysis of the results,
with emphasis on the details of theoretical modeling.  Section~\ref{expobs}
summarizes experiment design, single-station signal collection and recording,
signal combination, and observable extraction.  This section also reviews
the general approach to extracting parameters of astrometric and geodetic
interest from the observables.  The three succeeding sections
(\ref{gdel}-\ref{atmo}) consider the theoretical model in detail,
partitioning the interferometric delay model into three major components:
geometry, instrumentation (clocks), and atmosphere, and presenting
the best current models of each of these components.  The longest
section (\ref{gdel}) deals with the purely geometric portion of the delay,
wave front curvature, and gravitational bending, and considers time
definitions, tectonic motion, tidal and source structure effects, coordinate
frames, Earth orientation (universal time and polar motion), nutation,
precession, Earth orbital motion, and antenna offsets.  Sections~\ref{clk}
and \ref{atmo} discuss the non-geometric components of the model: clocks
and the atmosphere.  Section~\ref{curres} presents a brief overview of the
accomplishments of VLBI and its contributions to our understanding of
astrometry and geophysics during the past two decades.  Section~\ref{impr}
outlines model improvements that we anticipate will be required by more
accurate data in the future.

  The most convenient units for use in astrometric and geodetic
VLBI are millimeters, picoseconds, and nanoradians.  The first two are
connected by the speed of light = 299,792,458 m/s $\approx$1 mm/3 ps,
and will be used interchangeably.
Most commonly for the purposes of illustration, a 10,000-km baseline
will be considered.  A length change of 1 cm on such a baseline is
equivalent to an angular change of 1 nrad (1 part per billion, ppb)
or 0.2 milliarcseconds (mas); the equivalent conversion factors in
millimeters are 10 mm/nrad or 50 mm/mas.

\section{ EXPERIMENTS AND ANALYSIS }
\label{expobs}

  Interferometric measurements permit resolution on the order of the
wavelength of the radiation.  For radio waves at the typical
astrometric/geodetic frequencies of 2.3 GHz (S band) and 8.4 GHz
(X band), the wavelengths are 13 and 3.5 cm, respectively.  Some recent
activity is focusing on even higher frequencies
\citeaffixed{Lebach95}{$e.g.$}, and numerous interferometric observations
are planned at millimeter wavelengths, spanning the gap
between the radio and optical regions.  For experiments that are
dominated by random errors, gains can be achieved by
repeated measurements, thereby reducing the errors substantially.
Thus, the accuracy at X band may be expected to surpass 1 cm in many cases.

  Early radio interferometric experiments employed antennas that were
separated by distances on the order of 1 km.  These experiments had the
advantage that the stations shared a frequency standard, and signal
correlation was performed in real time \citeaffixed{Batty82}{$e.g.$}.
Such an arrangement is now known as a ``connected element interferometer''
to distinguish it from interferometers with separated elements
that are not in direct communication.  A connected element interferometer
is a close analogue of the Michelson stellar interferometer
which manipulates
signals with mirrors to produce a physical interference pattern at
the detector.  Frequency standards (``clocks'') and the technology
associated with measurements at centimeter wavelengths improved
dramatically as a result of radar development during World War II.
It became possible to separate the two components of the interferometer
with independent clocks and instrumentation, and to use high-speed data
recording techniques, thus eliminating the need for close physical proximity.
With such an arrangement there are no physical real-time interference
``fringes'', although the terminology of optical interferometry has
been carried over to the radio wavelength region.  Remote radio
interferometry is now commonplace.  With the two antennas separated
by long baselines, the experimental observables are no longer obtained
in real time, but are instead generated by subsequent analysis of
recorded information.  Such experiments have become known as ``very
long'' baseline interferometry or VLBI.  With the recent explosive
advances in communications, computing, and data storage technology,
connected element experiments over intercontinental distances may
again become standard in the future.

  Large dish antennas (20 to 100 m in diameter) that are used for radio
astronomy and VLBI are found in approximately 50 globally distributed
locations.  An equivalent number of sites have been occupied by smaller
transportable antennas during the past two decades.  The premier sites
are at relatively high altitudes and in arid climates, which were
selected to minimize the impact of atmospheric turbulence on
measurement quality.

\subsection{ Experiment design and data collection }
\label{expts}

  A diagram of a contemporary VLBI experiment is sketched in
Fig.~\ref{schemvlbi}.  Two antennas, separated by a baseline ${\bf{B}}$,
are simultaneously pointed at the same radio source, and detect the
incoming wave front propagating along unit vector~${\bf{\widehat k}}$.
\vbox{
\begin{figure}
\hskip 2.0in
\epsfysize=4.0in
\epsffile{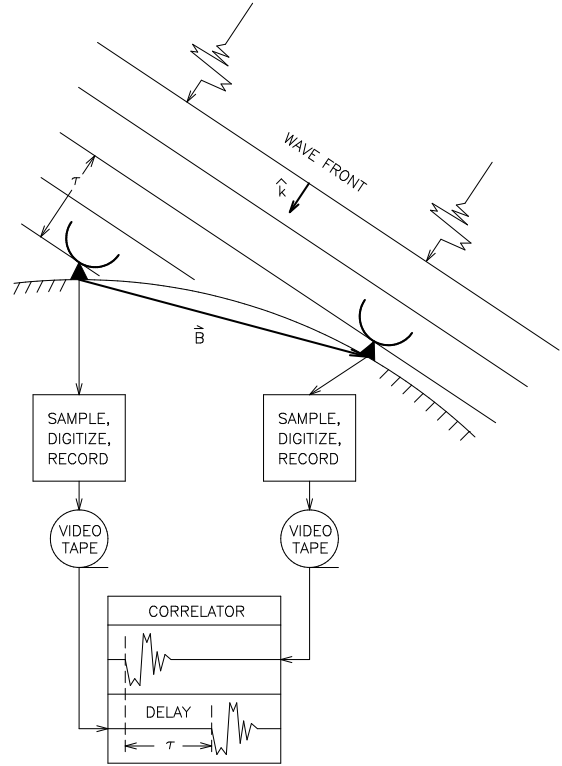}
\vskip 15pt
\caption{ Schematic diagram of a VLBI experiment.  Waveforms are shown
impinging from the direction $\widehat{\bf{k}}$ on two antennas separated
by a baseline ${\bf{B}}$ on the Earth's surface.  They are followed through
the data acquisition system to the point where the correlator determines
the delay $\tau$.  The signal waveforms are exaggerated for effect.  The
actual waveforms are random Gaussian processes. }
\label{schemvlbi}
\end{figure}
}
\noindent More than two antennas (stations) commonly participate
in an experiment,
but its fundamentals can be adequately illustrated by considering
only one of the many pairs of stations.  Prior to its arrival at the
first antenna, the signal has already been affected by the electromagnetic
and gravitational fields that it encountered along its journey through
interstellar space, the Solar System, and the Earth's atmosphere.
Between this time and the time of reception at the second antenna,
any motions of the Earth also make their influence felt.  All these
varied aspects need to be taken into account in analyses of the experiment.
We note that the VLBI technique is differential in the sense that the
observables depend on the $relative$ station locations and source positions,
but they are fairly insensitive to the stations' absolute geocentric
locations and the origin of right ascension (celestial longitude).  Other
techniques ($e.g.$, SLR, GPS) must supply such information, which minimally
consists of the geocentric position of one of the observing stations at a
given epoch, and the right ascension of one source.

  The information content of the VLBI observables depends critically
on the design of the experiment \citeaffixed{Shapiro76}{$e.g.$}.  The
observing schedule (station network, observation epochs, and local
azimuth and elevation angles of the selected sources) plays a crucial
role in determining the types and precision of parameters that can be
extracted in the subsequent data analyses.  At a bare
minimum there must be at least one observable for each parameter to
be estimated.  The experimenter must first decide how many parameters
he wishes to measure.  To consider a simple example: after having
fixed the station location and clock parameters of a reference station,
a typical astrometric or geodetic analysis will want to estimate, at a
minimum, for each of the remaining stations three site coordinates, a
zenith tropospheric delay parameter, and linear clock parameters
(offset and rate).  In addition to these station-specific parameters,
a number of ``global'' parameters are also usually estimated.  These
are common to the entire network, and describe the orientation of the
Earth (5 parameters) and the positions of the sources (2 angular
parameters per source); the right ascension of one source must be
fixed to establish the origin of the celestial reference frame.
Thus for an experiment with $N$ stations and $M$ sources, there
would be $N_p$ parameters to determine, where $ N_p ~=~ 6N ~+~ 2M
~-~ 1 $.  In recent years, analyses have tended to estimate
increasingly larger numbers of station-specific parameters for the
clocks and tropospheres, thereby increasing the coefficient of the
$N$ term many-fold.

  Note that the observable delay in arrival times is approximately
proportional to the scalar product of the baseline and signal
propagation vectors, and thus contains information concerning both
the station and source coordinates.  With sufficiently redundant
measurements, therefore, both the station and source coordinates
are accessible to estimation, subject to the limitations imposed by only
observing the scalar product.  Naturally, it is highly desirable to
have more observables than the minimum ($N_p$) required for a non-singular
solution in the analysis.  The additional observables permit reduction
of the formal statistical uncertainties in the estimated parameters.
They also allow the level of systematic errors to be characterized
using various tests of the repeatability of the results.
Such tests may include root-mean-square of the residuals (observed
minus theoretical) and repeatability of estimated parameters
derived from subsets of the data.  Excess observables also leave
open the option of estimating additional parameters
whose importance may become known only after the data are collected.
For all these reasons, the experiment design usually provides for
many more observables than the expected number of parameters.
In fact, a typical astrometric/geodetic experiment will have several
hundred observables per baseline.

  Given a network of stations, how are the source observations
chosen?  To answer this question, and to generate an optimal observing
schedule, the experimenter is guided by a few key principles.  First
the experiment must have sufficient spatial and temporal sampling in
order to permit unambiguous separation of the parameters of interest
during analysis.
As an example of the need for good spatial sampling, consider the
dry zenith tropospheric delay parameter $Z_{\rm d}$ and the station local
vertical coordinate $v$.  The measured delays $\tau$ will have
sensitivities to these parameters that scale approximately as
\begin{eqnarray}
\partial \tau / \partial Z{\rm _d} ~&\propto & ~1/\sin E  \\
\partial \tau / \partial v~~ ~&\propto & ~-\sin E ~,
\label{eqsens}
\end{eqnarray}
where $E$ is the elevation angle above the horizon.  These two parameters
will be correlated if $E$ is sampled only close to 45$^\circ$.  In order
to separate the parameters the experiment must observe sources over a wide
range of elevations.  In particular the low-elevation ($E < 10^\circ$)
measurements will be most sensitive to the troposphere, while those at
higher elevation angles will be most sensitive to the station vertical
coordinate.

  Next we consider the need for good temporal sampling.  Simply put, the
experiment must sample the parameter of interest more quickly than it
changes.  For periodic effects this implies that one must take at
least two samples within one full period of the effect of interest.
In astrometric and geodetic VLBI one is often interested in tides with
periods of 12 and 24 hours, 2 weeks, 1 month, a half a year, and one year.
Components of the Earth's nutation have periods ranging from days up
to 18.6 years.  In practice experiments collect data at the rate of many
observations per hour for a full 24-hour session.  Such sessions have been
carried out many times per year during the last two decades, allowing VLBI to
measure parameters which vary on all of the aforementioned time scales.
Day-long experiments also have the benefit of allowing observations through
the full range of source right ascension.  This minimizes problems that
would otherwise arise in connecting observations from one right ascension
zone to another.  As a result, VLBI can avoid the zonal RA errors that
have plagued optical observations which can only be performed at night.
Because VLBI measurements are done at microwave frequencies, they are
also much less sensitive to weather (clouds, rain) than optical techniques.

  The experimenter must next decide how long to observe each source.
While collecting more data bits improves the detection signal-to-noise
ratio (SNR), the desire to increase the SNR must be balanced against the
reality of finite recording bandwidths which are 50-100 Mbits/sec for
the Mark III VLBI system \cite{Rogers83}.  As a result the
optimal integration times typically range from 2 to 13 minutes.
A technique known as bandwidth synthesis \cite{Rogers70} provides
the highly accurate delays possible with a large radio frequency
bandwidth, while requiring a recording bandwidth that is only a small
fraction of the former.  This technique has been essential in overcoming
the limitations of currently available recording bandwidths.  For very
weak sources, much longer integration times may be required.  The
coherent integration time may be extended considerably with the
phase referencing technique \cite{Lestrade90}, in which a strong nearby
source is observed alternately with the weak source, and the phase
of the strong source is used as a reference.

  Lastly, we consider how best to use a given network of antennas.
For antennas separated by a few thousand kilometers or less it is
often practical to have all antennas simultaneously observing the
same source.  With such modest baselines, each station can observe
over a broad spatial extent.  However, for networks with baselines
significantly longer than an Earth radius, the desires for simultaneous
visibility of the source and for broad geometric coverage start to
be in conflict.  Because the area of the sky that is simultaneously
visible from both observing sites shrinks rapidly as baselines approach
an Earth diameter in length, options for observing strategies become
severely limited.  The solution has been to create ``sub-networks''.
This observing strategy assigns a subset of the stations in the network
to observe one source, while another subset of stations observes a
different source.  While this complicates the experiment logistics,
it has allowed the design of experiments with very strong geometries.

  Having considered the experiment design, we will now discuss how the
signal is detected and recorded.  For a Cassegrain antenna (see
Fig.~\ref{antpic} in Sec.~\ref{ant}), the incoming signal first strikes
the primary paraboloidal dish of the antenna, is then reflected up to the
hyperboloidal subreflector, and finally into the feed horn on the central
axis.  For a prime focus antenna, the signal goes directly from the
paraboloidal reflector to the feed at the focus position.  Past the feed
horn, the signal undergoes an initial stage of amplification, either by
transistor amplifiers such as field-effect or high-electron-mobility
transistors or by a traveling wave maser.  System temperatures typically
range from 20-100$\thinspace$K at S and X band.  After the first stage
of amplification the signal is heterodyned from radio frequency to a lower
intermediate frequency of several hundred MHz.
Next the signal is further heterodyned down to ``video'' frequencies.
There the signal is band limited to a width of a few MHz, sampled, and
digitized.  Lastly, it is formatted and recorded on digital video tape
(see the lower half of Fig.~\ref{schemvlbi}).

  The signal is digitized at a minimum level of
quantization to make the most effective use of the
limited recording bandwidth.  The digitized signals are recorded on
high capacity video tapes for a period of several minutes, yielding
a total of $\approx$10 gigabits of data.  In routine observing sessions
lasting 24 hours, this procedure is repeated many times, with as many as
a hundred distinct sources sampled at several distinct hour angles.  Note
that the independent station clocks must remain well enough synchronized
to obtain signal samples that will form a coherent interference
pattern.  The signal flux density is on the order of 1 Jansky
(Jy $= 10^{-26}$ W m$^{-2} $\thinspace Hz$^{-1}$), necessitating
antennas with large collecting areas and high sampling and recording
rates, as well as highly sensitive and stable detectors and frequency
standards.  Pointing and mechanical characteristics of the antenna
structure must be sufficiently responsive to permit movement between
sources widely separated in the sky within a few minutes.

  The fundamental measurement in VLBI times the arrival of the wave front
at the two ends of the baseline.  In order to take full advantage of the
timing precision possible with current frequency standards ($e.g.$, hydrogen
masers stable to 1 part in $10^{14}$) care must be taken to calibrate phase
shifts induced by the measurement instrumentation.  Such phase shifts can
corrupt the estimated
phase and hence the group delay of the incoming signal.  In order to
correct for these instrumental phase shifts a technique known as phase
calibration has been developed \cite{Rogers75,Fans76,Thomas78,Sgmn87}.
This technique compensates for instrumental phase errors
by generating a signal of known phase, injecting this signal into the
front end of the VLBI signal path, and examining the signal's phase after
it has traversed the instrumentation.  This calibration signal is embedded
in the broadband VLBI data stream as a set of low level monochromatic
``tones'' along with the signal from the radio source.  Typically
the calibration signal power is $< 1$\% of the broadband VLBI signal level.
The tones are extracted from the data at the time of subsequent processing.

\subsection{ Correlation: Producing the interference pattern }
\label{correli}

  After the signals at each antenna site are collected and recorded, the
next step initiates the analysis of a VLBI experiment.  The signals recorded
at all participating antennas are combined pairwise, thereby producing the
interference pattern.  This progression is shown
in Fig.~\ref{schemvlbi}.  The facilities for signal combination are
called correlators.  Several such installations are presently operating
in the U.~S.~A. (Haystack Observatory, Jet Propulsion Laboratory, U.~S.
Naval Observatory, Very Long Baseline Array), Germany (Bonn),
and Japan (Kashima).  They consist of special-purpose signal processing
hardware which is used to determine the difference in arrival times at
the two stations by comparing the recorded bit streams.
\citename{Moran76}~\citeyear{Moran76} and
\citename{Thomas80}~\citeyear{Thomas80,Thomas81,Thomas87}~discuss the
details of this process.  In brief,
the two bit streams representing the antenna voltages as functions of
time $t$, $V_1(t)$ and $V_2(t)$, are shifted in time relative to each
other until their cross-correlation function $R$ is maximized:
\begin{equation}
R \thinspace (\tau) = {1 \over T} \int \limits_0^T dt \thinspace V_1(t)
\thinspace {V^*_2}(t-\tau) ~.
\label{eqcorrho}
\end{equation}
$T$ is the averaging interval, $*$ denotes complex conjugate, and the time
$\tau$ corresponds to the difference in arrival times (the normalization
is arbitrary for the purposes of our present discussion).

  Following \citeasnoun{Thomas87}, a simplified diagram of the processing of
data bits in a VLBI correlator is shown in Fig.~\ref{correl}.  For
purposes of this illustration we have only shown 8-bit averaging,
one trial delay offset (known as a ``lag''), artificially small
delays, high fringe rates, and large amplitudes.  Various fixed
delays which occur in actual implementations are omitted in order
to emphasize only the essential steps.

\vbox{
\begin{figure}
\hskip 2.0in
\epsfysize=3.0in
\epsffile{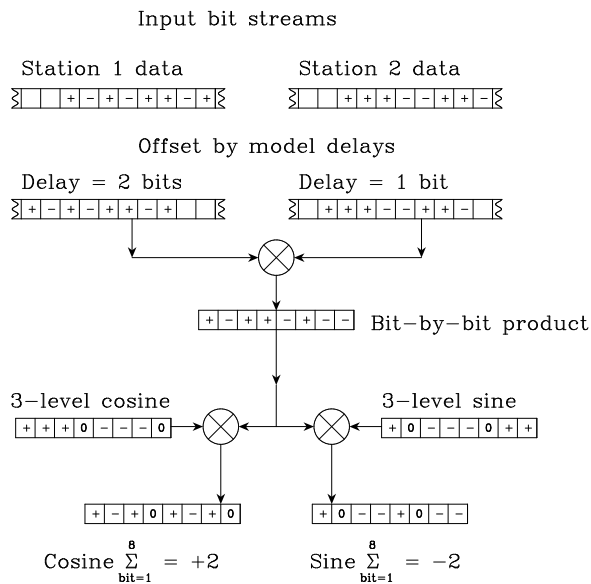}
\vskip 15pt
\caption{ Functional diagram of a VLBI correlator.  The input bit streams
are processed to the point of yielding the sine and cosine correlation
sums. }
\label{correl}
\end{figure}
}

  Starting at the top, Fig.~\ref{correl} shows a short segment of the bit
stream from each of two stations.  The data were recorded using 1-bit
sampling; when the signal voltage $V > 0$ the bit is ``+'', and when
$V < 0$ the bit is ``$-$''.  Some of the bit locations have been left
blank in this illustration, in order to show more clearly only the eight
bits that are followed through all steps of the example.  In the second
step, the bit streams have been delayed by 2 bits for station 1 and by
1 bit for station 2.  In reality these delays are typically on the order
of $10^5$ bits, where one bit is recorded every 0.25 $\mu$s.  In step 3,
the two bit streams have been cross-correlated.  The resulting product is
``+'' when the corresponding bits from the two streams match, and ``$-$''
when they differ.

  Since the Earth's rotational velocity induces Doppler shifts on the
order of $10^{-6}$ times the observing frequency, VLBI observations at
X band (8.4 GHz) will have their signals oscillating at several kHz.
VLBI signals are typically very weak (only 1 bit in $10^3$ or $10^4$ correlates
between the two stations), and one must therefore average over many bits to
detect the signal.  To achieve this the signals must first be
``counter-rotated''.  This involves multiplying the cross-correlated bits
by sine and cosine waves
to remove the above-mentioned Doppler shifts.  Thus counter-rotation
may also be thought of as a digital heterodyning of the cross-correlation
signal from kHz to near zero frequency.  In step 4, the bits are
counter-rotated by multiplying them with 3-level approximations of sine
and cosine waves.  (This counter-rotation can also be performed
prior to cross-correlating in step 3).
Step 5: having been counter-rotated, the signal may
now be averaged over many bits in order to allow the weak VLBI
signals to be detected.  For clarity the diagram shows averaging over
only eight bits.  The resulting cosine and sine sums are then
root-sum-squared and normalized to obtain the signal amplitude, and the
arctangent (sine/cosine) is taken to yield the signal's phase.  Note that
for the purposes of this illustration the signal amplitudes are
unrealistically large.

  Several aspects of the process described above distort the extracted
signal and must be accounted for.  First we note that 1-bit sampling
causes a loss in SNR and introduces a $2/\pi$ scale factor in the amplitudes
\cite{vanVlck66}.  The quantization of the delay to steps
of one bit changes both the amplitude and phase, while the use of a 3-level
approximation to the sine and cosine functions scales the amplitude.
\citename{Thomas80}~\citeyear{Thomas80,Thomas87}~and
\citeasnoun{Thmpsn86} discuss these
effects and their rectification in detail.

\subsection{ Post correlation: Generating the observables }
\label{postcorr}

  The correlation process is carried out in parallel for many (typically
14) frequency channels, with each channel producing average amplitudes
and phases every 1-2 seconds, as described in the previous section.
These results are stored for later analyses with post-correlation
software, which is discussed in detail by \citeasnoun{Lowe92}.  Briefly,
such code performs many functions in order to prepare the data for
the final processing by the modeling/estimation software.  Some of these
tasks are rejection of unreliable data by means of statistical tests,
establishment of reference times $t_0$ and frequencies $\omega _0$ for
the observables, phase calibration via the tones that were injected soon
after signal detection, and removal of any rudimentary model information
that was introduced during correlation.  The core task
of the post-correlation software is to take the set of phase samples
$\phi(\omega_i,t_j)$ from the various frequency channels $\omega_i$
and times $t_j$, and to fit the set of $\phi(\omega_i, t_j)$ with three
parameters: the phase $\phi_0$, the group delay $\tau_{\rm gd}$, and the
phase rate $\dot\tau_{\rm pd}$.  To accomplish this, the set of
$e^{i\phi(\omega_i,t_j)}$ are first Fourier transformed from the frequency
and time domain to the delay and delay rate domain respectively.
Figure~\ref{frngamp} shows the result of this operation for a
high-SNR observation.  Next the delay and delay rate domain data are
searched for peaks (at 763.784$\thinspace$ns, $-0.3878\thinspace$ps/s for
the example in Fig.~\ref{frngamp}).  This peak supplies the $a~priori$
values for a least squares fit which determines the observables
$\phi_0$, $\tau_{\rm gd}$, and $\dot \tau _{\rm pd}$ (phase, group
delay, and phase delay rate).

\vbox{
\begin{figure}
\vskip 15pt
\hskip 2.0in
\epsfysize=2.0in
\epsffile{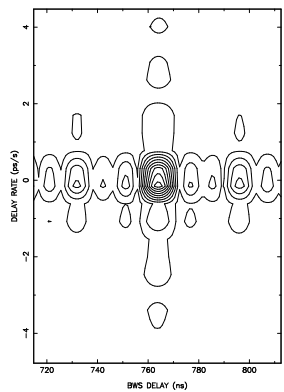}
\vskip 15pt
\caption{ Fourier transform of the phase $\phi(\omega_i,t_j)$ into the
delay, delay rate domain. }
\label{frngamp}
\end{figure}
}

  These phase-derived observables are determined (for phase $\phi$ and
circular frequency $\omega$) from a bilinear least-squares fit to the
measured phases $\phi(\omega,t)$:
\begin{equation}
\phi(\omega,t) = \phi_0(\omega_0,t_0) +
 {{\partial\phi} \over {\partial\omega}} (\omega - \omega_0) +
 {{\partial\phi} \over {\partial t}} (t - t_0) ~,
\label{delobsvs}
\end{equation}
where the phase, group delay, and phase rate are respectively defined as
\begin{equation}
\tau_{\rm pd} = \phi_0 / \omega, ~~~~~
\tau_{\rm gd} = {{\partial\phi} \over {\partial\omega}}, ~~~~~
\dot \tau_{\rm pd} = {{\partial\phi} \over {\partial t}} ~.
\end{equation}
Formal uncertainties of each observable are also produced (for the example
in Fig.~\ref{frngamp}, $\sigma_{\tau_{\rm gd}} = 0.010\thinspace$ns,
$\sigma_{\dot \tau_{\rm pd}} = 0.0008\thinspace$ps/s).
When data at two distinct frequencies are present ($e.g.$, S and X bands),
charged particle (ionospheric) effects are removed by applying a simple
model of dispersion, and combined (S/X) observables are formed (see
Sec.~\ref{ion}, Eq.~\ref{eqiontau}).  Problems with resolving
2$\pi$ ambiguities in the signal phase over large distances usually
preclude the direct use of the phase delay observable $\tau_{\rm pd}$.
This problem can be overcome in differenced observations of pairs of
nearby sources, for which ``phase connection'' is possible
\cite{Marcaide94}.
Under routine conditions, the uncertainties in the delay observables
are on the order of 10 picoseconds.  Numerous aspects of hardware,
instrumentation, and software contribute to this limit \cite{Rogers91}.
During correlation and post-correlation processing, the amount of data
is compressed by a factor of $10^9$, from terabits to kilobits.

  To summarize, the interferometer is capable of producing four data types:
phase delay, group delay, phase delay rate, and amplitude.  The time rate
of change of group delay cannot be measured accurately enough to be
useful for geodetic or astrometric purposes.  The delay models discussed
in Secs.~\ref{gdel}, \ref{clk}, and \ref{atmo} are directly applicable
to either group delay or to phase delay.  Amplitudes are not usually modeled
in astrometric/geodetic VLBI [for an exception see \citename{Charlot90a}
~\citeyear{Charlot90a,Charlot90b}].
Phase delay rate models may be built from the delay models as discussed
immediately below.

  The model for the time rate of change of phase delay must be
constructed from either analytical or numerical time derivatives.
If the latter route is taken, then only models of delay are needed.
The model phase delay rate $\dot \tau _{\rm pd} (t)$ is approximated
as a finite difference $R$ by
\begin{equation}
R = { \lbrack \tau _{\rm pd} (t + \Delta) - \tau _{\rm pd}
(t - \Delta) \rbrack } / ( 2 \Delta ) ~.
\label{eqpdr}
\end{equation}
In the limit $\Delta \rightarrow 0$, this expression for differenced phase
delay approaches the instantaneous time rate of change of phase delay
(fringe frequency) at time $t$.  In practice, $\Delta$ must be large
enough to avoid roundoff errors,
but should also be small enough to allow $R$ to be a
reasonably close approximation to the instantaneous delay rate.  A
suitable compromise appears to be a $\Delta$ in the vicinity of 0.1 second.
Fortunately, the capability to model interferometer performance accurately
is relatively insensitive to the choice of $\Delta$ over a fairly wide
range of values.  To be specific, if $\tau _{\rm pd}(t)$ is expanded to third
order in $\Delta$, the numerically evaluated rate becomes
\begin{equation}
R = \dot \tau _{\rm pd}(t) ~+~ \overdots{\tau}_{\rm pd}(t) ~ {{\Delta}^2 / 6} ~.
\end{equation}
Thus the error made in approximating the time derivative as $R$ is
proportional to $\overdots{\tau}_{\rm pd} {\Delta}^2$.  This term amounts
to a few times $10^{-16}$ s/s for a 10,000 km baseline and $\Delta=0.1$
s.  Comparisons of numerical and analytic derivatives have demonstrated
equivalence of the two methods to better than $10^{-15}$ s/s.

\subsection{ Modeling and parameter estimation }
\label{analmod}

  Prior to the detailed exposition in the next three sections of
the theoretical modeling required by geodetic and astrometric
VLBI, we present a brief summary of the VLBI modeling and parameter
estimation process, its history, and key references.  To form a
perspective, we start by listing in Table~\ref{errbudg} the approximate
relative importance of the various portions of the VLBI model.
Each tabulated maximum delay is given in distance
units, and represents the potential model error that would be caused by
its complete omission from the theoretical model.  Of the seven major
components in this table, the purely geometric delay clearly dominates,
since it can approach one Earth diameter.  Aberration
effects due to the Earth's orbital motion are four orders of magnitude
smaller.  The remaining components range over several orders of magnitude,
with Earth orientation, antenna structure, instrumentation, and atmospheric
effects being the most important.  Most portions of the station coordinate
and Earth orientation models become significant only in analyses of
data extending over time spans much longer than a typical observing session
of 24 hours.  The maximum delays arising from these models were therefore
estimated for data extending over one year.  Models of the last four
major categories, on the other hand, can exhibit their total variation
over the duration of one experiment, and their maximum contributions
were therefore estimated over a time span of one day.

  We also attempt to provide an ``error budget'' for the VLBI model by
tabulating rough estimates of modeling uncertainties in the last column
of Table~\ref{errbudg}.  Extreme situations are considered for these
estimates: Earth-sized baselines, extended sources, large antenna axis
offsets, uncalibrated instrumentation, and observations at low elevations.
It will be noted that many of the components can be modeled to accuracies
that are several orders of magnitude better than their maximum
contributions.  Thus, while this table indicates the general significance
of the physical effects, their ranking does not necessarily reflect the
uncertainties of their contributions to VLBI observables.  Interactions
between theory and experiment are quite important in interpretation of
VLBI measurements: empirical analyses point out inadequate parts of the
model, which can then be improved.  Such iteration has taken
place to varying degrees for different model components.  Presently,
each of the three basic aspects of a VLBI measurement (source, intervening
medium, and receiver) are seen to impose limits to VLBI resolution and
accuracy that are roughly of comparable magnitude.

  The effort expended in improvement of various model components has by
no means been proportional to their relative contributions to the
observables.  For example, the ionosphere can be nearly totally
eliminated as an error source if the measurements are performed
at two widely separated frequencies.  In recent years, much more
attention has focused on station motion effects, despite their
smaller size.  While recent methods of treating Earth orientation
and troposphere errors have produced substantial reductions of
their contributions to the VLBI error budget, they nevertheless
remain as two of the leading error sources in contemporary
VLBI analyses.  Two other important errors are due to incomplete
modeling of the time-dependent structures of the
radio sources and receiving antennas.  Both may amount to
tenths of a milliarcsecond (mas) (1 nrad, many mm).  For day-long
Earth-based measurements, the dominant error is due to the
troposphere: commonly used models of atmospheric propagation can be
incomplete at nearly the 10-20$\thinspace$mm level.  In fact, present
troposphere modeling is probably less accurate than the formal precision
of the VLBI observables.  Details of tidal motions on the Earth's
surface have also not been fully characterized at the few mm level.
For longer periods (on the order of 1 year), aperiodic time-dependent
processes come into play.  At present, these are likewise not well enough
understood $a~priori$ at the nrad or mm level.  The error budget of
Table~\ref{errbudg} leads to a (conservative) total modeling uncertainty
of approximately 30 mm.  More complete models of the Earth's tidal response
and atmosphere, the physical processes in quasars, and the mechanical
response of large antenna structures, will be required to realize
the full potential accuracy of the VLBI technique and to reduce this
limit to below 10 mm.  It is hoped that this review provides a good
point of departure for such model improvements.

  The $a~priori$ model is refined by estimating model parameter
corrections which best fit the data.  In astrometric and geodetic
applications of radio interferometry, the values of the observables from
many different radio sources are processed by a multiparameter
least-squares estimation algorithm to extract refined model parameters.
As the accuracy of the observables improves with advances in instrumentation,
increasingly complete models of the delays and delay rates need to be
developed, which will be the thrust of this review.

  Linear least-squares methods are employed to extract the best
estimates of model parameters from the VLBI observables.  Fortunately,
nonlinearities in this mathematical problem are small, and the $a~priori$
knowledge of most aspects of the model is sufficiently good to permit
efficient estimation via linearized least squares.  The only operational
complication is caused by the sheer volume of data: more than two
million pairs of delay and delay rate observations have been
accumulated in experiments through 1997.  This necessitates
careful bookkeeping procedures to keep computing requirements manageable
when tens or hundreds of thousands of parameters are estimated in
simultaneous analyses of all extant data.  The most important of these
is the separation of the parameters into ``local'' and ``global'' categories
in order to minimize the need to invert very large matrices.  For example,
in processing data which extend over many observing sessions, the clock
and atmosphere model parameters apply only to individual sessions, while
station and source coordinates might be determined by the entire data set.
Specialized least-squares or filtering methods are usually employed in
VLBI parameter estimation.  Particular attention needs to be paid to
numerical stability, and to retaining correlation information in
experiments that are widely separated in time.  The ``square root
information filter'' (SRIF) \cite{Bierman77} is one such formalism
for achieving these goals.

  Severe limitations are imposed on the information content of the
VLBI observables by the fact that the geometric delay measures only the
scalar product of the baseline and source direction vectors.  Thus the
data determine the $relative$ directions of the sources and baseline,
and only the orientation of the whole Earth relative to the sources is
strongly determined.  Weak ties of the baseline to geographic features
of the Earth are provided by effects such as the Earth's motion in its
orbit (aberration), tidal displacements, and tropospheric delay.  These
fix the baseline in the terrestrial frame at an uncertainty level of a
few tenths of an arcsecond.  A more rigid link can only be provided by
the introduction of a reference station with known coordinates ($e.g.$,
from satellite techniques) in the terrestrial reference frame, and
the adoption of a conventional origin for celestial orientation ($e.g.$,
the right ascension of one reference source).

  A further complication in analysis is introduced by degeneracy in the
parameter space of the multi-faceted model.  Linear combinations of a
subset of the parameters describing one portion of the model may produce
identical changes in the observables to those described by parameters of
another portion of the model.  The tidal displacements of a global network
of stations provide an illustrative example: certain linear combinations of
station tidal motions are equivalent to the three components of rotation of
the Earth as a whole.  All such potential degeneracies must be identified
and accounted for in parameter estimation procedures.

  The account of the physical models presented here is intended to be
helpful in providing a general understanding of the analysis of VLBI
observations.  Since modeling is intimately connected to the software
that is used for data analysis, we take advantage of our own experience
and adhere to the outline of the mathematical description of model
implementation for the multiparameter estimation code
``Masterfit/Modest'' \cite{Fans83,Sovers96}
in the succeeding sections.  The theoretical foundation of this
large-scale computer implementation was laid by J.~G. Williams
~\citeyear{Wllms70a},
who also developed many of the original algorithms.  Software and model
development has continued at the Jet Propulsion Laboratory since
the 1970s.  Independent code of similar scope and ancestry is the
``Calc/Solve'' package, which was developed at the Goddard
Space Flight Center (GSFC) during the same time period \cite{GSFC81}.
This software has been used in the National Aeronautics and Space
Administration (NASA) Crustal Dynamics and Space Geodesy Projects
(CDP and SGP), National Oceanic and Atmospheric Administration (NOAA)
International Radio Interferometric Surveying (IRIS), and National Earth
Orientation Service (NEOS) analyses.  It is an outgrowth of the original
algorithms of \citeasnoun{Hinter72} and \citeasnoun{Robrtsn75}, and has
been evolving predominantly at the Goddard Space Flight Center.
Numerous mutations of Calc/Solve over the years have resulted in at
least three related, but not identical, packages used for VLBI analyses
in Germany \cite{Campbell88}, Japan \cite{Kunimori}, and Spain
\cite{Zarraoa89}.  More recently, independent software has
been developed in Ukraine \cite{Yatskiv91}, France
\cite{Gontier}, Norway \cite{Andrsn95}, and Russia \cite{Petrov95}.
There have been occasional comparisons between the various codes
\cite{SovMa85,Gontier,Rius92}.
These comparisons have been beneficial in exposing discrepancies,
correcting software problems, and clarifying the physical models.
Such discrepancies have been surprisingly minor, given the complexity
of these large computer programs.

\section{ GEOMETRIC DELAY MODELS }
\label{gdel}

   The geometric delay is defined as the difference in time of arrival
of a signal
at two geometrically separate points which would be measured by perfect
instrumentation, perfectly synchronized, if there were a perfect vacuum
between the observed extragalactic or Solar-System source and the
Earth-based instrumentation.  For Earth-fixed baselines, this delay is
at most the light time of one Earth radius (20 milliseconds) due to
non-transparency of the Earth.  It can change rapidly (by as much as
3 $ \mu $s per second) as the Earth rotates.  While VLBI experiments
are occasionally carried out with more than ten participating stations,
the correlator generates observable delays and their time rates of
change independently for each baseline connecting a pair of stations.
Without loss of generality, the delay model can thus be developed for
a single baseline involving only two stations, which is what we
present here.  The geometric component is by far the
largest component of the observed delay.  The main complexity of this
portion of the model arises from the numerous coordinate transformations
that are necessary to relate the celestial reference frame used for
locating the radio sources to the terrestrial reference
frame in which the station locations are represented.

   We use the term ``celestial reference frame'' to denote a reference
frame in which there is no net proper motion of the extragalactic radio
sources which are observed by the interferometer. This is only an
approximation to some truly ``inertial'' frame.  Currently, this celestial
frame is a Solar-System-barycentric (SSB), equatorial frame with the
equator and equinox of Julian date 2000 January 1.5 (J2000.0) as defined
by the 1976 International Astronomical Union (IAU) conventions,
including the 1980 nutation series \cite{Sdlmn92,Kaplan81}.
In this equatorial frame, some definition of the origin of right
ascension must be made.  The right ascension is nearly arbitrary
(neglecting small aberrational, tidal, and gravitational bending effects
due to the Earth's changing position within the Solar System).  Any
definition differs by a simple rotation from any other definition.
The important point is that consistent conventions must be used
throughout the model development.  The need for this consistency has
recently led to a definition of the origin of right ascension in terms
of the positions of extragalactic radio sources \cite{Ma97}.
Interferometric observations of both natural radio sources and
spacecraft at planetary encounters will then connect the optical
reference frame and planetary ephemeris with the radio reference frame
\cite{Newhall86,Folkner94,Lestrade95}.

   Unless otherwise stated, we will mean by ``terrestrial reference
frame'' a frame tied to the mean surface features of the Earth.
The most common such frame, which we use here, is a right-handed version
of the Conventional International Origin (CIO) reference system with the
pole defined by the 1903.0 pole.  In practice, the tie is most simply
realized by defining the position of one of the interferometric observing
stations, and then determining the positions of the other stations under
this constraint.  This tie further requires that the determinations of
Earth orientation agree on the average with measurements of the Earth's
orientation by the International Earth Rotation Service (IERS)
~\citeyear{IERS96a} [and its predecessor,
Bureau International de l'Heure (BIH)~\citeyear{BIH83}]
over some substantial time interval ($\approx$~years).
Such a procedure, or its functional equivalent, is necessary to tie the
measurement to the Earth, since the interferometer is sensitive only to
the baseline vector.  With the exception of minor tidal and
tropospheric effects, the VLBI technique does not have
any preferred origin relative to the structure of the Earth. The
rotation of the Earth does, however, provide a preferred direction in space
which can be associated indirectly with its surface features.

   In contrast, geodetic techniques which involve the use of artificial
satellites or the Moon are sensitive to the center of mass of the Earth
\citeaffixed{Vigue92}{$e.g.$} as well as its spin axis.  Thus, such
techniques require only a definition of the origin of longitude.  Laser
ranging to the retroreflectors on the Moon allows a realizable definition
of a terrestrial frame, accurately positioned relative to a celestial
frame which is tied to the planetary ephemerides \cite{Folkner94}.
The required collocation of the laser and
VLBI stations is being provided by Global Positioning Satellite
(GPS) measurements of baselines between VLBI and laser sites
starting in the late 1980s \citeaffixed{RayM91}{$e.g.$}.
Careful definitions and experiments of this sort are required to realize
a coordinate system of centimeter accuracy.

   Except for sub-centimeter relativistic complications caused by the
locally varying Earth potential (as discussed below), construction
of the VLBI model for the observed delay can be summarized in 7 steps as:

\newcounter{model}
\begin{list}%
{\arabic{model}.}{\usecounter{model}\setlength{\rightmargin}{\leftmargin}}
\item Specify the proper locations of the two stations as measured
in an Earth-fixed frame at the time that the wave front intersects
station 1.  Let this time be the proper time $ t_{1}^{\prime} $
as measured by a clock in the Earth-fixed frame.
\item Modify the station locations for Earth-fixed effects such as solid
Earth tides, tectonic motion, and other local station and antenna motion.
\item Transform these proper station locations to a geocentric celestial
coordinate system with its origin at the center of the Earth, and moving
with the Earth's center of mass (but not rotating).  This is a composite
of 12 separate rotations.
\item Perform a Lorentz transformation of these proper station
locations from the geocentric celestial frame to a frame at rest relative
to the center of mass of the Solar System, and rotationally aligned with
the celestial geocentric frame.
\item In this Solar-System-barycentric (SSB) frame, compute the proper
time delay for the passage of the specified wave front from station 1
to station 2.  Correct for source structure.  Add the effective change in
proper delay caused by the differential gravitational retardation of the
signal within the Solar System.
\item Perform a Lorentz transformation of this SSB geometric delay
back to the celestial geocentric frame moving with the Earth.
This produces the adopted
model for the geometric portion of the observed delay.
\item To this geometric delay, add the contributions due to clock
offsets (Sec.~\ref{clk}), to tropospheric delays, and to the effects
of the ionosphere on the signal (Sec.~\ref{atmo}).
\end{list}

   As indicated in step 5, the initial calculation of delay is
carried out in a frame at rest relative to the center of mass of
the Solar System (SSB frame).  First, however, steps 1 through 4
are carried out in order to relate proper locations in the
Earth-fixed frame to corresponding proper locations in the SSB frame.
Step 4 in this process transforms
station locations from the geocentric celestial frame to the SSB
frame.  This step incorporates special-relativistic effects to all
orders of the velocity ratio $v/c$.  From a general relativistic
point of view, this transformation is a special relativistic
transformation between proper coordinates of two local frames
(geocentric and SSB) in relative motion.  For both frames,
the underlying gravitational potential can be taken approximately
as the sum of locally constant potentials caused by all masses
in the Solar System.  The complications caused by small local
variations in the Earth's potential are discussed below.
The initial proper delay is then computed (step 5) in the SSB frame
on the basis of these SSB station locations and an $a~priori$
SSB source location.  A small proper-delay correction is then
applied to account for the differential gravitational retardation
introduced along the two ray paths through the Solar System,
including retardation by the Earth's gravity.  A final Lorentz
transformation (including all orders of $v/c$) then transforms the
corrected SSB proper delay to a model for the observed delay in the
celestial geocentric reference frame.  Note that, rather than
transforming between frames, some formulations account for special
relativistic effects as an $ad~hoc$ term expanded to order $(v/c)^2$.
While this may be convenient for calculating the delay, it obscures
the important physical insights that are gained by a clear definition
of the reference frame used for each step.

   Since the Earth's gravitational potential $U_{\rm E}$ varies slightly
through the Earth ($\Delta U_{\rm E}/c^2 \approx 3.5 \times 10^{-10}$ from
center to surface), the specification of proper distance for a baseline
passing through the Earth is not as straightforward with
respect to the Earth's potential as it is with respect to the essentially
constant potentials of distant masses.  To overcome this difficulty,
VLBI-derived station locations are now customarily specified in terms of
the ``TDT (Terrestrial Dynamic Time) spatial coordinates'' that are
used in Earth-orbiter models.
A proper length that corresponds to a modeled baseline can be obtained
through appropriate integration of the local metric
\cite{SSal91}.  Such proper lengths deviate slightly
($\leq$~3~mm) from baselines modeled on the basis of the TDT convention in
the worst case (a full Earth diameter).  In practice, such a conversion is
not necessary if baseline measurements obtained by different investigators
are reported and compared in terms of TDT spatial coordinates.

   The model presented here has been compared \cite{TrhftT91}
with the ``1-picosecond'' relativistic model for VLBI
delays developed by \citeasnoun{SSal91}.  When reduced to
the same form, these two models are identical at the picosecond level,
term by term, with one exception.  \cite{TrhftT91} show that
a correction is needed to the Shahid-Saless $et~al.$ modeling of the
atmospheric delay in the SSB frame.  This correction changes the
Shahid-Saless $et~al.$ result by as much as 10 picoseconds.  Comparisons
of this formulation with the Goddard Space Flight Center implementation
of the ``consensus'' model \cite{Eubanks91,IERS96b} show
agreement to better than 1 ps, suggesting that the atmospheric delay
problem has been solved in the consensus model.  The remainder of this
section provides the details for the first six steps of the general
outline given above.

\subsection{ Wave front arrival time difference }
\label{arrtim}

The fundamental part of the geometric model is the calculation (step 5
above) of the time interval for the passage of a wave front from station 1
to station 2.  This calculation is actually performed in a coordinate
frame at rest relative to the center of mass of the Solar System.  The
SSB coordinate system is used because source positions, planetary orbits,
and motions of interplanetary spacecraft are most naturally modeled in
this frame.  This part of the
model is presented first to provide a context for the subsequent sections,
all of which are heavily involved with the details of time definitions and
coordinate transformations.  We will use the same subscript and superscript
notation which is used in Sec.~\ref{emot} to refer to the station locations
as seen by an observer at rest relative to the center of mass of the Solar
System.

   First, we calculate the proper time delay that would be observed if
the wave front were planar.  This calculation is next generalized to a
curved wave front, and finally we take into account the incremental effects
which result from the wave front propagating through the various
gravitational potential wells in the Solar System.

\subsubsection{ Plane wave front }
\label{pwf}

  Consider the case of a plane wave moving in the direction
${\bf{\widehat k}}$ with station 2 having a velocity
${\bbox{\beta}}_2$ as shown in Fig.~\ref{figpw}.  As mentioned
above, distance and time are to be represented as proper coordinates
in the SSB frame.  The speed of light $c$ is set equal to 1 in the
following formulation.  The proper time delay is the time it takes
the wave front to move the distance $l$ at speed $c$.  This distance
is the sum of the two solid lines perpendicular to the wave front
in Fig.~\ref{figpw}:
\begin{equation}
t_2-t_1 =
{\bf{\widehat k}} \cdot [ {\bf{r}}_2 (t_1) - {\bf{r}}_1 (t_1 )]
+ {\bf{\widehat k}} \cdot {\bbox{\beta}}_2 [ t_2 - t_1 ] ~.
\end{equation}
Note that station 2
has moved since $t_1$: the second term represents the distance
that station 2 moves before receiving the signal at $t_2$.
This leads to the following expression for the geometric delay:
\begin{equation}
t_2-t_1 =
{ { {\bf{\widehat k}} \cdot [ {\bf{r}}_2 (t_1) - {\bf{r}}_1 (t_1)] }
\over { 1- {\bf{\widehat k}} \cdot {\bbox{\beta}}_2 } } ~.
\label{eqgdel}
\end{equation}
The baseline vector in the SSB frame, ${\bf{r}}_2 (t_1) - {\bf{r}}_1 (t_1)$,
is computed from proper station locations Lorentz transformed from the
geocentric celestial frame, using Eq.~(\ref{eqdr}) in Sec.~\ref{emot}.
The source unit vector in the direction of signal propagation
(source to receiver)
\begin{equation}
{\bf{\widehat k}} = \left( \matrix {
-\cos \alpha \cos \delta \cr
-\sin \alpha \cos \delta \cr
-\sin \delta \cr } \right)
\label{celvec}
\end{equation}
is calculated from the SSB frame angular source coordinates $\alpha$
(right ascension) and $\delta$ (declination).

\vbox{
\begin{figure}
\vskip 15pt
\hskip 2.0in
\epsfysize=2.0in
\epsffile{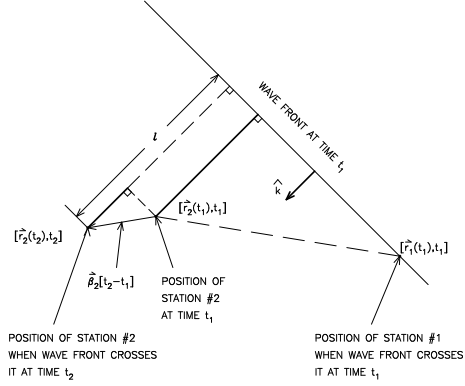}
\vskip 15pt
\caption{ Geometry for calculating the transit time of a plane wave
front.  Some refinements have been added to the schematic diagram
of Fig.~1. }
\label{figpw}
\end{figure}
}

\subsubsection{ Curved wave front }
\label{cwf}

In the case of a signal generated by a radio source within the Solar
System it is necessary to include the effect of the curvature of the
wave front.  As depicted in Fig.~\ref{figcw}, let a source irradiate
two Earth-fixed stations whose positions are given by
$ {\bf{r}}_{1,2} (t) $ relative to the Earth's center. The position
of the Earth's center in the SSB frame $({\bf{R}} _c (t_1))$
as a function of signal reception time $t_1$ at station 1 is
measured relative to the position of the emitter at the time of emission
($t_e $).  While this calculation is actually done in the Solar System
barycentric coordinate system, the development that follows is by no
means restricted in applicability to that frame.
\vbox{
\begin{figure}
\vskip 15pt
\hskip 1.5in
\epsfysize=3.0in
\epsffile{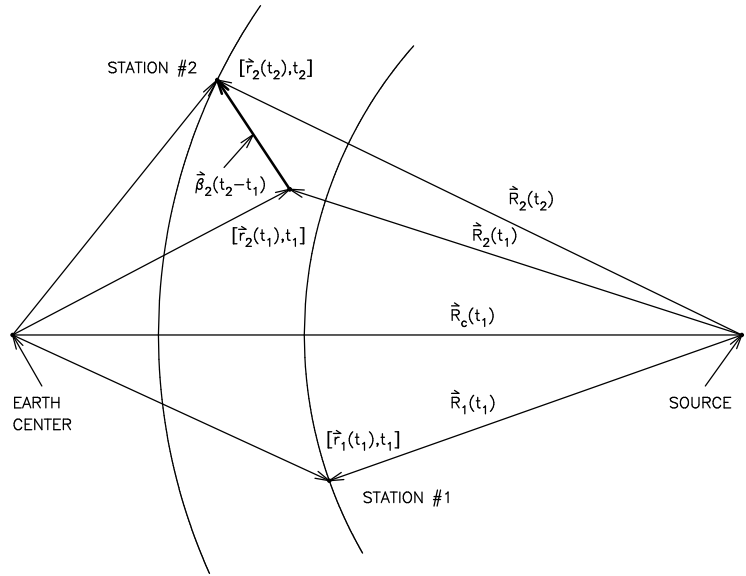}
\vskip 15pt
\caption{ Geometry for calculating the transit time of a curved wave front.
It is only necessary to consider curvature effects for sources closer than
$\approx$30 light years. }
\label{figcw}
\end{figure}
}

   Suppose that a wave front emitted by the source at time $ t_e $
reaches station 1 at time $ t_1 $ and arrives at station 2 at time
$t_2$.  The geometric delay in this frame will be given by
\begin{equation}
\tau = t_2 - t_1 =
| {\bf{R}}_2 ( t_2 ) | - | {\bf{R}}_1 ( t_1 ) | ~,
\end{equation}
where all distances are again measured in units of light travel time.
If we approximate the velocity of station 2 by
\begin{equation}
{\bbox{\beta}}_2 = { { {\bf{R}}_2 (t_2 ) - {\bf{R}}_2 (t_1 ) }
\over { t_2 - t_1 } }
\end{equation}
and use the relation ($i$=1,2)
\begin{equation}
{\bf{R}}_i (t_1) = {\bf{R}}_c (t_1) + {\bf{r}}_i (t_1)
\end{equation}
we obtain:
\begin{eqnarray}
\tau &=& | {\bf{R}}_c (t_1) + {\bf{r}}_2 (t_1) + {\bbox{\beta}} _2
\tau | - | {\bf{R}}_c (t_1) + {\bf{r}}_1 (t_1) | \nonumber \\
&=& R_c (t_1) ~\bigl[ ~ | {\bf{\widehat R}} _c + {\bbox{\varepsilon}} _2
| - | {\bf{\widehat R}} _c  + {\bbox{\varepsilon}} _1 | ~ \bigr] ~,
\label{eqcurv}
\end{eqnarray}
where
\begin{eqnarray}
{\bbox{\varepsilon}} _1 &=& {\bf{r}}_1 (t_1) / R_c (t_1) \nonumber \\
{\bbox{\varepsilon}} _2 &=& \lbrack {\bf{r}}_2 (t_1) + {\bbox{\beta}}_2 \tau
\rbrack / R _c (t_1) ~.
\label{eqeps}
\end{eqnarray}
For $ {\varepsilon}_1 $ and $ {\varepsilon}_2 \leq 10^{-4} $, only terms
of order $ {\varepsilon}^3 $ need to be retained in expanding the
expression for $\tau$ in Eq.~(\ref{eqcurv}) for computational purposes
on machines that employ 15-16 decimal digit representations.  This gives:
\begin{equation}
\tau = { { { \bf{\widehat R}} _c {\bf{\cdot}} ~\lbrack {\bf{r}}_2 (t_1)
- {\bf{r}}_1 (t_1) \rbrack } \over { \lbrack 1 - {\bf{\widehat R}}
_c {\bf{\cdot}} {\bbox{\beta}} _2 \rbrack } } ~+~
{ { R_c \Delta _c ( \tau ) } \over { 2 ~ \lbrack 1 - {\bf{\widehat R}} _c
{\bf{\cdot}} {\bbox{\beta}} _2 \rbrack } } ~,
\label{eqexpcurv}
\end{equation}
where to order $ \varepsilon ^3 $
\begin{equation}
\Delta _c (\tau) = \bigl[\varepsilon _2^2 - \varepsilon _1^2 \bigr] - \bigl[
{ ( {\bf{\widehat R}} _c {\bf{\cdot}} {\bbox{\varepsilon}} _2 ) } ^2 +
{ ( {\bf{\widehat R}} _c {\bf{\cdot}} {\bbox{\varepsilon}} _1 ) } ^2 +
{ ( {\bf{\widehat R}} _c {\bf{\cdot}} {\bbox{\varepsilon}} _2 ) } ^3 -
{ ( {\bf{\widehat R}} _c {\bf{\cdot}} {\bbox{\varepsilon}} _2 ) }
\varepsilon _2^2 -
{ ( {\bf{\widehat R}} _c {\bf{\cdot}} {\bbox{\varepsilon}} _1 ) } ^3 +
{ ( {\bf{\widehat R}} _c {\bf{\cdot}} {\bbox{\varepsilon}} _1 ) }
\varepsilon _1^2 \bigr] ~.
\label{eqexpcurvd}
\end{equation}
The first term in Eq.~(\ref{eqexpcurv}) is just the plane wave approximation,
$i.e.$, as
$ R_c \rightarrow \infty $, $ {\bf{\widehat R}} _c \rightarrow
{\bf{\widehat k}} $, with the second term in brackets
in Eq.~(\ref{eqexpcurvd}) approaching zero as $ { r^2 } / R_c $. Given that
the ratio of the first term to the second term is $\approx {{r} / {R_c}}$,
wave front curvature is not calculable using sixteen-decimal-digit
arithmetic if $R > 10^{16} r$.  For Earth-fixed baselines that are
as long as an Earth diameter, requiring that the effects of curvature be
less than 1 ps $\approx$ 0.3 mm implies that the above formulation
(Eq.~\ref{eqexpcurvd}) must be used for $R < 3\times10^{14}$ km,
or approximately
30 light years.  At the same accuracy level, fourth and higher order terms
in $\varepsilon$ become important for $R < 2\times10^{5}$ km, or
approximately inside the Moon's orbit.  \citeasnoun{Fuku94} has presented a
formulation which should be applicable for Earth-based VLBI with radio
sources as close as the Moon.

  The procedure for the solution of Eq.~(\ref{eqexpcurv}) is iterative for
$ \varepsilon < 10^{-4} $, using the following:
\begin{equation}
\tau _n ~=~ \tau _0 ~+~ { { R_c \Delta _c ( \tau _{n-1} ) } \over { 2
\lbrack 1 ~-~ {\bf{\widehat R}} {\bf{\cdot}} {\bbox{\beta}} _2 \rbrack } } ~,
\end{equation}
where
\begin{equation}
\tau _0 ~=~ \tau _{\rm plane~wave } ~.
\end{equation}
  For $ \varepsilon > 10^{-4} $, directly iterate on
Eq.~(\ref{eqcurv}) itself, using the procedure:
\begin{equation}
\tau _n =
R_c | {\bf{\widehat R}} _c + {\bbox{\varepsilon}} _2 ( \tau _{n-1} ) |
- R_c | {\bf{\widehat R}} _c + {\bbox{\varepsilon}} _1 | ~,
\end{equation}
where again $\tau _0$ is the plane wave approximation.

\subsubsection{ Gravitational delay }
\label{gravdel}

  As predicted by Einstein's theory of general relativity
(\citename{Einst11}, 1911, 1916),
an electromagnetic signal propagating in a gravitational potential is
retarded relative to its travel time in field-free space.
\citename{Shapiro64}~\citeyear{Shapiro64,Shapiro67} pointed out that
this produces both a deviation from
a straight-line path and a time delay.  For VLBI this implies that
the computed differential arrival time value of the signals at
${\bf{r}}_1 (t_1)$ and ${\bf{r}}_2 (t_2)$ must be corrected for
gravitational effects.  For Earth-based experiments the dominant
contribution comes from the Sun.  Gravitational potential effects
and curved wave front effects can be calculated independently of each
other since the former are a small perturbation ($\approx$~8.5
microradians or $\leq 1^{''}\!.75 $), even for Sun-grazing rays.

  Our formulation of the relativistic light travel time is based on
\possessivecite{Moyer71} implementation of the work of \citeasnoun{Tausner66}
and \citeasnoun{Holdr67}.  For the (exaggerated) geometry illustrated in
Fig.~\ref{geodes}, the required correction to $coordinate$ time delay
due to the $p$th gravitating body is
\begin{equation}
{ \Delta _{{\rm G}p} = { { ( 1 + \gamma _{_{\rm PPN}} ) \mu_p } \over {c^3} }
{ { \Biggl[ ~{\rm{ln}} \Biggl( { { r_s + r_2 (t_2) + r_{s2} } \over
{ r_s + r_2 (t_2 ) - r_{s2} } } \Biggr) - {\rm{ln}} \Biggl(
{ { r_s + r_1 (t_1) + r_{s1} } \over { r_s + r_1 (t_1) - r_{s1} } }
\Biggr) ~\Biggr] } } } ~,
\label{eqbend}
\end{equation}
where $ r_{si}~(i=1,2)$ is the separation of station $i$ from source $s$:
\begin{equation}
r_{si} = | {\bf{r}}_i (t_i) - {\bf{r}}_s (t_e) | ~.
\end{equation}
The gravitational constant $\mu_p$ is
\begin{equation}
\mu_p = Gm _p ~,
\end{equation}
where $G$ is the universal gravitational constant, and $ m _p $ is
the mass of body $p$.  A higher-order term ($\propto {\mu_p}^2/c^5$)
contributes to $\Delta _{{\rm G}p}$ only for observations extremely close
to the Sun's limb \cite{IERS92}.  Here $\gamma _{_{\rm PPN}}$ is the
$\gamma$ factor in the parametrized post-Newtonian gravitational
theory \citeaffixed{Misner73}{$e.g.$}.  In general relativity
$\gamma _{_{\rm PPN}} = 1$, but it can be allowed to be an estimated
parameter to permit experimental tests of general relativity, as
originally proposed by \citeasnoun{Shapiro64}.  A dramatic demonstration
of the importance of general relativity in estimating the delay
can be made by ``turning off'' this added delay by setting
$\gamma _{_{\rm PPN}} = -1$ in the model.  This can increase the
discrepancy between theoretical and observed delays by an order of
magnitude because the Sun's gravity induces a bending of 20 nrad
(4 milliarcseconds) even for ray paths 90$^\circ$ removed from its
position relative to the Earth.
\vbox{
\begin{figure}
\vskip 15pt
\hskip 1.0in
\epsfysize=2.5in
\epsffile{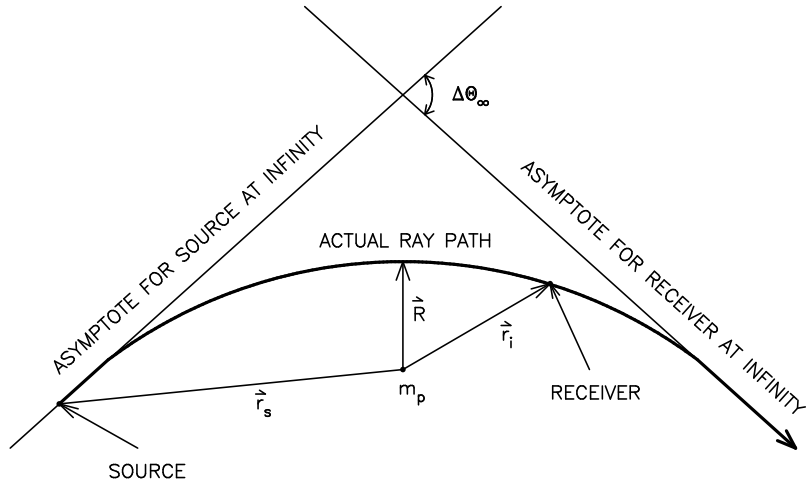}
\vskip 15pt
\caption{ Schematic representation of the geodesic connecting the source
and receiver in the presence of a gravitational mass. }
\label{geodes}
\end{figure}
}

  Depending on the particular source-receiver geometry in a VLBI
experiment, a number of useful approximations are possible for
the correction $\Delta _{{\rm G}p}$ of Eq.~(\ref{eqbend}).  Dropping
the time arguments in Eq.~(\ref{eqbend}), we have:
\begin{equation}
\Delta _{{ \rm G}p} = { { ( 1 + \gamma _{_{\rm PPN}} ) \mu_p } \over {c^3} }
\thinspace {\rm{ln}} \Biggl[ \Biggl( { { r_s + r_2 + r_{s2} } \over
{ r_s + r_1 + r_{s1} } } \Biggr) \Biggl(
{ { r_s + r_1 - r_{s1} } \over { r_s + r_2 - r_{s2} } }
\Biggr) \Biggr] ~.
\label{eqbend1}
\end{equation}
   This formulation is appropriate for the most general geometry, in
which $ r_s \approx r_i \approx r_{si} $.  For the practical case of
Earth-based VLBI with distant sources and closely spaced VLBI receivers,
however, $ { | {\bf{r}}_2 - {\bf{r}}_1 | } / r_1 \rightarrow 0, r_i /
r_s \rightarrow 0 $.  The gravitational time delay $\Delta_{{\rm G}p}$ may
then be expanded in terms of $r_i / r_s , ~r_{si} / r_s$.  Making use
of the relationship
\begin{equation}
r_{si} = { \lbrack {\bf{r}}_s^2 - 2 {\bf{r}}_s \cdot {\bf{r}}_i
+ {\bf{r}}_i^2 \rbrack } ^{1/2}
\approx r_s - {\bf{r}}_i \cdot {\bf{\widehat r}} _s
\end{equation}
leads to
\begin{equation}
\Delta_{{\rm G}p} = { { ( 1 + \gamma _{_{\rm PPN}} ) \mu_p} \over { c^3 } }
\thinspace {\rm{ln}} { \Biggl( { { r_1 + {\bf{r}}_1 \cdot {\bf{\widehat r}} _s }
\over { r_2 + {\bf{r}}_2 \cdot {\bf{\widehat r}} _s } } \Biggr) }
\label{eqbend2}
\end{equation}
for $ r_i / r_s \rightarrow 0 $.  If we define
${\bf{r}}_2 = {\bf{r}}_1 + \Delta {\bf{r}}$, and require that
${ | \Delta {\bf{r}} | } / r_1 \rightarrow 0$ (short baselines) then
\begin{equation}
r_2 ( 1 + {\bf{\widehat r}} _2 \cdot {\bf{\widehat r}} _s )
\approx r_1 ( 1 + {\bf{\widehat r}} _1 \cdot {\bf{\widehat r}}
_s ) + \Delta {\bf{r}} \cdot ( {\bf{\widehat r}} _1
+ {\bf{\widehat r}} _s ) ~.
\end{equation}
Substituting into Eq.~(\ref{eqbend2}) and expanding the logarithm, we obtain:
\begin{equation}
\Delta_{{\rm G}p} = { { { - ( 1 + \gamma_{_{\rm PPN}} ) \mu_p } \over { c^3} } }
\thinspace { { ( {\bf{r}}_2 - {\bf{r}}_1 ) \cdot
( {\bf{\widehat r}} _1 + {\bf{\widehat r}} _s ) } \over
{ r_1 ( 1 + {\bf{\widehat r}} _1 \cdot {\bf{\widehat r}} _s ) } } ~.
\label{eqbend3}
\end{equation}

  These three formulations of Eqs.~(\ref{eqbend1}, \ref{eqbend2}
or \ref{eqbend3}) are computationally appropriate for different
situations.  For Earth-based VLBI observations, the
correction $\Delta_{{\rm G}p}$ is calculated for each of the major bodies
in the Solar System (Sun, planets, Earth, and Moon).  The immense
central mass of the Galaxy (on the order of $10^{11}$ solar masses)
contributes an additional delay.  Assuming a mass concentration of
2/3 in the nucleus and the current estimate of our distance from the
center of 8.5 kpc \cite{Binney}, it can be
estimated that the gravitational influence of the Galactic center on
electromagnetic signals exceeds that of the Sun by a factor of
$\approx$40.  Geometrically, this causes a bending of 4 arc seconds when
ray paths approach within $\approx 10^\circ$ of the center (for the closest
routinely observed radio source).  However, because of the very slow
rotation of the Solar System about the Galactic center (240 million
years), the time variation of this bending is extremely small.
Therefore, for practical purposes, Galactic bending merely produces
a quasi-static distortion of the sky, and can be absorbed into the
source coordinates which form the basis of our model of the celestial
sphere.

  Before the correction $\Delta_{{\rm G}p}$ can be applied to a proper delay
computed according to Eq.~(\ref{eqgdel}), it must be converted from a
coordinate-delay correction to a proper-delay correction appropriate
to a near-Earth frame.  For such proper delays, the gravitational
correction is given \citeaffixed{Hellings86}{$e.g.$} to good approximation by
\begin{equation}
\Delta _{{\rm G}p} ^{\prime} = \Delta _{{\rm G}p} -
 ( 1 + \gamma _{_{\rm PPN}} ) U \tau ~,
\label{eqbndprp}
\end{equation}
where $\tau$ is the proper delay given by Eq.~(\ref{eqgdel}), and where
$U = \mu_p/{r_p c^2}$ is the negative of the gravitational potential of
the given mass
divided by $c^2$, as observed in the vicinity of the Earth ($U$ is a
positive quantity).  The $U\tau$ term is a consequence of the relationship
of coordinate time to proper time, and the $\gamma _{_{\rm PPN}} U \tau$ term
is a consequence of the relationship of coordinate distance to proper
distance.

  The total gravitational correction used is
\begin{equation}
\Delta_G ^{\prime} = \sum\limits_{p=1}^N \Delta_{{\rm G}p}^{\prime} ~,
\end{equation}
where the summation over $p$ is for the $N$ major bodies in the Solar
System.  For the Earth, the ${(1+\gamma_{_{\rm PPN}})U\tau}$ term in
Eq.~(\ref{eqbndprp}) is omitted if one wishes to work in the
``TDT spatial coordinates'' that are used in reduction of Earth-orbiter
data.  The scale factor $( 1 + \gamma _{_{\rm PPN}} ) U$ is approximately
$19.7\times10^{-9}$ for the Sun.  A number of other conventions are
possible.  One of these, which includes the ${(1+\gamma_{_{\rm PPN}})U\tau}$
term for the Earth evaluated at the Earth's surface, yields an additional
scale factor of $1.4\times10^{-9}$.  In either case, the model delay
is decreased relative to the proper delay.  Consequently, all inferred
``measured'' lengths increase by the same fraction relative to proper
lengths ($i.e.$, by 19.7 parts per billion and 21.1 ppb in the two cases).

  Some care must be taken in defining the positions given by ${\bf{r}}_s$,
${\bf{r}}_2 (t_2 )$, and ${\bf{r}}_1 (t_1)$.  We have chosen the origin
to be the position of the gravitational mass at the time of closest
approach of the received signal to that object.  The position ${\bf{r}}_s$
of the source relative to this origin is the position of that source at
the time $t_e$ of the emission of the received signal.  Likewise, the
position ${\bf{r}}_i (t_i)$ of the $i{\rm{th}}$ receiver is its position
in this coordinate system at the time of reception of the signal.  Even
with this care in the definition of the relative positions, we are making
an approximation, and implicitly assuming that such an approximation is
no worse than the approximations used by \citeasnoun{Moyer71} to obtain
Eq.~(\ref{eqbend}).

  Some considerations follow, concerning the use of appropriate times to
obtain the positions of the emitter, the gravitational object, and the
receivers.  For a grazing ray emitted by a source at infinity, using the
position of the $p$th gravitating body at the time of reception of the
signal at station~1 ($t_r$) rather than at the time of closest approach
of the signal to $m_p$ ($t_a$) can cause a substantial error on baselines
with dimensions of the Earth, as shown by the following calculation.  From
Fig.~\ref{trntim}, the distance of closest approach $R$ changes
during the light transit time $t_{\rm tr}$ of a signal from a gravitational
object at a distance $ R_{\rm EG} $ by
\begin{equation}
\Delta R \approx R_{\rm EG} \thinspace \dot{\alpha} \thinspace t_{\rm tr}
= { \dot{\alpha} R_{\rm EG}^2 } / c ~,
\end{equation}
where ${\dot{\alpha}}$ is the time rate of change of the angular position
of the deflecting body as observed from Earth.  Since the deflection is
\begin{equation}
\Theta \approx 2 { { ( 1 + \gamma _{_{\rm PPN}} ) \mu_p }
\over { c^3 } } \Biggl( { c \over R } \Biggr)
\end{equation}
it changes by an amount
\begin{equation}
\Delta \Theta = - \Theta \Biggl( { { \Delta R } \over
{ R } } \Biggr) = - \Theta \Biggl( { { {\dot \alpha } R_{\rm EG}^2 } \over
{ c R } } \Biggr)
\label{eqltrns}
\end{equation}
during the light transit time.

\vbox{
\begin{figure}
\vskip 15pt
\hskip 1.0in
\epsfysize=2.5in
\epsffile{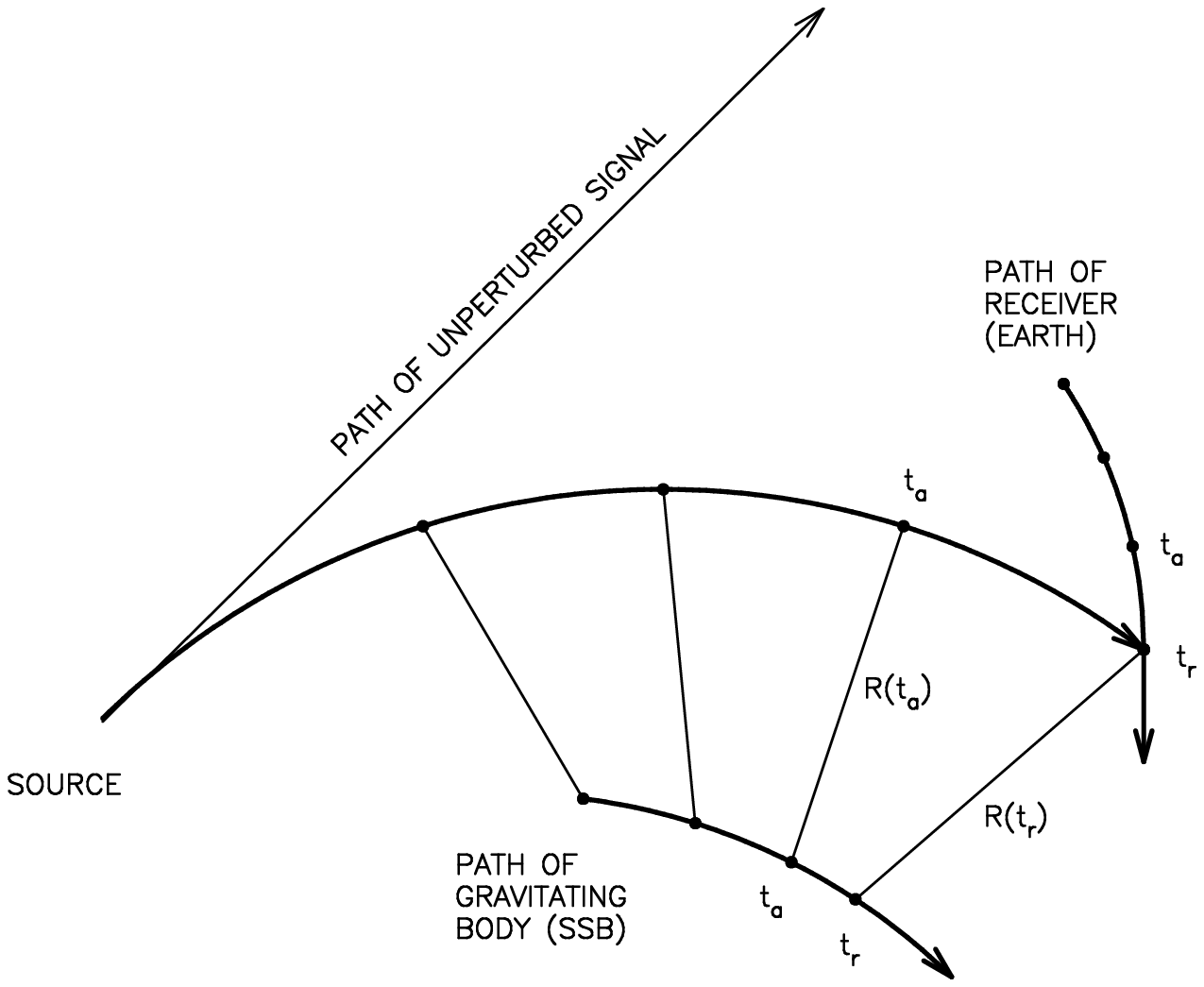}
\vskip 15pt
\caption{ Schematic representation of the motion of a gravitating object
(Sun) during the transit time of a signal from the point of closest approach
to reception by an antenna on Earth.  Both move between the time $t_a$
at closest approach $R(t_a)$ and the time $t_r$ at reception.  This motion
needs to be taken into account for the highest accuracy. }
\label{trntim}
\end{figure}
}

  We consider the two bodies of largest mass in the Solar System: the Sun
and Jupiter.  For rays that just graze the surfaces, their respective
deflections $\Delta \Theta$ are 8470 and 79 nanoradians.  From Solar
System ephemerides, the barycentric angular velocities ${\dot{\alpha}}$
are estimated to be $\approx$~0.05 and 17 nrad/s for the Sun and Jupiter.
(The Sun's motion in the barycentric frame has approximately the orbital
period of Jupiter, $\approx$12 years, with a radius on the order of the
Sun's radius).  Using approximate radii and distances from Earth to
estimate $ R_{\rm EG} $ and $ \Theta $, Eq.~(\ref{eqltrns}) gives 30 nrad
for Jupiter; the corresponding value for the Sun is only 0.05 nrad.  For
a baseline whose length equals the radius of the Earth, $\delta ( \Delta
\Theta ) R_{\rm E}$ is thus approximately 200 and 0.3 mm for Jupiter and
the Sun, respectively.  The effect is much smaller for the Sun in spite
of its much larger mass, due to its extremely slow motion in the
barycentric frame.

  In view of the very rapid decrease of gravitational deflection with
increasing distance of closest approach, it is extremely improbable that
a random VLBI observation will involve rays passing close enough to a
gravitating body for this correction to be of importance.  Exceptions
are experiments that are specifically designed to measure planetary
gravitational delay \citeaffixed{TrhftL91}{$e.g.$}.  In order to
guard against such an unlikely situation in routine work, and to
provide analysis capability for special experiments, it is prudent
to perform the transit-time correction for all planets for all
observations.  To obtain the positions of the gravitational objects,
we employ an iterative procedure, using the positions and velocities
of the objects at signal reception time.  If ${\bf{R}} (t_r)$ is the
position of the gravitational object at signal reception time $t_r$
then that object's position ${\bf{R}} (t_a)$ at the time $t_a$ of
closest approach of the ray path to the object ${\bf{R}}_a$ was
\begin{equation}
{\bf{R}} (t_a) = {\bf{R}} ( t_r ) - (t_r - t_a) {\bf{\overline V}} ~,
\end{equation}
\begin{equation}
t_r - t_a = { | {\bf{R}}_c | } / c ~.
\end{equation}
This correction is done iteratively, using the velocity ${\bf{V}} (t_r)$
as an approximation of the mean velocity ${\bf{\overline V}}$.
Because $v/c \approx 10^{-4}$, the iterative solution,
\begin{equation}
{\bf{R}}_n (t_a) = {\bf{R}} (t_r) ~-~ | {\bf{R}} _{n-1} (t_a) | ~
{\bf{V}} (t_r) / c
\end{equation}
rapidly converges to the required accuracy.

\subsection{ Time information }
\label{time}

  Before continuing with the description of the geometric model, some
definitions must be introduced concerning time-tag information in the
experiments, and the time units which will appear as arguments below.
A general reference for time definitions is the {\it Explanatory Supplement}
\cite{Sdlmn92}.  The data epoch for VLBI observables is taken from the UTC
(Coordinated Universal Time) time tags in the data stream at station 1,
UTC$_1$.  This time is converted to Terrestrial Dynamic Time TDT,
with the conversion consisting of the following components:
\begin{eqnarray}
{\rm TDT} = ({\rm TDT}-{\rm TAI}) &+& ({\rm TAI}-{\rm UTC}_{\rm IERS} ) +
({\rm UTC}_{\rm IERS} - {\rm UTC}_{\rm STD} )
 \nonumber \\
&+& ({\rm UTC}_{\rm STD} - {\rm UTC}_1 ) + {\rm UTC}_1 ~.
\label{eqtdt}
\end{eqnarray}
The four offsets in Eq.~(\ref{eqtdt}) thus serve to convert the station 1
time tags to TDT.  Their meanings are as follows:

\newcounter{ttrns}
\begin{list}%
{\arabic{ttrns}.}{\usecounter{ttrns}\setlength{\rightmargin}{\leftmargin}}
\item TDT$-$TAI is 32.184 seconds by definition;
 TAI (Temps Atomique International) is atomic time.
\item TAI$-$UTC$_{\rm IERS}$ is the offset between atomic and coordinated
 time.  It is a published integer second offset (accumulated leap seconds)
 for any epoch after 1~January, 1972.  Prior to that time,  it is a more
 complicated function, which will not be discussed here since normally
 no observations previous to the mid-1970s are modeled.  The International
 Earth Rotation Service (IERS), its predecessor, Bureau International de
 l'Heure (BIH), and Bureau International des Poids et Mesures (BIPM) are
 the coordinating bodies responsible for upkeep and publication of
 standard time and Earth rotation quantities.
\item UTC$_{\rm IERS}-$UTC$_{\rm STD}$ is the offset in UTC between the
 coordinated time scales maintained by the IERS (BIPM) and
 secondary standards maintained by numerous national organizations.
 For VLBI stations in the U.~S. these secondary standards are those
 of the National Institute of Standards and Technology in Boulder,
 Colorado, and the U.~S. Naval Observatory in Washington, DC.  These
 offsets can be obtained from BIPM Circular T \citeaffixed{BIPM97}{$e.g.$}.
\item UTC$_{\rm STD}-$UTC$_1$ is the (often unknown) offset between UTC
 kept by station 1 and the secondary national standard.  This may be
 as large as several microseconds, and is not precisely known for all
 experiments.  It can be a source of modeling error:  an error
 $\Delta t$ in epoch time causes an error of $\approx B \omega_{\rm E}
 \Delta t = 7.3 \times 10^{-5} $ mm per km of baseline $B$ per $\mu$s
 of clock error $\Delta t$, where $\omega_{\rm E}$ is the rotation rate of
 the Earth.  Even in the extreme case of a 10,000 km baseline, however,
 this amounts to only 0.7 mm per $\mu$s, while present-day clock
 synchronization is usually at least an order of magnitude better,
 at the 0.1-$\mu$s level.
\end{list}

  Coordinated Universal Time and TDT are not convenient for modeling Earth
orientation.  This is normally done in terms of UT1, a quantity that is
proportional to the angle of rotation of the Earth.  The $a~priori$ offset
UT1$-$UTC and the position of the Earth's rotation pole can be obtained by
interpolation of the IERS Bulletin A smoothed values.  However, any other
source of UT1$-$UTC and pole position could be used provided it is
expressed in a left-handed coordinate system (see Sec.~\ref{utpm}).
Part of the documentation for any particular set of results needs to
include a clear statement of what values of UT1$-$UTC and pole position
were used in the data reduction process.

  For the Earth model based on the IAU conventions,
the following definitions are employed throughout \cite{Kaplan81}:
\newcounter{tdef}
\begin{list}%
{\arabic{tdef}.}{\usecounter{tdef}\setlength{\rightmargin}{\leftmargin}}
\item Julian date at epoch J2000.0 = 2451545.0.
\item All time arguments denoted by $T$ below are measured in Julian
 centuries of 36525 days of the appropriate time relative to the epoch
 J2000.0, $i.e.$, $ T~=~{ { (JD - 2451545.0) } / { 36525 } } $.
\item For the time arguments used to obtain precession and nutation, or
 to refer to the Solar System ephemeris, Barycentric Dynamic Time (TDB,
 Temps Dynamique Barycentrique) is used.  This is related to Terrestrial
 Dynamic Time (TDT, Temps Dynamique Terrestre) by the following
 approximation:
\begin{equation}
{\rm TDB} = {\rm TDT} + 0.^{\rm s} 001658 \sin (g + 0.0167 \sin (g) ) ~,
\label{eqtdb}
\end{equation}
where
\begin{equation}
g = 2 \pi { { (357.^{\circ } 528 + 35999.^{\circ} 050 ~T ) } /
{ 360^{\circ} } }
\label{eqmnanom}
\end{equation}
is the mean anomaly of the Earth in its orbit.
\end{list}

  For present analyses of Earth-based VLBI observations, Eq.~(\ref{eqtdb})
is adequate.  \citeasnoun{Moyer81} gave a more accurate relation
between TDB and TDT by accounting for the dominant gravitational
effects of the major planets.  \citeasnoun{Hirayama} and
\citeasnoun{Fairh90} have extended this work to the nanosecond level.
In the future, TDT and TDB will be replaced by two new time scales,
${\rm TCG}$ and ${\rm TCB}$, geocentric and barycentric coordinate time
\cite{Fuku86}.  These will eliminate the need to
rescale spatial coordinates that was discussed in Sec.~\ref{gravdel}.

\subsection{ Station locations }
\label{staloc}

  Coordinates of the observing stations are expressed in the Conventional
International Origin (CIO 1903) reference system, with the reference
point for each antenna defined as in Sec.~\ref{ant}.  The present accuracy
level of space geodesy makes it imperative to account for various types
of crustal motions.  Among these deformations are tectonic motions, solid
Earth tides, ocean effects, and alterations of the Earth's surface due to
local geological, hydrological, and atmospheric processes.  Mismodeled
effects will manifest themselves as temporal changes of the Earth-fixed
baseline.  It is therefore important to model all crustal motions as
completely as possible.  The current level of mismodeling of these motions
is probably one of the leading sources of systematic error (along with
the troposphere) in analyses of VLBI data.

  Evaluation of the time dependence of station locations is most simply
done by estimating a new set of coordinates in the least-squares process
for each VLBI observing session ($e.g.$, Fig.~\ref{pltmot} in
Sec.~\ref{aprot}).  Subsequent fits to these results can then produce
estimates of the linear time rate of change of the station location.
For rigorous interpretation of the statistical significance of the results,
correlations of coordinates estimated at different epochs must be fully
taken into account when calculating uncertainties of the rates.
The advantage of this approach is that the contribution of each session
to the overall time rate may be independently evaluated, since it is
clearly isolated.  Any nonlinear ($e.g.$, seasonal) variations are
easily detected.  Also, no tectonic model
information is imposed on the solution.  An alternative approach is to
model long-period tectonic motion directly, by introducing time rates of
change of the station coordinates as parameters.  The model is linear,
with the position of station $i$ at time $t$, ${\bf{r}}_i = (x_i,y_i,z_i)$,
expressed in terms of its velocity $\dot {\bf{r}}_i$ as
\begin{equation}
{\bf{r_i}} = {\bf{r}}_i^0 + \dot {\bf{r}}_i (t - t_0) ~.
\end{equation}
Here $t_0$ is a reference epoch, at which the station position is
${\bf{r}}_i^0 = (x_i^0, y_i^0, z_i^0)$.

\subsubsection{ Tectonic plate motion }
\label{tect}

  As alternatives to estimating linear time dependence of the station
coordinates from VLBI experiments, several standard models of tectonic
plate motion are available.  They all describe the motion as a
rotation of a given rigid plate (spherical cap) about its rotation
pole on the surface of a spherical Earth.  Time dependence of the
Cartesian station coordinates of station $i$ which resides on
plate $j$ is expressed as
\begin{equation}
{\bf{r}}_i = {\bf{r}}_i^0 + ( {\bbox{\omega}}^j {\bf{\times}}
 {\bf{r}}_i^0 ) (t-t_0) ~,
\end{equation}
where ${\bbox{\omega}}^j = (\omega_{\rm x}^j,~\omega_{\rm y}^j,
~\omega_{\rm z}^j)$ is the angular velocity vector of the plate.

  These tectonic motion models are based on paleomagnetic data spanning
millions of years, but they also provide an excellent quantitative
characterization of present-day plate motions.  This attests to the
smooth character of tectonic motion over immense time scales.
Although the possibility of slow relative motions of the Earth's
land masses was first suggested by Wegener before World War I \cite{Wegen}
mantle convection was identified as the driving mechanism only much
later, and the reality of this phenomenon was not generally accepted
until the 1960s \cite{Menard}.  The first quantitative
global tectonic motion model is due to \citeasnoun{MnstJ78}, and
was also the first to be used in VLBI analyses.  It is denoted AM0-2
in the original paper.  More recent models, denoted Nuvel-1 and NNR-Nuvel-1,
are due to \citeasnoun{DeMets90} and \citeasnoun{ArgG91}, respectively.
In Nuvel-1, the Pacific plate is stationary,
\vbox{
\begin{figure}
\hskip 0.5in
\epsfysize=3.0in
\epsffile{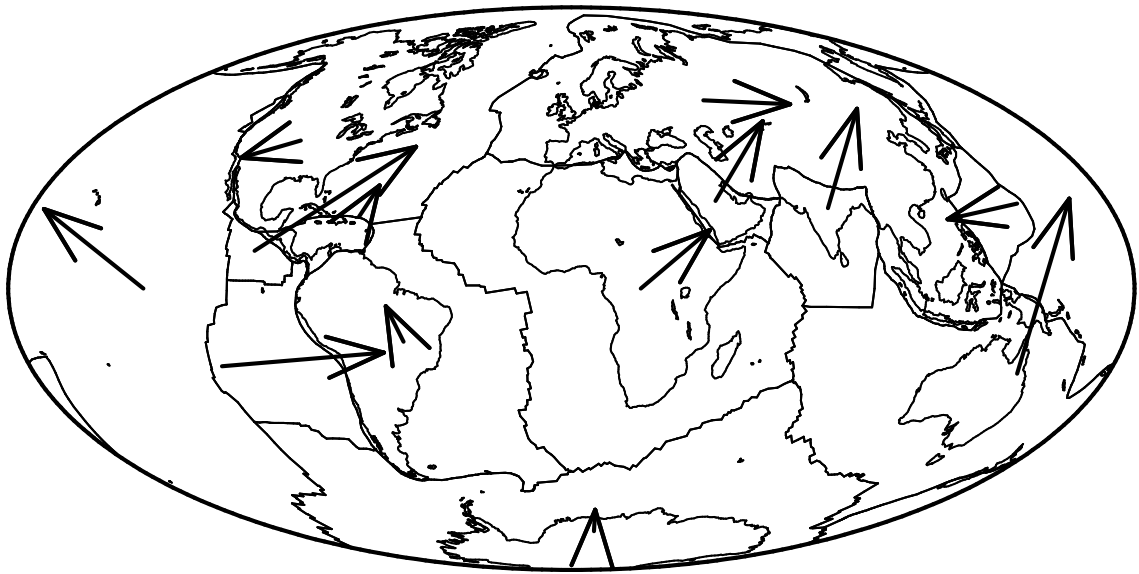}
\vskip 15pt
\caption{ Schematic directions and magnitudes of the motions of the
major tectonic plates in the NNR-Nuvel-1A model.  The longest
arrows represent motions of approximately 10 cm/yr. }
\label{nuvplt}
\end{figure}
}
\noindent while NNR-Nuvel-1 is
based on the imposition of a no-net-rotation (NNR) condition.
Inadequate knowledge of the internal mechanics of the Earth makes
uncertain any absolute determination of plate rotation relative to
the deep interior.  Hence the NNR condition is customarily imposed:
${\bf{v}}\times{\bf{r}}$ integrated over the Earth's surface is
constrained to be zero, where ${\bf{v}}$ is the velocity at point
${\bf{r}}$ on one of the rigidly rotating tectonic plates.
With some notable exceptions, the Nuvel-1 models give rates that are very
close to those of the AM0-2 model.  The AM0-2 India plate has been split
into two: Australia and India, and there are four additional
plates: Juan de Fuca, Philippine, Rivera and Scotia.  A recent revision
of the paleomagnetic time scale has led to a rescaling of the Nuvel-1
rates.  These ``Nuvel-1A'' and ``NNR-Nuvel-1A'' model rates are equal to
the Nuvel-1 and NNR-Nuvel-1 rates, respectively, multiplied by a factor
of 0.9562 \cite{DeMets94}.  Table~\ref{nuv1a} shows the angular
velocities of the 16 tectonic plates in the NNR-Nuvel-1A model.
Thirteen major plate motions are also shown in Fig.~\ref{nuvplt}, with
the velocities depicted approximately to scale.  The largest arrows
represent motions of $\approx$ 10 cm/yr.  A more detailed example of
the effect of tectonic motion on relative station locations is given
in Fig.~\ref{pltmot} of Sec.~\ref{aprot}, which shows the
convergence of the North American and Pacific plates.

\subsubsection{ Tidal station motion }
\label{tid}

  The Earth is not perfectly rigid, and its crust deforms in response
to the gravitational attraction of other massive bodies.  The most
important of these are the Sun and Moon, whose periodic orbits induce
time-varying displacements of the Earth's crust.  Such motions, with
periodicities ranging from hours to years, are called tidal effects.
Tides produce station displacements that can be far larger than those
caused by tectonic motions in a year, necessitating their inclusion in
models of VLBI observables.  The tidal displacements can be classified
into several categories, of which contemporary VLBI models normally
include four.  In the standard terrestrial coordinate system, these
tidal effects modify the station location ${\bf{r}}_0$ by an amount
\begin{equation}
{\bf{\Delta}} = {\bf{\Delta}}_{\rm sol} + {\bf{\Delta}}_{\rm pol} +
{\bf{\Delta}}_{\rm ocn} + {\bf{\Delta}}_{\rm atm} ~,
\label{eqtideff}
\end{equation}
where the four terms are due to solid Earth tides (the direct primary
effect), pole tide, ocean loading, and atmosphere loading (secondary
effects), respectively.  Other Earth-fixed
effects ($e.g.$, glacial loading) can be incorporated by extending the
definition of ${\bf{\Delta}}$ to include additional terms.
All four tidal effects are most easily calculated in some variant
of a VEN (Vertical, East, North) local geodetic coordinate system.
Application of the transformation ${\bf V}{\bf W}$ (given in the next
section, Eqs.~\ref{eqv}) transforms them to the geocentric Earth-fixed
coordinate frame.

\paragraph{ Solid Earth tides }
\label{soltid}

  Calculating the shifts of the positions of the observing stations
caused by solid Earth tides is rather complicated due to the solid tides'
coupling with the ocean tides, and the effects of local geology.
Some of these complications are addressed below ($e.g.$, ocean loading).
An isolated simple model of Earth tides is the multipole response model
developed by \citeasnoun{Wllms70b}, who used \citeasnoun{Melch66} as
a reference.
Let ${\bf{R}}_p$ be the position of a tide-producing body, and ${\bf{r}}_0$
the station position. To allow for a phase shift $\psi$ of the tidal
effects from a nominal value of 0,
the phase-shifted station vector ${\bf{r}}_s$ is calculated from
${\bf{r}}_0$ by applying a matrix ${\bf \rm L}$, describing a right-handed
rotation through an angle $ \psi $ about the Z axis of date,
$ {\bf{r}}_s = {\bf \rm L} {\bf{r}}_0 $. This lag matrix ${\bf \rm L}$ is
\begin{equation}
{\bf \rm L} = \left( \matrix { ~~\cos \psi & \sin \psi &0\cr -\sin \psi
& \cos \psi &0\cr 0 &0 &1 \cr } \right) ~.
\end{equation}
A positive value of $ \psi $ implies that the peak response on an Earth
meridian occurs at a time $\delta t = \psi / \omega_{\rm E}$ after that
meridian plane containing $ {\bf{r}}_0 $ crosses the tide-producing object,
where $\omega_{\rm E}$ is the angular rotation rate of the Earth
(Eq.~\ref{erot}).
No significant departures from a zero phase shift have yet been detected:
the peak response occurs when the meridian plane containing ${\bf{r}}_s$
also includes ${\bf{R}}_p$.  Some theoretical models predict out-of-phase
components due to anelasticity.

  The tidal potential at ${\bf{r}}_s$ due to the perturbing body at
${\bf{R}}_p$ may be expressed as a multipole expansion
\begin{eqnarray}
U_{\rm tidal} &=& { {G m_p} \over {R_p} } \Bigl [ \Bigl ( { {r_s} \over {R_p} }
\Bigr )^2 P_2 (\cos\theta) + \Bigl ( { {r_s} \over {R_p} } \Bigr )^3
P_3 (\cos\theta) \Bigr ]  \nonumber \\
&=& ~~U_2~ ~+~ ~U_3 ~,
\label{Utidal}
\end{eqnarray}
where only the quadrupole and octupole terms have been retained.
Here, $G$ is the gravitational constant, $m_p$ is the mass of the
perturbing body, $P_2$ and $P_3$ are Legendre polynomials, and $\theta$
is the angle between ${\bf{r}}_s$ and ${\bf{R}}_p$.  While the
quadrupole displacements are on the order of 500 mm, the mass and
distance ratios of the Earth, Moon, and Sun limit the octupole
terms to a few mm.  An estimate of the
retardation correction (employing the position of the tide-producing
mass at a time earlier than that of the observation by an amount
equal to the light-travel time) shows that this correction is well
below 1 mm, and can therefore be neglected.  We calculate tidal effects
in the terrestrial CIO 1903 frame.  While $r_s$ is given in this frame,
$R_p$ is typically taken from the planetary ephemeris which is defined
in the SSB frame.  The Lorentz transform from the SSB to the geocentric
celestial frame given the Earth's orbital velocity of $10^{-4} c$ may be
neglected because the maximum solid tide is 500 mm
($\times 10^{-4} = 0.05$ mm).  However, the rotation between the celestial
frame and the terrestrial frame is large and must be accounted for:
\begin{equation}
{\bf{R}}_{p_t} ~=~ {\bf Q}^{-1} {\bf{R}}_{p_c} ~,
\end{equation}
where {\bf Q} is the rotational transformation between frames defined by
Eq.~(\ref{eqq}).  The subscripts $t$ and $c$ on ${\bf{R}}_p$ indicate when
the vector is expressed in the terrestrial or celestial frame, respectively.
The perturber unit vector ${\bf{\widehat R}}_p$ has the form of the
signal propagation vector from the radio source (Eq.~\ref{celvec}),
with the right ascension and declination of the perturbing body
($\alpha_p, \delta_p$) replacing the source coordinates ($\alpha,
\delta$) and the vector direction reversed:
\begin{equation}
{\bf{\widehat R}} _p = {\bf Q}^{-1} ~\left( \matrix {
\cos \delta _p \sin \alpha _p   \cr
\cos \delta _p \cos \alpha _p   \cr
\sin \delta _p  \cr } \right) ~.
\label{ptbunt}
\end{equation}
The station unit vector takes a similar form in terms of the longitude
and geocentric latitude:
\begin{equation}
{\bf{\widehat r}} _s = \left( \matrix {
\cos \phi _s \sin \lambda _s   \cr
\cos \phi _s \cos \lambda _s   \cr
\sin \phi _s  \cr } \right) ~.
\label{stnunt}
\end{equation}
After the SSB coordinates of the perturbing body are rotated into the
terrestrial frame of the station, the scalar product of the unit vectors
yields the angle needed to calculate the Legendre polynomials in
Eq.~(\ref{Utidal}), $\cos \theta = {\bf{\widehat r}} _s \cdot
{\bf{\widehat R}} _p$.

  Once the tidal potential is obtained, the next step is to determine the
displacement experienced by a station located at $r_s$.
In a local geodetic VEN coordinate system on an elliptical Earth,
the solid tidal displacement vector ${\bbox{\delta}}_{\rm sol}$ is
\begin{equation}
{\bbox{\delta}}_{\rm sol} = \sum\limits _i { \lbrack ~\delta_{1}^{(i)} ,
 ~\delta_{2}^{(i)}  , ~\delta_{3}^{(i)} \rbrack } ^T ~,
\end{equation}
where the $\delta_{j}^{(i)} (i=2,3)$ are the quadrupole and octupole
displacements.  The components of ${\bbox{\delta}}_{\rm sol}$ are
obtained from the tidal potential as
\begin{eqnarray}
\delta_{1}^{(i)} &=& h_i U_i / g   \\
\delta_{2}^{(i)} &=& l_i\cos{\phi_s} \Bigl ( { { \partial {U_i} } \over
 { \partial {\lambda_s} } } \Bigr ) /g   \\
\delta_{3}^{(i)} &=& l_i \Bigl ( { { \partial {U_i} } \over { \partial
 {\phi_s} } } \Bigr ) /g ~,
\end{eqnarray}
where $h_i (i=2,3)$ are the radial (quadrupole and octupole) Love
numbers, $l_i (i=2,3)$ the corresponding tangential Love numbers, and
$\lambda_s$ and $\phi_s$ are the station longitude and geocentric latitude,
and $g$ the surface acceleration due to gravity,
\begin{equation}
g = G m_{\rm E} / r_s^2 ~.
\end{equation}
Values of the various Love numbers range from 0 to 1, since the Earth is
neither perfectly rigid nor perfectly elastic.  For the highest accuracy,
distinction must be made between displacements in the radial $vs.$
vertical, and tangential $vs.$ horizontal directions, which differ to
the extent of the Earth's flattening (about 1 part in 300).

  Reverting to Cartesian coordinates, some algebra then produces the
following expressions for the quadrupole and octupole components of
${\bbox{\delta}}_{\rm sol}$ in terms of the coordinates of the station
($x_s, y_s, z_s$) and the tide-producing bodies ($X_p, Y_p, Z_p$):
\begin{equation}
\delta_{1}^{(2)} = (h_2/g) \sum\limits _p { { 3 {\mu}_p r_s^2 }
 \over { R_p^5 } }
\Bigl[ { { { ( {\bf{r}}_s \cdot {\bf{R}}_p ) } ^2 } \over
{ 2 } } - { { r_s^2 R_p^2 } \over { 6 } } \Bigr]
\label{eqg1tid}
\end{equation}
\begin{equation}
\delta_{2}^{(2)}  = (l_2/g) \sum\limits _p { { 3 {\mu}_p r_s^3 }
 \over { R_p^5 } }
( {\bf{r}}_s \cdot {\bf{R}}_p ) ( x_s Y_p - y_s X_p )
/ \sqrt { x_s^2 + y_s^2 }
\end{equation}
\begin{equation}
\delta_{3}^{(2)}  = (l_2/g) \sum\limits _p { { 3 {\mu}_p r_s^2 }
 \over { R_p^5 } }
{ ( {\bf{r}}_s \cdot {\bf{R}}_p ) }
{ \biggl [ \sqrt { x_s^2 + y_s^2 } ~Z_p - { { z_s}
\over { \sqrt {x_s^2 + y_s^2 } } }
{ ( x_s X_p + y_s Y_p ) } \biggr ] }
\end{equation}
\begin{equation}
\delta_{1}^{(3)} = (h_3/g) \sum\limits _p { { {\mu}_p r_s^2 }
 \over { 2 R_p^7 } } ( {\bf{r}}_s \cdot {\bf{R}}_p )
\biggl[ { { 5 ( {\bf{r}}_s \cdot {\bf{R}}_p ) } ^2 }
 - { 3 r_s^2 R_p^2 }  \biggr]
\end{equation}
\begin{equation}
\delta_{2}^{(3)}  = (l_3/g) \sum\limits _p { { 3 {\mu}_p r_s^3 }
 \over { 2 R_p^7 } }
\biggl [ 5 ( {\bf{r}}_s \cdot {\bf{R}}_p ) ^2  - r_s^2 R_p^2
\biggr ] ( x_s Y_p - y_s X_p ) / \sqrt { x_s^2 + y_s^2 }
\end{equation}
\begin{equation}
\delta_{3}^{(3)}  = (l_3/g) \sum\limits _p { { 3 {\mu}_p r_s^2 }
 \over { 2 R_p^7 } }
\biggl [ 5 ( {\bf{r}}_s \cdot {\bf{R}}_p ) ^2  - r_s^2 R_p^2 \biggr ]
{ \biggl [ \sqrt { x_s^2 + y_s^2 } ~Z_p - { { z_s}
\over { \sqrt {x_s^2 + y_s^2 } } }
{ ( x_s X_p + y_s Y_p ) } \biggr ] } ~,
\label{eqg3tid}
\end{equation}
where $\mu_p$ is the ratio of the mass of the disturbing body, $p$,
to the mass of the Earth, and
\begin{equation}
{\bf{R}}_p = { \lbrack X_p , Y_p , Z_p \rbrack } ^T
\end{equation}
is the vector from the center of the Earth to that body.  The
summations are over tide-producing bodies, of which only the Sun and
the Moon are normally included.  Recent work considers low-level
planetary contributions \cite{HartmS94,HartmW94,Wllms95},
of which Venus' appear to be
the most important.  \citeasnoun{Mathews95} have recently re-examined
the basic definitions underlying the derivation of the tidal potential.
They conclude that the most reasonable definition is one that uses the
reference ellipsoid, and thus implies that the Love numbers have a
slight latitude dependence.

  The above formulation implicitly assumes that the Love numbers
$h_i$ and $l_i$ are independent of the frequency of the tide-generating
potential.  A more accurate treatment entails a harmonic expansion of
Eqs.~(\ref{eqg1tid})-(\ref{eqg3tid}) and use of a different set of $h_i$,
$l_i$ for each frequency component.  Each harmonic term is denoted by
its historical (Darwin) name, if one exists, and the Doodson code
\cite{IERS92} ($e.g$., $K_1$ has Doodson number = 165555).  The Dood\-son
notation classifies the tidal components according to increasing speed.
The correction to the Love number which scales the solid tidal radial
displacement for the $k$th harmonic term at station $s$ is given by
\begin{equation}
\delta h_2^{sk} = 3 \thinspace \sqrt{5/24\pi} \thinspace \delta h_2^k
\thinspace H_k \thinspace \sin\phi_s \cos\phi_s \sin (\lambda_s + \theta_k) ~,
\end{equation}
where $\delta h_2^k$ is the difference between the nominal quadrupole
Love number $h_2$ = 0.609 and the frequency dependent Love number
\cite{Wahr79}, $H_k$ is the amplitude of the $k$th harmonic term in the tide
generating expansion from \citeasnoun{Cartwrt73}, $\phi_s$ is the
geocentric latitude of the station, $\lambda_s$ is the East longitude
of the station and $\theta_k$ is the $k$th harmonic tide argument.
The Love numbers and tidal amplitudes of the six dominant nearly diurnal
tides are listed in Table~\ref{frqtidh2}.  They yield purely vertical
station displacements \cite{Naudet94} (in mm)
\begin{eqnarray}
 \psi _1~(166554)~~~~\delta \Delta \delta_1 &=& ~~~~0.37 \sin 2\phi_s
    \sin (\lambda_s + \alpha_G + l')  \\
         (165565)~~~~\delta \Delta \delta_1 &=&  ~~-1.84 \sin 2\phi_s
    \sin (\lambda_s + \alpha_G - \Omega ) \\
     K_1~(165555)~~~~\delta \Delta \delta_1 &=& -12.68 \sin 2\phi_s
    \sin (\lambda_s + \alpha_G)  \\
         (165545)~~~~\delta \Delta \delta_1 &=& ~~~~0.24 \sin 2\phi_s
    \sin (\lambda_s + \alpha_G + \Omega)  \\
     P_1~(163555)~~~~\delta \Delta \delta_1 &=& ~~~~1.32 \sin 2\phi_s
    \sin [\lambda_s + \alpha_G -2(\Omega + F - D) ]  \\
     O_1~(145555)~~~~\delta \Delta \delta_1 &=& ~~~~0.62 \sin 2\phi_s
    \sin [\lambda_s + \alpha_G -2(\Omega + F) ] ~,
\end{eqnarray}
where $\phi_s, \lambda_s$, and $\alpha_G$ are the station geocentric latitude
and longitude and Greenwich RA, respectively.  The astronomical arguments
$l', F, D, \Omega$ (mean anomaly of the Sun, mean argument of the
latitude of the Moon, mean elongation of the Moon from the Sun, and the
mean longitude of the ascending lunar node) are defined in
Sec.~\ref{tercel}.  These displacements are then summed and the total is
used as the first order correction to each station's vertical displacement.
Horizontal corrections are presently ignored.  Note that the largest
correction, the $K_1$ term, is identical to that already recommended
in 1983 by the MERIT standards (\citename{Melbourne83}, 1983, 1985).
A fairly complete recent treatment of the frequency dependence of
tidal response has been given by \citename{Mathews97}~\citeyear{Mathews97}.
With such amendments, the solid tide model presented here should be adequate
at the mm level.

  To convert the locally referenced displacement, ${\bbox{\delta}}_{\rm sol}$,
which is expressed
in the local VEN coordinate system, to the Earth-fixed frame, two rotations
must be performed.  The first, ${\bf W}$, rotates by an
angle $\phi_{s({\rm gd})}$ (station geodetic latitude) about the y axis to
an equatorial system.  The second, ${\bf V}$, rotates about the resultant z
axis by an angle, $-\lambda_s$ (station longitude), to bring the
displacements into the standard geocentric coordinate system.  The
result is
\begin{equation}
{\bf{\Delta}}_{\rm sol} = {\bf V}{\bf W} {\bbox{\delta}}_{\rm sol} ~,
\label{eqbigdel}
\end{equation}
where
\begin{equation}
{\bf W} = \left( \matrix { \cos \phi_{s({\rm gd})} &0
&- \sin \phi_{s({\rm gd})}\cr
0 &1 &0\cr \sin \phi_{s({\rm gd})} &0 & ~~\cos \phi_{s({\rm gd})}\cr } \right)
,~~~~~~ {\bf V} = \left( \matrix { \cos \lambda _s &- \sin \lambda _s &0 \cr
\sin \lambda _s & ~~\cos \lambda _s &0 \cr 0 &0 &1\cr } \right) ~.
\label{eqv}
\end{equation}

\medskip\medskip\noindent
We calculate geodetic latitudes according to \possessivecite{Bowring76}
formulation which properly accounts for stations that are at some altitude
above the reference ellipsoid:
\begin{equation}
\phi_{s({\rm gd})} ~=~ \tan ^{-1} \Biggl[
 {
 {z_s ~+~ e_2^2~R_{\rm E}~\sin^3\theta}~(1 - f)
 \over
 {r_{{\rm s}p_s} ~-~ e_1^2~R_{\rm E}~\cos^3\theta}
 }  \Biggr] ~,
\label{eqglat}
\end{equation}
\noindent
where $r_{{\rm s}p_s} = ({x_s}^2+{y_s}^2)^{1/2}$ is the station radius from
the Earth's spin axis, $f ~(\approx 1 / 300)$ is the geoid flattening
factor, $e_1^2 = 2 f - f^2$, and $e_2^2 = e_1^2 / (1 - e_1^2)$
are the first and second eccentricities squared, and
\begin{equation}
\theta ~=~ \tan ^{-1} \Biggl[
 { z_s  \over { r_{{\rm s}p_s}~(1 - f) } }  \Biggr] ~.
\label{eqglatt}
\end{equation}
Simpler formulas which assume that the station is on the ellipsoid make
errors of order $10^{-7} ~\rm radians$ for a mid-latitude station at 1 km
altitude.  The difference between geodetic and geocentric latitude can
affect the tide model by as much as $f$ times the tidal effect, or
$\approx~1\thinspace$mm.

\paragraph{ Pole tide }
\label{poltid}

  One of the significant secondary tidal effects is the displacement
of a station by the elastic response of the Earth's crust to shifts
in the spin axis orientation.  The spin axis is known to describe
an approximately  circular path of $\approx$~20-m diameter at the
north pole with an irregular period of somewhat less than one year
(see Fig.~\ref{pmplt} in Sec.~\ref{aprot}).  Depending on
where the spin axis pierces the crust at the instant of a VLBI
measurement, the ``pole tide'' displacement will vary from time to
time.  This effect must be included if centimeter accuracy is desired.

  \citeasnoun{Yoder84} and \citeasnoun{Wahr85} derived an expression
for the displacement
of a point at geodetic latitude $\phi_{s({\rm gd})}$ and longitude $\lambda_s$
due to the pole tide:
\begin{eqnarray}
{\bbox{\delta}}_{\rm pol} = - \thinspace { { \omega_{\rm E}^2 R_{\rm E} } \over
{ g } } \biggl[ ~\sin \phi_{s({\rm gd})}
\cos \phi_{s({\rm gd})} ( p_x \cos \lambda_s + p_y \sin \lambda_s ) &~h~&
{\bf{\widehat r}}  \nonumber  \\
+ \cos 2 \phi_{s({\rm gd})} ( p_x \cos \lambda_s + p_y \sin \lambda_s ) &~l~&
{\bf{\widehat {\bbox{\phi}}}} \nonumber \\
+ \sin \phi_{s({\rm gd})} ( - p_x \sin \lambda_s + p_y \cos \lambda_s ) &~l~&
{\bf{\widehat {\bbox{\lambda}}}} ~\biggr] ~.
\label{eqpoltid}
\end{eqnarray}
Here $\omega_{\rm E}$ is the rotation rate of the Earth, $R_{\rm E}$ the
equatorial radius of the Earth, $g$ the acceleration due to gravity at
the Earth's surface, and $h$ and $l$ the customary Love numbers.
Displacements of the instantaneous spin axis from the long-term
average spin axis along the x and y axes are given by $p_x$ and $p_y$.
Eq.~(\ref{eqpoltid}) shows how these map into station displacements
along the unit vectors in the radial (${\bf{\widehat r}}$),
latitude (${\bf{\widehat {\bbox{\phi}}}}$), and longitude
(${\bf{\widehat {\bbox{\lambda}}}}$) directions. With the standard values
$\omega_{\rm E} \approx 7.292 \times 10^{-5}$ rad/sec,
$R_{\rm E} \approx$ 6378 km, and $g \approx$ 980.7 $\rm{cm}$/$\rm{sec} ^2$,
the factor $\omega_{\rm E}^2 R$/$g$ = 0.00346.
Since the maximum values of $p_x$ and $p_y$ are on the order of
10 meters, and $h \approx 0.6$, $l \approx 0.08$,
the maximum displacement due to the pole tide is 1 to 2 cm,
depending on the location of the station ($\phi_{s({\rm gd})}$, $\lambda_s$).
The locally referenced displacement ${\bbox{\delta}}_{\rm pol}$ is
transformed via the matrix ${\bf V}{\bf W}$ of Eqs.~(\ref{eqv}) to give the
displacement ${\bf{\Delta}}_{\rm pol}$ in the standard geocentric
coordinate system.

\paragraph{ Ocean loading }
\label{ocnld}

This section is concerned with another of the secondary tidal effects,
namely the elastic response of the Earth's crust to ocean tides,
which moves the observing stations to the extent of a few cm.
Such effects are commonly called ``ocean loading.''
The current standard model of ocean loading is general enough
to accommodate externally derived parameters describing the tide phases
and amplitudes at a number of frequencies.  The locally referenced
three-dimensional displacement ${\bbox{\delta}}_{\rm ocn}$
(components $\delta_j$) due to ocean loading is related to the frequencies
$\omega_i$, amplitudes $\xi_{ij}$, and phases $\varphi_{ij}$.
In a local Cartesian coordinate system (usually with unit vectors in the
vertical, East, and North directions) at time $t$,
\begin{equation}
\delta _j = \sum \limits _{i=1}^N \xi _{ij} \cos (\omega _i t +
V_i - \varphi _{ij} ) ~.
\end{equation}
  The quantities $\omega _i$ (frequency of tidal constituent $i$)
and $V_i$ (astronomical argument of constituent $i$) depend
only on the ephemeris information (positions of the Sun and Moon).
The algorithm of Goad \cite{IERS89,IERS96b} is usually used to calculate
these two quantities.  On the other hand the amplitude $\xi _{ij}$ and
Greenwich phase lag $\varphi _{ij}$ of component $j$ are determined
by the particular model assumed for the deformation of the Earth.
The local displacement vector is transformed via Eqs.~(\ref{eqv}) to
the displacement ${\bf{\Delta}}_{\rm ocn}$ in the standard geocentric frame.

   In present models, the local displacements and their phases,
$\xi _{ij}$ and $\varphi _{ij}$, are calculated from ocean tidal
loading models with as many as 11 frequencies.
The eleven components are denoted, in standard notation:
$K_2$, $S_2$, $M_2$, and $N_2$ (all with approximately 12-hour
periods), $K_1$, $P_1$, $O_1$, $Q_1$ (24 h), $M_{\rm f}$ (14 day),
$M_{\rm m}$ (monthly), and $S_{\rm sa}$ (semiannual).

   Three ocean loading models have been used over the years.
They differ in the displacements that are calculated and the number of
components that are considered, as well as in the numerical values
that they yield for $\xi _{ij}$ and $\varphi _{ij}$.
Scherneck's results \citeyear{Schrnk83,Schrnk91} are the most complete in
the sense of considering both vertical and horizontal displacements and all
eleven tidal components.  They have now been adopted for the
IERS standards \citeyear{IERS96b}.  Goad's model \citeyear{Goad83}
was adopted in the MERIT
and early IERS standards \citeyear{IERS89}, but only considers vertical
displacements.  \possessivecite{Pagiat82} model \cite{Pagiat90},
based on \citeasnoun{PagiatLV82}, considers only six tidal
components ($S_2$, $M_2$, $N_2$, $K_1$, $P_1$, and $O_1$).

  An extension of the 1991 Scherneck model \cite{Schrnk93} accounts
for modulation of the eleven tidal frequencies by multiples of $N^\prime$,
which corresponds to the lunar nodal period (18.6 years).
On the assumption that these additional terms yield
ocean loading amplitudes which are in the same ratio to each main
loading term as the companion tides are to the main tides, the
additional station displacements can be written as
\begin{equation}
\delta _j^\prime = \sum \limits _{i=1}^N \sum \limits _k
  r_{ki} \xi _{ij} \cos \bigl[ (\omega _i + n_{ki} \omega_{N^\prime})t +
  V_i + n_{ki} N^\prime - \varphi _{ij} ) \bigr] ~,
\end{equation}
where the $k$ summation extends over all integer multiples $n_{ki}$
of the lunar node $N^\prime$, and $r_{ki}$ is the ratio of the
tidal amplitude of each companion $k$ to the tidal amplitude of
the parent $i$.  Of 26 such components listed by \citeasnoun{Cartwrt73},
20 are estimated to be significant in contributing to the largest
ocean loading displacements at the 0.01 mm level.  Table~\ref{compan} shows
the multiples $n_{ki}$ and amplitude ratios $r_{ki}$ for these 20 components.

  In pushing the limits of Earth modeling to below 1 cm accuracy
in the mid-1990s, ocean loading station displacements are one
aspect of the models that are undergoing close scrutiny.
Initial studies indicate that ocean loading amplitudes can be
derived from VLBI experiments at an approximate accuracy level
of 1-2 mm \cite{Sovers94}.  When estimating parameters, great care
must be used in order to avoid singularities due to the identity
of certain components of station displacements due to nominally
different physical effects
(``confounding of parameters'').  Since some components of ocean loading,
solid Earth tides, and ocean tidally induced UTPM variations have the
same frequencies, certain linear combinations of the station displacements
that they cause are identical (see Secs.~\ref{soltid} and \ref{ocnutpm}).

\subsubsection{ Non-tidal station motion }
\label{nontid}

  Many other processes take place in the Earth's atmosphere and in its
crust that affect the location of an observing station on time scales
ranging from seconds to years, and distance scales ranging from local
to global.  Present knowledge of such processes is somewhat sketchy, but
both theoretical and experimental research are starting to provide useful
results.  Local processes include the effects of ground water and
snow cover redistribution \cite{Chao96}, and magma chamber
activity in volcanically active areas \cite{Webb95}.
Effects that have more widespread repercussions are
atmospheric loading and post-glacial relaxation.  In contrast to the
tidal effects, whose time dependence is fixed by the precisely known
motions of the bodies of the Solar System, the non-tidal processes
do not have well-known, periodic time dependencies.  The two ``global''
effects, atmospheric loading and post-glacial relaxation, are discussed
briefly in the next two sections.  Two others have not yet been
investigated in detail: non-tidal ocean surface height variations
\cite{Gaspar97}, and surface displacements caused by pressure variations
at the core-mantle boundary \cite{Fang96}.

\paragraph{ Atmosphere loading }
\label{atmld}

  By analogy with the ocean tides that were considered
in the previous section, a time-varying atmospheric pressure distribution
can also induce crustal deformation.  \citeasnoun{Rabbel86} first estimated
the effects of atmospheric loading on VLBI baseline determinations, and
concluded that they may amount to many millimeters of seasonal variation.
In contrast to ocean tidal effects, analysis of atmospheric loading
does not benefit from the presence of a well-understood periodic driving
force.  Otherwise, estimation of atmospheric loading via Green's function
techniques is analogous to methods used to calculate ocean loading
effects.  Rabbel and Schuh recommend a simplified
form of the dependence of the vertical crust displacement on
pressure distribution.  It involves only the instantaneous pressure
at the site in question, and an average pressure anomaly over a circular
region $C$ of radius $R=2000$ km surrounding the site.  The expression
for the vertical displacement (mm) is
\begin{equation}
\Delta r = -0.35 \Delta p -0.55 \bar p ~,
\label{eqpload}
\end{equation}
where $\Delta p$ is the local pressure anomaly ($p - p_{_{\rm STD}}$,
mbar) relative to the standard pressure
\begin{equation}
p_{_{\rm STD}} = 1013.25~\exp (-h/8.567) ~.
\label{eqpht}
\end{equation}
The pressure anomaly $\bar p$ (mbar) is averaged over the 2000-km
circular region mentioned above.  The site altitude $h$ (km) should
be calculated relative to the standard reference ellipsoid, using
quantities defined in the solid Earth tide section above
(Eq.~\ref{eqglat}):
\begin{equation}
h ~=~ { {r_{{\rm s}p}} \over {\cos \phi_{\rm gd}} } -
  { R_{\rm E} \over \sqrt{ 1~-~ e_1^2~ \sin^2\phi_{\rm gd}} } ~.
\label{flatht}
\end{equation}
Note that the reference point for this displacement is the site
location at its standard pressure.  The locally referenced $\Delta r$
is transformed to the standard geocentric coordinate system via the
transformation of Eqs.~(\ref{eqv}).

  Such a rudimentary model has been recently incorporated into
VLBI analyses.  A mechanism for characterizing the pressure anomaly
$\bar p$ expresses the two-dimensional surface pressure distribution
(relative to the standard pressure) surrounding a site as a quadratic
polynomial
\begin{equation}
p(x,y) = \Delta p + A_1 x + A_2 y + A_3 x^2 + A_4 xy + A_5 y^2 ~,
\end{equation}
where $x$ and $y$ are the local East and North distances of the point
in question from the VLBI site.  The pressure anomaly may then be
evaluated by the simple integration
\begin{equation}
\bar p ~= \int\!\!\int_C dx\,dy~p(x,y)~\bigg/\int\!\!\int_C dx\,dy ~,
\end{equation}
giving
\begin{equation}
\bar p = \Delta p + ( A_3 + A_5 ) R^2 / 4
\end{equation}
and
\begin{equation}
\Delta r = -0.90 \Delta p -0.14 ( A_3 + A_5 ) R^2 ~.
\label{eqavload}
\end{equation}
A quadratic fit to any available area pressure measurements can then
determine the coefficients $A_{1-5}$.  Future advances in understanding
the atmosphere-crust elastic interaction can probably be accommodated
by adjusting the coefficients in Eq.~(\ref{eqpload}).  As an initial
step along these lines, a station-dependent factor $f$ is introduced
to scale the local coefficient in Eq.~(\ref{eqavload}):
\begin{equation}
\Delta r = -0.90 ( 1 ~+ f) \Delta p -0.14 ( A_3 + A_5 ) R^2 ~.
\label{eqploadf}
\end{equation}
This may account for differing geographical features surrounding different
sites.  In particular, $f$ may depend on the fraction of ocean within the
2000 km radius.  Some recent analyses have produced empirical estimates
of atmosphere loading coefficients for a number of sites
\cite{Manabe91,vanDamH94,MacMil94,vanDam94}.  It is not yet clear whether
the site-to-site
variation of these coefficients is free from other systematic errors.

\paragraph{ Post-glacial relaxation }
\label{glac}

  Thick glacial ice sheets covering Scandinavia, Greenland, and Canada
melted $\approx$10,000 years ago.  The removal of their considerable
weight pressing on the Earth's crust is believed to result in relaxation
(``rebound'') that continues at present \cite{TshngPlt91}.
The magnitude of this motion is estimated to be as large as several
mm/yr, predominantly in the vertical direction, at sites in and near
the locations of ancient glaciers and ice sheets.
Current theory of deglaciation effects
is not yet sufficiently developed to produce definitive results.
The parameters describing deglaciation history, as well as the rheological
properties of the Earth's mantle, are not accurately known.  Nevertheless,
it appears that reasonable pa\-ra\-me\-ter choices yield some agreement
with empirical measurements of baselines in the affected areas
\cite{Mitrov93,Peltier95}.

  In summary, models have been presented that describe the four major
time-dependent station motions (solid, pole, and ocean tides, and
atmospheric loading).  Each of the locally referenced displacement vectors
is then transformed to the standard geocentric coordinate system via
rotations like Eq.~(\ref{eqbigdel}).  After this transformation, the
final station location in the terrestrial frame is
\begin{equation}
{\bf{r}}_t = {\bf{r}}_0 +
{\bf{\Delta}}_{\rm sol} + {\bf{\Delta}}_{\rm pol} +
{\bf{\Delta}}_{\rm ocn} + {\bf{\Delta}}_{\rm atm} ~.
\label{eqstnloc}
\end{equation}

\subsection{ Earth orientation }
\label{tercel}

The Earth is approximately an oblate spheroid, spinning in the presence of
two massive moving objects, the Sun and the Moon.  These are positioned such
that their time-varying gravitational effects not only produce tides on the
Earth, but also subject it to torques.  In addition, the Earth is covered by
a complicated fluid layer, and is not perfectly rigid internally.  As a
result, the orientation of the Earth is a very complicated function of time,
which can be represented to first order as the composite of a time-varying
rotation rate, a wobble, a nutation, and a precession.  The exchange of
angular momentum between the solid Earth and the fluids on its surface,
as well as between its crust and deeper layers, is not readily predictable,
and thus must be continually experimentally determined
\cite{LeMouel}.  Nutation and precession are fairly well
modeled theoretically.  At the accuracy with which VLBI can determine
baseline vectors, however, even these models are not completely adequate.

  Currently, the rotational transformation ${\bf Q}$ of coordinates from the
terrestrial frame to the celestial geocentric frame is composed of 6
separate transformations (actually 12 rotations, since the nutation,
precession, and ``perturbation'' transformations ${\bf N}$, ${\bf P}$,
and ${\bf \Omega}$ each consist of 3 rotations) applied to a vector in
the terrestrial system:
\begin{equation}
{\bf Q} = {\bf \Omega} {\bf P}{\bf N}{\bf U}{\bf X}{\bf Y} ~.
\label{eqq}
\end{equation}
In order of appearance in Eq.~(\ref{eqq}), the transformations are the
perturbation rotation, precession, nutation, UT1, and the $x$ and $y$
components of polar motion.  All are discussed in detail in the following
four sections.  With this definition of ${\bf Q}$, if ${\bf{r}}_t$ is a station
location expressed in the terrestrial system, $e.g.$, the result of
Eq.~(\ref{eqstnloc}), then that location expressed in the geocentric
celestial system is
\begin{equation}
{\bf{r}}_c = {\bf Q} {\bf{r}}_t ~.
\label{eqrcrt}
\end{equation}

  This particular formulation evolved from the historical development
of astrometry, and is couched in that language.  With modern measurement
techniques, such a formulation is esthetically unsatisfactory, but it
currently offers a practical way for merging VLBI results into the
long historical record of astrometric data.  The precession
and nutation matrices (${\bf P}$ and ${\bf N}$) depend on two angles
(nutation in celestial longitude $\Delta \psi$
and in obliquity $\Delta \varepsilon$) that vary
slowly in the celestial reference frame.  On the other hand, the UTPM
matrices (${\bf U}$, ${\bf X}$, and ${\bf Y}$) depend on three angles that
vary slowly
in a terrestrial reference frame co-rotating with the Earth.  Thus the
desire to have expressions that vary slowly with time led to a description
that uses 5 angles to specify the orientation of the Earth, rather than
the absolute minimum of 3 angles.  Recently improving measurement
accuracy demands high-frequency terms in ${\bf U}{\bf X}{\bf Y}$,
some of which duplicate
parts of ${\bf N}$, and thus negate the advantages of the historical
decomposition.  Esthetically it would be much more pleasing to separate
${\bf Q}$ into two rotation matrices:
\begin{equation}
{\bf Q} = {\bf Q}_1 {\bf Q}_2 ~,
\end{equation}
where $ {\bf Q}_2 $ are those rotations to which the Earth would be subjected
if all external torques were removed (approximately ${\bf U}{\bf X}{\bf Y}$
above), and
where $ {\bf Q}_1 $ are those rotations arising from external torques
(approximately $ {\bf \Omega} {\bf P}{\bf N} $ above).  Even then, the tidal
response of the
Earth prevents such a separation from being perfectly realized.
Eventually, the entire problem of obtaining the matrix ${\bf Q}$ and the
tidal effects on station locations may be solved numerically.
Note that the matrices appearing in the transformation of Eq.~(\ref{eqq})
are not the same as those historically used in astrometry.  Since we
rotate the Earth from ``of date'' to J2000.0 (rather than the celestial
sphere from J2000.0 to coordinates of date), the matrices ${\bf \Omega}$,
${\bf P}$,
and ${\bf N}$ are transposes of the conventional transformations.  We will
now consider each of the rotations ${\bf Y}$, ${\bf X}$, ${\bf U}$,
${\bf N}$, ${\bf P}$, and ${\bf \Omega}$ in detail.

\subsubsection{ UT1 and polar motion }
\label{utpm}

  The Earth's instantaneous spin axis traces a quasi-circular,
quasi-periodic path approximately 20 m in diameter with a ``period''
somewhat less than one year, which is known as polar motion (see
Fig.~\ref{pmplt} in Sec.~\ref{aprot}).  The first transformation, ${\bf Y}$,
is a right-handed rotation about the $x$ axis of the terrestrial frame
by an angle $\Theta _2$.  Currently, the terrestrial frame is the CIO
1903 frame, except that the positive $y$ axis is at 90 degrees east
(toward Bangladesh).  The $x$ axis is coincident with the 1903.0 meridian
of Greenwich, and the $z$ axis is the 1903.0 standard pole.
\begin{equation}
{\bf Y} = \left( \matrix { 1 &0 &0 \cr 0 & ~~~\cos \Theta _2 & \sin \Theta _2
\cr
0 &-\sin \Theta _2 & \cos \Theta _2 \cr } \right) ~,
\end{equation}
where $ \Theta _2 $ is the $y$ pole position published by IERS.

  The next rotation in sequence is the right-handed rotation (through an angle
$ \Theta _1 $ about the $y$ axis) obtained after the previous rotation
has been applied:
\begin{equation}
{\bf X} = \left( \matrix { \cos \Theta _1 &0 &-\sin \Theta _1 \cr
0 &1 &0 \cr \sin \Theta _1 &0 & ~~~\cos \Theta _1 \cr } \right) ~.
\end{equation}
In this rotation, $ \Theta _1 $ is the IERS $x$ pole position.
Note that we have incorporated in the matrix definitions the
transformation from the left-handed system used by IERS to the right-handed
system we use.  Note also that instead of IERS data used as a pole
definition, we could instead use any other source of polar motion data
provided it was represented in a left-handed system.  The only effect would
be a change in the definition of the terrestrial reference system.

  The application of ${\bf X}{\bf Y}$ to a vector in the terrestrial
system of
coordinates expresses that vector as it would be observed in a coordinate
frame whose $z$ axis was along the Earth's ephemeris pole.  The third rotation,
${\bf U}$, is about the resultant $z$ axis obtained by applying
${\bf X}{\bf Y}$.  It is a
rotation through the angle $-H$, where $H$ is the hour angle of the true
equinox of date ($i.e.$, the dihedral angle measured westward between the
$xz$ plane defined above and the meridian plane containing the true equinox
of date):
\begin{equation}
{\bf U} = \left( \matrix { \cos H &- \sin H &0 \cr \sin H & ~~~\cos H &0 \cr
0 &0 &1 } \right) ~.
\end{equation}
The equinox of date is the point defined on the celestial equator by the
intersection of the mean ecliptic with that equator.  It is that intersection
where the mean ecliptic rises from below the equator to above it (ascending
node).  This angle $H$ is composed of two parts:
\begin{equation}
H = h _{\Upsilon} + \alpha_{\rm E} ~,
\end{equation}
where $h _ \Upsilon$ is the hour angle of the mean equinox of date, and
$ \alpha_{\rm E} $ (equation of the equinoxes) is the difference in hour angle
of the true equinox of date
and the mean equinox of date, a difference which is due to the nutation
of the Earth.  This set of definitions is cumbersome and couples nutation
effects with Earth rotation.  However, in order to provide a direct
estimate of conventional UT1 (universal time) it is necessary to endure
this historical approach.

  UT1 is defined to be such that the hour angle of the mean
equinox of date is given by the following expression \cite{Aoki82,Kaplan81}:
\begin{eqnarray}
h _{\Upsilon} = {\rm UT1} & + & 6^{\rm h} ~41^{\rm m} ~50^{\rm s} .54841
~+~ 8640184^{\rm s} .812866 ~T        \nonumber \\
& + & 0^{\rm s} .093104 ~ T^2 ~-~ 6^{\rm s} .2 \times 10^{-6} ~T^3 ~,
\end{eqnarray}
where the quantity
\begin{equation}
T = ({\rm{Julian}}~{\rm UT1}~{\rm{date}} ~-~ 2451545.0 ) / 36525.0
\end{equation}
represents the number of Julian centuries since J2000.0.  An equivalent
expression which is normally used is
\begin{eqnarray}
h _{\Upsilon} = &86400^{\rm s} ({\rm UT1}~ {\rm{Julian~day~fraction}}) +\
 67310^{\rm s} .54841  ~~~~~~~~~~~~~~ \nonumber \\
&+ 8640184^{\rm s} .812866 ~ T + 0^{\rm s} .093104 ~ T^2 - 6^{\rm s} .2 \times
10^{-6} ~T^3 ~.
\end{eqnarray}
This expression produces a time UT1 which tracks the Greenwich hour
angle of the real Sun to within $16^{\rm m}$.  However, it really is sidereal
time, modified to fit our intuitive desire to have the Sun directly
overhead at noon on the Greenwich meridian.  Historically, differences
of UT1 from a uniform measure of time, such as atomic time (UTC or
TAI), have been used in specifying the orientation of the Earth
(see the plot of UT1$-$TAI in Fig.~\ref{rotplt} of Sec.~\ref{aprot}).

  By the very definition of ``mean of date'' and ``true of date'', nutation
causes a difference in the hour angles of the mean equinox of date and the
true equinox of date.  This difference, called the ``equation of the
equinoxes'', is denoted by $\alpha_{\rm E}$ and is obtained as follows:
\begin{equation}
\alpha_{\rm E} = \tan ^{-1} { \Biggl( { { y _{\Upsilon ^{\prime}} } \over
{ x _{\Upsilon ^{\prime}} } } \Biggr) } = \tan ^{-1}
{ \Biggl( { { N ^{-1} _{21} }
\over { N ^{-1} _{11} } } \Biggr) } = \tan ^{-1} { \Biggl ( { { N _{12} }
\over { N _{11} } } \Biggr) } ~,
\label{eqeq}
\end{equation}
where the vector
\begin{equation}
\left ( \matrix { x _{\Upsilon} ^{\prime} \cr \noalign{\smallskip}
y _{\Upsilon} ^{\prime} \cr \noalign{\smallskip}
z _{\Upsilon} ^{\prime} } \right ) = { N ^{-1} } { \left ( \matrix {
1 \cr 0 \cr 0 } \right ) }
\end{equation}
is the unit vector, in true equatorial coordinates of date, toward the mean
equinox of date.
In mean equatorial coordinates of date, this same unit vector is just
$(1,0,0) ^T $.  The matrix ${{\bf N}^{-1} }$ is the inverse
(or equally, the transpose) of the transformation matrix ${\bf N}$,
which will be defined below in Eq.~(\ref{eqnutm}), to effect the
transformation from true equatorial
coordinates of date to mean equatorial coordinates of date.
To a very good approximation,
\begin{equation}
\alpha_{\rm E} = \Delta \psi \cos(\overline \varepsilon + \Delta \varepsilon) ~.
\end{equation}

  The IERS Conventions \citeyear{IERS92} had recommended a slightly different
expression, which uses the mean obliquity $\overline \varepsilon$ rather
than the true obliquity $\overline \varepsilon + \Delta \varepsilon$,
and can change $\alpha_{\rm E}$ by a few nanoradians:
\begin{equation}
\alpha_{\rm E} = \Delta \psi \cos{\overline \varepsilon} ~.
\end{equation}
More recently, the IERS \citeyear{IERS96b} recommends including two
additional terms so that the equation of the equinoxes involves the
longitude of the lunar node
$\Omega$ (to be defined in Sec.~\ref{nut}):
\begin{equation}
\alpha_{\rm E} = \Delta \psi \cos{\overline \varepsilon} ~+
0.^{\prime\prime}00264 \sin\Omega ~+
0.^{\prime\prime}000063 \sin 2 \Omega ~.
\end{equation}
This expression should be used in analyses done after 1~January, 1997
(including re-analyses of old data).  That particular date is chosen
to minimize the discontinuity of $\alpha_{\rm E}$ ($\Omega = \pi$ on
26~February, 1997).  One last note on the equation of the equinoxes:
augmentation of the $a~priori$
precession constant will shift the position of the celestial ephemeris
pole and thus require an adjustment to the equation of the equinoxes
\begin{equation}
\Delta \alpha_{\rm E} = \Delta p_{\rm LS} \cos{\overline \varepsilon} ~-~
 \Delta p_{\rm PL} ~,
\end{equation}
where $\Delta p_{\rm LS}$ and $\Delta p_{\rm PL}$ are the $a~priori$ shifts in
lunisolar and planetary precession constants, respectively.

  It is convenient to apply ${\bf U}{\bf X}{\bf Y}$ as a group.
To parts in $10^{12}$, ${\bf X}{\bf Y}={\bf Y}{\bf X}$.  However, to
the same accuracy
${\bf U}{\bf X}{\bf Y} \ne {\bf X}{\bf Y}{\bf U} $.  Neglecting terms
of $O(\Theta^2)$
(which produce station location errors of approximately 0.006 mm):
\begin{equation}
{\bf U}{\bf X}{\bf Y} = \left( \matrix { \cos H &- \sin H &-\sin \Theta _1
\cos H - \sin \Theta _2 \sin H \cr
\sin H & ~~\cos H &-\sin \Theta _1 \sin H + \sin \Theta _2 \cos H \cr
~\sin \Theta _1 &- \sin \Theta _2 & ~~1 \cr } \right) ~.
\end{equation}

  Over relatively short time spans, Earth rotation might be modeled
as a time-linear function, by analogy with tectonic motion of the
stations over longer periods.  If the x, y components of polar motion
and UT1 are symbolized by
$\Theta_{1-3}$, and the reference time is $t_0$, then this model is
\begin{equation}
\Theta_i = {\Theta}_i^0 + {\dot{\Theta}}_i (t - t_0) ~,
\end{equation}
where $\Theta_i^0$ are the values of UTPM at the reference epoch.

\paragraph{ Tidal UTPM variations }
\label{tidutpm}

  Tidally induced shifts of mass in the solid Earth, oceans, and
atmosphere carry angular momenta which must be redistributed
in a manner that conserves the total angular momentum.  This leads
to variations in the orientation and rotation rate of the Earth:
modification of polar motion and UT1.  Such small effects emerged
above the detection threshold in space geodesy in the early 1990s.
Modeling them is important if sub-centimeter accuracy is to be
attained in the interpretation of VLBI measurements.

  Just as various tidal forces affect the station locations
(Secs.~\ref{soltid}~-~\ref{ocnld}), they also affect polar motion and UT1
($\Theta_{1-3}$).  Equations similar to (\ref{eqtideff}) may be written
for each of the three components of Earth orientation:
\begin{equation}
\Theta_i = \Theta_{i0} ~+~ {{\Delta\Theta_i}_{\rm sol}} ~+~
 {{\Delta\Theta_i}_{\rm ocn}} ~+~ {{\Delta\Theta_i}_{\rm atm}} ~,
\end{equation}
where $\Theta_i$ (i=1,3) symbolizes each of the three components
of UTPM, $\Theta_{i0}$ is its value in the absence of tidal effects,
and the three $\Delta$ terms are the respective contributions of
solid Earth, ocean, and atmospheric tides.  The next two sections
describe the current models of solid and ocean tide contributions
that are presently included in VLBI analyses.  At present not enough
is known about atmospheric tidal effects, and only loose upper limits
can be placed on their contribution.

\paragraph{ Solid Earth tide UTPM variations }
\label{solutpm}

The pioneering work in tidal effects on Earth orientation was that of
\cite{YoderWP81}.  It was limited to UT1, and included solid tidal
and some ocean tidal effects.  Their calculated $\Delta {\rm UT1}$ can
be represented as
\begin{equation}
\Delta {\rm UT1} = \sum \limits _{i=1} ^{N} \Biggl[ A_i \sin \biggl(
\sum \limits _{j=1} ^{5} k _{ij} \alpha _j \biggr) \Biggr] ~,
\end{equation}
where $N$=41 is chosen to include all terms with periods from 5 to 35
days.  There are no longer-period contributions until a period of 90 days
is reached.  However, these long-period terms \citeaffixed{IERS96b}{$e.g.$}
are already included in the
results reported by the current Earth-orientation measurement services.
The values for $ k _{ij} $ and $ A_i $, along with the periods involved,
are given in Table 4.37.1 of the {\it Explanatory Supplement} \cite{Arch92a}.
The $ \alpha _j $ are just the five fundamental arguments
defined in Eqs.~(\ref{eqalp1}~-~\ref{eqalp5}) as $l$, $l^{\prime}$, $F$,
$D$, and $\Omega$, respectively.

\paragraph{ Ocean tide UTPM variations }
\label{ocnutpm}

  Redistribution of the angular momentum produced by ocean tides affects
the Earth's rotation pole position and velocity.  This effect was first
quantified by \citeasnoun{Brosche82}, \citeasnoun{Baader83}, and
\citename{Brosche89}~\citeyear{Brosche89,Brosche91}.  The dominant effects
on polar motion and UT1 occur at diurnal, semidiurnal, fortnightly,
monthly, and semiannual tidal periods.  Assuming that the periods
longer than one day are adequately accounted for either in modeling
which combines solid and ocean tidal effects (not strictly
true with the Yoder model), or are already present in the $a~priori$
UTPM series, only the diurnal and semidiurnal frequencies need to
be modeled.  Further limiting the model to tidal components with
apparent amplitudes larger than 1 $\mu$s gives eight components.

  For unified notation, again define $\Theta_l$ ($l$=1,3) = x, y polar
motion and UT1, respectively.  Then the ocean tidal effects
$\Delta\Theta$ can be written as
\begin{equation}
\Delta\Theta_l = \sum \limits _{i=1} ^{N} \Biggl[
 A_{il} \cos \biggl( \sum \limits_{j=1}^{5} k_{ij} \alpha_j +
   n_i (h_\Upsilon + \pi) \biggr) +
 B_{il} \sin \biggl( \sum \limits_{j=1}^{5} k_{ij} \alpha_j +
   n_i (h_\Upsilon + \pi) \biggr) \Biggr] ~,
\end{equation}
$A_{il}$ and $B_{il}$ are the cosine and sine amplitudes that may
be calculated from theoretical tidal models (as in the work of Brosche)
or empirically determined from data
\cite{SJG93,Herr94,Watkins94,Gipson96}.
Table~\ref{shptid} lists the eight dominant terms in the model.
These numerical coefficients are taken from the empirical results of
\citeasnoun{SJG93}, and are representative of a number of recent
empirical models.  Note that the $K_1$ terms contain large arbitrary
conventional components, and are thus not solely due to ocean effects.
Theoretical calculations of polar motion ocean effects have been made
by \citeasnoun{ChaoR96}, yielding the amplitudes $A_{il}$ and $B_{il}$
using models based on TOPEX/Poseidon altimetry \cite{Fu94}.
The theoretical results agree well with empirical determinations, and
confirm the dominant role of ocean tides in excitation of short-period
UTPM variations.

  The ocean tidal UTPM effects are also modulated by the 18.6-year
lunar node variation $N^\prime$.  As in the case of ocean loading
station displacements (Sec.~\ref{ocnld}), the contributions
${\Delta\Theta_l}^{\prime}$ of the companion tides to $\Delta\Theta_l$
can be written as
\begin{eqnarray}
{\Delta\Theta_l}^{\prime}  = \sum \limits _{i=1} ^{N}
 \sum \limits _k r_{ki} \Biggl[
 &A_{il} \cos \biggl( \sum \limits_{j=1}^{5} k_{ij} \alpha_j +
   n_i (h_\Upsilon + \pi) + n_{ki} \omega_{N^\prime})t +
   n_{ki} N^\prime \biggr) +  \nonumber \\
 &B_{il} \sin \biggl( \sum \limits_{j=1}^{5} k_{ij} \alpha_j +
   n_i (h_\Upsilon + \pi) + n_{ki} \omega_{N^\prime})t +
   n_{ki} N^\prime \biggr) \Biggr] ~,
\end{eqnarray}
where the strengths of the companion tides $r_{ki}$ are found in
Table~\ref{compan}.

  Polar motion as seen in the reference frame $B$ which rotates with the
Earth is identical to nutation in the space-fixed celestial frame $S$.
A rotational frequency $\omega_B$ = 1 cycle per sidereal day (cpsd)
is identical to $\omega_S$ = 0, while $\omega_B$ = 0 corresponds to
$\omega_S$ = $-$1 cpsd.  Generally, nutations at frequency $\omega_S$
correspond to polar motions with frequencies $\omega_B$ = $-$1 + $\omega_S$.
The retrograde diurnal parts of the polar motion terms with coefficients
$A_{i\thinspace(1,2)}$ and $B_{i\thinspace(1,2)}$
corresponding to the tidal components listed in Table~\ref{shptid}
($i$ = 5 to 8) are thus equivalent to components of the nutation model,
and due care must be taken when both classes of parameters are estimated.

\subsubsection{ Nutation }
\label{nut}

  With the completion of the UT1 and polar motion transformations,
we are left with a station location vector, $ {\bf{r}} _{\rm date} =
{\bf U}{\bf X}{\bf Y} {\bf{r}} _t $.  This is the station location
relative to true
equatorial celestial coordinates of date.  The last set of
transformations are nutation ${\bf N}$, precession ${\bf P}$, and the
perturbation
rotation ${\bf \Omega}$, applied in that order.  These transformations give
the station location ${\bf{r}} _c$ in celestial equatorial coordinates:
\begin{equation}
{\bf{r}}_c = {\bf \Omega} {\bf P}{\bf N} { {\bf{r}} _{\rm date} } ~.
\end{equation}

  The first of these transformations, the matrix ${\bf N}$, is a
product of three separate rotations \cite{Melbourne68}:
\newcounter{ntrans}
\begin{list}%
{\arabic{ntrans}.}{\usecounter{ntrans}\setlength{\rightmargin}{\leftmargin}}
\item {\bf A}($\varepsilon$): true equatorial coordinates of date to ecliptic
coordinates of date,
\begin{equation}
{\bf A}( \varepsilon ) =
\left( \matrix { 1 &0 &0\cr 0 &~~~\cos \varepsilon &\sin \varepsilon\cr 0
&-\sin \varepsilon &\cos \varepsilon\cr } \right) ~,
\end{equation}
where $\varepsilon$ is the true obliquity of the ecliptic of date,
\item ${\bf C}^T ( \Delta \psi )$: nutation in longitude from ecliptic
coordinates of date to mean ecliptic coordinates of date,
\begin{equation}
{\bf C}^T ( \Delta \psi ) = \left( \matrix { ~~~\cos \Delta \psi &\sin
\Delta \psi &0\cr -\sin \Delta \psi &\cos \Delta \psi &0\cr 0 &0 &1 \cr }
\right) ~,
\end{equation}
where $\Delta \psi$ is the nutation in ecliptic longitude, and
\item ${\bf A}^T ( \overline \varepsilon )$: ecliptic coordinates of date
to mean equatorial coordinates.
\end{list}

  In ecliptic coordinates of date, the mean equinox is at an angle
$\Delta\psi$.  The angle $\Delta \varepsilon = \varepsilon -
\overline{\varepsilon}$ is the nutation in obliquity, and
$\overline{\varepsilon}$ is the mean obliquity (the dihedral angle between the
plane of the ecliptic and the mean plane of the equator). ``Mean'' as used in
this section implies that the short-period ($T \leq 18.6$ years) effects of
nutation have been removed.  Actually, the separation between nutation and
precession is rather arbitrary, but historical. The net rotation is
\begin{equation}
{\bf N} = {\bf A}^T ( \overline{\varepsilon} ) {\bf C}^T ( \Delta \psi )
{\bf A} ( \varepsilon )
\label{eqnutm}
\end{equation}
\begin{eqnarray*}
= \left( \matrix { ~~~~~~~~~\cos \Delta \psi &
\cos \varepsilon \sin \Delta \psi ~~~~~~~~~ &
\sin \varepsilon \sin \Delta \psi ~~~~~~~~~\cr
- \cos \overline{\varepsilon} \sin \Delta \psi &
~~\cos \overline{\varepsilon} \cos \varepsilon
\cos \Delta \psi ~+~ \sin \overline{\varepsilon} \sin \varepsilon &
~~\cos \overline{\varepsilon} \sin \varepsilon \cos \Delta \psi ~-~
\sin \overline{\varepsilon} \cos \varepsilon \cr
- \sin \overline{\varepsilon} \sin \Delta \psi &
~~\sin \overline{\varepsilon} \cos \varepsilon
\cos \Delta \psi ~-~ \cos \overline{\varepsilon} \sin \varepsilon &
~~\sin \overline{\varepsilon} \sin \varepsilon
\cos \Delta \psi ~+~ \cos \overline{\varepsilon} \cos \varepsilon \cr }
\right) ~.
\end{eqnarray*}

\medskip\medskip\noindent
It should again be pointed out (see Sec.~\ref{tercel} above) that
this is the reverse of the customary astronomical nutation.

  The 1980 IAU nutation model \cite{Sdlmn82,Kaplan81} is
used to obtain the values of $\Delta \psi$ and
$\varepsilon-\overline{\varepsilon}$.  The mean obliquity is
obtained from \citeasnoun{Lieske77} or from Kaplan \citeyear{Kaplan81}:
\begin{equation}
\overline{\varepsilon} = 23^{\circ} ~26^{\prime} ~21.^{\prime\prime}448
~-~46.^{\prime\prime} 8150 ~T ~-~5.^{\prime\prime}9\times10^{-4} T^2
~+~1.^{\prime\prime} 813\times10^{-3} T^3 ~,
\end{equation}
\begin{equation}
T = ({\rm{Julian}}~{\rm TDB}~{\rm{date}} ~-~ 2451545.0 ) / 36525.0 ~.
\label{eqtcen}
\end{equation}
The nutation in longitude $ \Delta \psi $ and in obliquity $ \Delta
\varepsilon = \varepsilon - \overline{\varepsilon}$ can be represented
by a series expansion of the sines and cosines of linear combinations
of five fundamental arguments.  The latter are nearly linear in time
\cite{Kaplan81,Hohenk92} (in radians):
\newcounter{farg}
\begin{list}%
{\arabic{farg}.}{\usecounter{farg}\setlength{\rightmargin}{\leftmargin}}
\item the mean anomaly of the Moon:
\begin{equation}
\alpha _1 = l = 2.35554839 + 8328.69142288T + 1.5180\times10^{-4}T^2
 + 3.1\times10^{-7}T^3 ~,
\label{eqalp1}
\end{equation}
\item the mean anomaly of the Sun:
\begin{equation}
\alpha _2 = l ^{\prime} = 6.24003594 + 628.30195602T - 2.80\times10^{-6}T^2
 - 5.8\times10^{-8}T^3 ~,
\end{equation}
\item the mean argument of latitude of the Moon:
\begin{equation}
\alpha _3 = F = 1.62790193 + 8433.46615832T - 6.4272\times10^{-5}T^2
 + 5.3\times10^{-8}T^3 ~,
\end{equation}
\item the mean elongation of the Moon from the Sun:
\begin{equation}
\alpha _4 = D = 5.19846951 + 7771.37714617T - 3.341\times10^{-5}T^2
 + 9.2\times10^{-8}T^3 ~,
\end{equation}
\item the mean longitude of the lunar ascending node on the ecliptic:
\begin{equation}
\alpha _5 = \Omega = 2.18243862 - 33.75704593T + 3.614\times10^{-5}T^2
 + 3.9\times10^{-8}T^3 ~.
\label{eqalp5}
\end{equation}
\end{list}

  With these fundamental arguments, the 1980 IAU nutation quantities
can then be represented by
\begin{equation}
\Delta \psi = \sum\limits_{j=1}^{N} \Biggl[ ~ ( A_{0j} + A_{1j}
T ) \sin~\biggl( \sum\limits_{i=1}^{5} k_{ji} \alpha _i (T) \biggr)
~ \Biggr]
\label{eqdelps}
\end{equation}
and
\begin{equation}
\Delta \varepsilon = \sum\limits_{j=1}^{N} \Biggl[ ~ ( B_{0j} + B_{1j}
T ) \cos~\biggl( \sum\limits_{i=1}^{5} k_{ji} \alpha _i (T) \biggr)
~ \Biggr] ~,
\label{eqdelep}
\end{equation}
where the various values of $ \alpha _i $, $ k_{ji} $, $ A_j $, and
$ B_j $ are tabulated in Table 3.222.1 of the {\it Explanatory Supplement}.

  Inadequacies of the standard nutation model can be corrected
by adding various classes of terms to the nutations
$\Delta \psi$ and $\Delta \varepsilon$ in Eqs.~(\ref{eqdelps})
and (\ref{eqdelep}).  These include the out-of-phase
nutations, the free-core nutations \cite{Yoder83} with period
$\omega_{\rm f}$ (nominally 430 days), and the ``nutation tweaks''
$\delta \psi$ and $\delta \varepsilon$, which are
arbitrary constant increments of the nutation angles
$\Delta \psi$ and $\Delta \varepsilon$.  Unlike the usual
nutation expressions, the tweaks have no time dependence.
The out-of-phase nutations $\Delta \psi^o$ and $\Delta \varepsilon^o$,
which are not included in the
1980 IAU nutation series, are identical to Eqs.~(\ref{eqdelps}) and
(\ref{eqdelep}), with the replacements {sin~$\leftrightarrow$~cos}:
\begin{equation}
\Delta \psi^o = \sum \limits_{j=1}^N \Biggl[ ( A_{2j} + A_{3j} T )
 \cos \biggl( \sum \limits_{i=1}^5 k_{ji} \alpha _i (T)
\biggr) \Biggr]
\end{equation}
and
\begin{equation}
\Delta \varepsilon^o = \sum \limits_{j=1}^N \Biggl[ ( B_{2j} + B_{3j} T
) \sin \biggl( \sum \limits_{i=1}^5 k_{ji} \alpha _i (T)
\biggr) \Biggr] ~.
\end{equation}

Expressions similar to these are adopted for the free-core nutations:
\begin{equation}
\Delta \psi^{\rm f} = ( A_{00} + A_{10} T ) \sin ( \omega_{\rm f} T ) +
( A_{20} + A_{30} T ) \cos ( \omega_{\rm f} T )
\end{equation}
and
\begin{equation}
\Delta \varepsilon^{\rm f} = ( B_{00} + B_{10} T ) \cos ( \omega_{\rm f} T ) +
( B_{20} + B_{30} T ) \sin ( \omega_{\rm f} T ) ~.
\end{equation}
Since the free-core nutation is retrograde, $\omega_{\rm f}$ is negative.
The nutation model thus contains a total of 856 parameters:
$A_{ij}$ ($i$=0,3; $j$=1,106) and $B_{ij}$ ($i$=0,3; $j$=1,106)
plus the free-nutation amplitudes $A_{i0}$ ($i$=0,3), $B_{i0}$ ($i$=0,3).
The only nonzero $a~priori$ amplitudes are the $A_{0j}$, $A_{1j}$,
$B_{0j}$, $B_{1j}$ ($j$=1,106).

  The nutation tweaks are just constant additive factors to
the angles $\Delta \psi$ and $\Delta \varepsilon $:
\begin{equation}
\Delta \psi \rightarrow \Delta \psi + \delta \psi ~~~~~ {\rm and}~~~~~
\Delta \varepsilon \rightarrow \Delta \varepsilon + \delta \varepsilon ~.
\end{equation}

  Deficiencies in the IAU nutation model became clearly evident in the
1980s \cite{Herr86}.  Several methods of correcting them are
in current use.  The first possibility is to use empirically determined
values of $\delta\psi$, $\delta\varepsilon$ that are available from the
IERS (see also Fig.~\ref{nutplt} in Sec.~\ref{aprot}).  With this choice
the nutation angles are determined by interpolating among results from
other VLBI experiments near the date of interest.  An alternative is to
estimate $\delta\psi$ and $\delta\varepsilon$ directly from the data.

  Another avenue for improving the $a~priori$ nutation model is
to select one of the published replacements of the 1980 IAU series.
Early work by Zhu $et~al.$~\citeyear{Zhu89,Zhu90} refined the 1980 IAU theory
of nutation both by re-examining the underlying Earth model and
by incorporating experimental results.  \citeasnoun{Herr91} has extended
the work of Zhu $et~al.$ and used geophysical parameters from
\citeasnoun{Mathews91} to generate the ZMOA 1990-2 (Zhu, Mathews, Oceans,
Anelasticity) nutation series.  \citeasnoun{Kinosh90} have
re-examined the rigid-Earth nutation theory, and attempted to include
all terms larger than 0.005 mas, in particular planetary terms not
present in any previous theories.  The 263 lunisolar terms
have been corrected for the Earth's non-rigidity \cite{Schy93}.
Other additions and corrections to the Kinoshita-Souchay model are
found in \citeasnoun{Schy96}, \citename{Wllms94}~\citeyear{Wllms94,Wllms95},
and \citeasnoun{Hartm96}.

  The Kinoshita-Souchay planetary contributions to $\Delta \psi$
and $\Delta \varepsilon$ are
\begin{equation}
\Delta \psi = \sum\limits_{j=1}^{N} \Biggl[ ~ S_{\psi j}
       \sin~\biggl( \sum\limits_{i=1}^{10} k_{ji} \beta _i (T) \biggr)
~+~ C_{\psi j}
       \cos~\biggl( \sum\limits_{i=1}^{10} k_{ji} \beta _i (T) \biggr)
~ \Biggr]
\label{eqksps}
\end{equation}
and
\begin{equation}
\Delta \varepsilon = \sum\limits_{j=1}^{N} \Biggl[ ~ S_{\varepsilon j}
       \sin~\biggl( \sum\limits_{i=1}^{10} k_{ji} \beta _i (T) \biggr)
~+~ C_{\varepsilon j}
       \cos~\biggl( \sum\limits_{i=1}^{10} k_{ji} \beta _i (T) \biggr)
~ \Biggr] ~,
\label{eqksep}
\end{equation}
where the astronomical arguments are symbolized by $\beta _i$;
the last four $\beta_i$ are identical with the $\alpha_i$ defined above
($\beta_7 = D = \alpha_4$, $\beta_8 = F = \alpha_3$, $\beta_9 = l =
\alpha_1$, $\beta_{10} = \Omega = \alpha_5$), while the first five
are mean heliocentric longitudes of the planets (in units of radians):

\begin{eqnarray}
~&1.~~{\rm Venus:} ~~~~ \beta _1 ~&= l_{\rm V} =~ 3.176146697 ~+~
 1021.3285546 ~T ~,   ~\\
~&2.~~{\rm Earth:} ~~~~ \beta _2 ~&= l_{\rm E} =~ 1.753470314 ~+~
  628.30758492 ~T ~,  ~\\
~&3.~~{\rm Mars:} ~~~~~ \beta _3 ~&= l_{\rm M} =~ 6.203480913 ~+~
  334.06124315 ~T ~,  ~\\
~&4.~~{\rm Jupiter:} ~~ \beta _4 ~&= l_{\rm J} =~ 0.599546497 ~+~
   52.96909651 ~T ~,  ~\\
~&5.~~{\rm Saturn:} ~~~ \beta _5 ~&= l_{\rm S} =~ 0.874016757 ~+~
   21.32990954 ~T ~,
\label{plnarg}
\end{eqnarray}
\noindent and the sixth is the accumulated general precession:
\begin{equation}
\beta _6 = p_{\rm A} =~ 0.02438175 ~T ~+~ 5.38691\times10^{-6} ~T^2 ~.
\label{agp}
\end{equation}
\noindent
It should be noted that the paper of Kinoshita and Souchay gives
expressions for the lunisolar tidal arguments that are slightly at
variance with the IAU formulas presented above in
Eqs.~(\ref{eqalp1}~-~\ref{eqalp5}).  These differences may
be of significance in high-accuracy modeling studies.

  Since the present standard model of nutation is known to be in error
by amounts that are large in comparison to present measurement capabilities
(see Fig.~\ref{nutplt} in Sec.~\ref{aprot}), the International Astronomical
Union considers it important to formulate and adopt an improved nutation
model by the end of the century.  A working group is presently considering
variants of the ZMOA and Kinoshita-Souchay models in this connection.
In the meantime, the IERS~\citeyear{IERS96a} maintains empirical time series
of the nutation angles $\Delta \psi$, $\Delta \varepsilon$ for use in
applications that require higher accuracy than the 1980 IAU model.

\subsubsection{ Precession }
\label{prec}

The next transformation in going from the terrestrial frame to the
celestial frame is the rotation ${\bf P}$.  This is the precession
transformation from mean equatorial coordinates of date to the equatorial
coordinates of the reference epoch ($e.g.$, J2000.0).
As was the case with the nutation
matrix of Eq.~(\ref{eqnutm}), this is a rotation whose sense is opposite to
that of the conventional astrometric precession.  It is a composite of
three rotations discussed in detail by \citeasnoun{Melbourne68} and
\citeasnoun{Lieske77}:
\begin{equation}
{\bf R}(-Z) =
\left( \matrix { ~~~\cos Z& \sin Z& 0\cr -\sin Z& \cos Z& 0\cr
~~0& 0& 1\cr } \right) ~,
\end{equation}
\begin{equation}
$$ ~~~{\bf Q}(\Theta) =
\left( \matrix { ~~~\cos \Theta& 0& \sin \Theta\cr
~~0& 1& 0\cr -\sin \Theta& 0& \cos \Theta\cr } \right) ~,
\end{equation}
\begin{equation}
{\bf R}(-\zeta) =
\left( \matrix { ~~~\cos \zeta& \sin \zeta& 0\cr
-\sin \zeta& \cos \zeta& 0\cr ~~0& 0& 1\cr } \right) ~.
\end{equation}
\begin{equation}
{\bf P} = {\bf R}(-\zeta) {\bf Q}(\Theta) {\bf R}(-Z)
\end{equation}
\begin{eqnarray*}
= \left( \matrix {{\phantom{+}}\cos\zeta \cos\Theta \cos Z - \sin\zeta \sin Z &
 {\phantom{+}}\cos\zeta \cos\Theta \sin Z + \sin\zeta \cos Z &
 {\phantom{+}}\cos\zeta \sin\Theta \cr
-\sin\zeta \cos\Theta \cos Z - \cos\zeta \sin Z &
 -\sin\zeta \cos\Theta \sin Z + \cos\zeta \cos Z &
 -\sin\zeta \sin\Theta \cr
-\sin\Theta \cos Z & -\sin\Theta \sin Z & \cos\Theta \cr } \right) ~.
\end{eqnarray*}
The auxiliary angles $\zeta$, $\Theta$, $Z$ depend on precession
constants, obliquity, and time as
\begin{eqnarray}
\zeta ~&=&~ 0.5mT ~+~ 0^{\prime\prime}.30188 ~ T^2
 ~+~ 0^{\prime\prime}.017998 ~T^3  \label{eqzeta} \\
Z  ~&=&~ 0.5mT ~+~ 1^{\prime\prime}.09468 ~T^2
 ~+~ 0^{\prime\prime}.018203 ~T^3~  \label{eqzed} \\
\Theta ~&=&~ ~~~~~nT ~-~ 0^{\prime\prime}.42665 ~T^2
 ~-~ 0^{\prime\prime}.041833 ~T^3 ~,
\label{eqtheta}
\end{eqnarray}
where the speeds of precession in right ascension and declination
are, respectively,
\begin{eqnarray}
m &=& p_{\rm LS} \cos \bar \varepsilon_0 - p_{\rm PL}  \\
n &=& p_{\rm LS} \sin \bar \varepsilon_0
\end{eqnarray}
and $p_{\rm LS}$ = the lunisolar precession constant
(including the geodesic precession), $p_{\rm PL}$ = the
planetary precession constant, $\bar \varepsilon_0$ = the obliquity
at J2000.0, and $T$ (Eq.~\ref{eqtcen})
is the time in centuries past J2000.0.  Nominal values
at J2000.0 are $p_{\rm LS}$ = $5038^{\prime\prime}.7784$/cy, $p_{\rm PL}$ =
$10^{\prime\prime}.5526$/cy; these yield the expressions given by
\citeasnoun{Lieske77} and \citename{Kaplan81}~\citeyear{Kaplan81}:
\begin{eqnarray}
\zeta  ~&=&~ 2306^{\prime\prime}.2181 ~T ~+~ 0^{\prime\prime}.30188
 ~ T^2 ~+~ 0^{\prime\prime}.017998 ~T^3  \\
\Theta ~&=&~ 2004^{\prime\prime}.3109 ~T ~-~ 0^{\prime\prime}.42665
 ~T^2 ~-~ 0^{\prime\prime}.041833 ~T^3  \\
Z ~&=&~ 2306^{\prime\prime}.2181 ~T ~+~ 1^{\prime\prime}.09468
 ~T^2 ~+~ 0^{\prime\prime}.018203 ~T^3  ~,
\end{eqnarray}

  Direct estimates of precession corrections can be obtained from
observations.  The most recent such results \cite{Charprc95,WltrSov96}
indicate that the current IAU nominal value
of $p_{\rm LS}$ is in error by $-$3.0$\pm$0.2 milliarcseconds per year.
The precession matrix completes the standard model for the orientation
of the Earth.

\subsubsection{ Perturbation rotation }
\label{pert}

  For exploring departures from the canonical precession, nutation,
and UTPM transformations, the standard model for the rotation of the
Earth as a whole may be modified by small incremental rotations
about the resulting axes.  Define such a perturbation rotation matrix as
\begin{equation}
{\bf \Omega} ~=~ {\bf \Delta} _x {\bf \Delta} _y {\bf \Delta} _z ~,
\end{equation}
where
\begin{equation}
{\bf \Delta} _x ~=~ \left( \matrix { 1 &0 &0\cr 0 &1 & \delta \Theta _x\cr
0 & -\delta \Theta _x &1\cr } \right)
\end{equation}
with $ \delta \Theta _x $ being a small angle rotation about the $x$ axis,
in the sense of carrying $y$ into $z$;
\begin{equation}
{\bf \Delta} _y ~=~ \left( \matrix { 1 &0 & -\delta \Theta _y\cr
0 &1 &0\cr \delta \Theta _y &0 &1\cr } \right)
\end{equation}
with $ \delta \Theta _y $ being a small angle rotation about the $y$ axis,
in the sense of carrying $z$ into $x$; and
\begin{equation}
{\bf \Delta} _z ~=~ \left( \matrix { 1 & \delta \Theta _z &0\cr
-\delta \Theta _z &1 &0\cr 0 &0 &1\cr } \right)
\end{equation}
with $ \delta \Theta _z $ being a small angle rotation about the $z$ axis,
in the sense of carrying $x$ into $y$.
For angles on the order of 1 arc second we can neglect terms on the order
$ \delta \Theta ^2 R_{\rm E} $ as they give effects on the order of 0.15 mm.
Thus, in that approximation
\begin{equation}
{\bf \Omega} ~=~ \left( \matrix { 1 & ~~\delta \Theta _z & -\delta \Theta _y\cr
-\delta \Theta _z &1 & ~~\delta \Theta _x\cr ~~\delta \Theta _y
& -\delta \Theta _x &1\cr } \right) ~.
\end{equation}
Each of the rotation angles can be expressed as a function of time:
\begin{equation}
\delta \Theta _i ~=~ \delta \Theta _i ( T ) ~=~ \delta \Theta _{i0}
+ \delta \dot \Theta _i T +  f_i (T) ~,
\end{equation}
which is the sum of an offset, a time-linear rate, and some higher order
or oscillatory terms.  In particular,
a non-zero value of $\delta \dot \Theta _y$ is equivalent to a change in
the precession constant, and $\delta \dot \Theta _x$ is equivalent to the
time rate of change of the obliquity $\varepsilon$.  Setting
\begin{equation}
\delta \Theta _x = \delta \Theta _y = \delta \Theta _z = 0
\end{equation}
gives the effect of applying only the standard rotation matrices.

  Starting in the terrestrial frame with the Earth-fixed vector
${\bf{r}} _0$ to which tidal effects ${\bf \Delta}$ are added, we have
shown (in Secs.~\ref{staloc} through \ref{tercel} above) how we obtain
the same vector ${\bf{r}}_c$ expressed in the geocentric
celestial frame:
\begin{equation}
{\bf{r}}_c ~=~ {\bf \Omega} {\bf P} {\bf N} {\bf U} {\bf X} {\bf Y}
( {\bf{r}}_0 + {\bf{\Delta}} ) ~.
\end{equation}

\subsection{ Earth orbital motion }
\label{emot}

  We now wish to transform these station locations from a geocentric
celestial reference frame moving with the Earth to a celestial reference
frame which is at rest relative to the center of mass of the Solar System.
The reason for this apparent complication is that
the Solar System barycentric (SSB) frame is particularly useful for
expressing the source positions, formulating gravitational retardation,
and performing trajectory calculations of interplanetary spacecraft.
These SSB station locations will then be used to calculate the geometric
delay (see Sec.~\ref{arrtim}).  We will then transform the resulting time
interval back to the frame in which the time delay is actually measured
by the interferometer $-$ the geocentric celestial frame moving with the
Earth.

  Let ${\displaystyle \Sigma ^{\prime}}$ be a celestial geocentric frame
moving with vector velocity ${\bbox{\beta}} c$ relative to a frame
${\displaystyle \Sigma}$ at rest relative to the
Solar System center of mass.  Due to the Earth's orbital motion, $\beta$
is on the order $10^{-4}$.  Further, let $ {\bf{r}} (t)$ be the position
of a point ($e.g.$, station location) in space as a function of time, $t$,
as measured in the ${\displaystyle \Sigma}$ (SSB) frame.  In the
${\displaystyle \Sigma ^{\prime}}$
(geocentric) frame, there is a corresponding position
$ {\bf{r}} ^{\prime} (t ^{\prime} )$ as a function of time, $t ^{\prime} $.
We normally observe and model $ {\bf{r}} ^{\prime} (t ^{\prime} ) $
as shown in Secs.~\ref{staloc} through \ref{tercel}.  However, in
order to calculate the geometric delay in the SSB frame
${\displaystyle \Sigma}$, we will need the transformations of ${\bf{r}} (t)$
and $ {\bf{r}} ^{\prime} (t ^{\prime} )$, as well as of $t$ and
$ t ^{\prime}$, as we shift frames of reference. Measuring positions
in units of light travel time, we have from \citeasnoun{Jackson} the
Lorentz transformation:
\begin{equation}
{\bf{r}} ^{\prime} (t ^{\prime}) = {\bf{r}} (t) + (\gamma - 1)
[{\bf{r}}(t) \cdot {\bbox{\beta}}] {\bbox{\beta}} / { \beta ^2 }
- \gamma {\bbox{\beta}} t
\end{equation}
\begin{equation}
t ^{\prime} = \gamma \lbrack t - {\bf{r}} (t) \cdot {\bbox{\beta}} \rbrack ~,
\end{equation}
and the inverse transformation:
\begin{equation}
{\bf{r}} (t)  = {\bf{r}} ^{\prime} (t ^{\prime} ) + ( \gamma - 1 )
[{\bf{r}} ^{\prime} (t ^{\prime} ) \cdot {\bbox{\beta}}] {\bbox{\beta}}
/ \beta ^2 + \gamma {\bbox{\beta}} t ^{\prime}
\label{eqlorr}
\end{equation}
\begin{equation}
t = \gamma \lbrack t ^{\prime} + {\bf{r}} ^{\prime} (t ^{\prime}) \cdot
{\bbox{\beta}} \rbrack ~,
\label{eqlort}
\end{equation}
where in this section
\begin{equation}
\gamma = { (1 - {\bbox{\beta}}^2) } ^{ -{1/2} } ~.
\end{equation}

   Let $ t_1 $ represent the time measured in the SSB frame
${\displaystyle \Sigma}$, at which a wave front crosses antenna 1
at position $ {\bf{r}} _1 (t_1) $.  Let $ {\bf{r}} _2 (t_1) $
be the position of antenna 2 at this same time, as measured in the
SSB frame.  Also, let $ t_2 $
be the time measured in this frame at which that same wave front intersects
station 2.  This occurs at the position $ {\bf{r}} _2 ( t_2 ) $.
Following Sec.~\ref{arrtim}, Eq.~(\ref{eqgdel}) we can calculate
the geometric delay $ t_2 - t_1 $.  Transforming this time interval
back to the geocentric ${\displaystyle \Sigma ^{\prime}}$ frame, we obtain
\begin{equation}
{ t_2 } ^{\prime} - t_1 ^{\prime} = \gamma ( t_2 - t_1 ) -
\gamma \lbrack {\bf{r}} _2 ( t_2 ) - {\bf{r}} _1 ( t_1 ) \rbrack
\cdot {\bbox{\beta}} ~.
\end{equation}
Assume further that the motion of station 2 (with barycentric velocity
${\bbox{\beta}} _2$) is rectilinear over this time interval.
This assumption is not strictly true but, as discussed below, the
resulting error is much less than 1 mm in calculated delay.  Thus
\begin{equation}
{\bf{r}} _2 (t_2) = {\bf{r}} _2 (t_1) + {\bbox{\beta}} _2 (t_2 -
t_1) ~,
\end{equation}
which gives
\begin{equation}
{\bf{r}} _2 (t_2) - {\bf{r}}_1 (t_1) = {\bf{r}}_2 (t_1) -
{\bf{r}}_1 (t_1) + {\bbox{\beta}}_2 (t_2 - t_1 )
\end{equation}
and
\begin{eqnarray}
{ t_2 } ^{\prime} - t_1 ^{\prime} &=& \gamma ( t_2 - t_1 ) ~-~ \gamma
\lbrack {\bf{r}}_2 (t_1) - {\bf{r}}_1 (t_1) \rbrack \cdot
{\bbox{\beta}} ~-~ \gamma {\bbox{\beta}} _2 \cdot {\bbox{\beta}}
\lbrack t_2 - t_1 \rbrack   \nonumber \\
&=& \gamma ( 1 - {\bbox{\beta}} _2 \cdot {\bbox{\beta}} )
( t_2 - t_1 ) ~-~ \gamma \lbrack {\bf{r}}_2 (t_1) - {\bf{r}}_1
(t_1) \rbrack \cdot {\bbox{\beta}} ~.
\end{eqnarray}
This is the expression for the geometric delay that would be observed in
the geocentric ${\displaystyle \Sigma ^{\prime}}$ frame in terms of the
SSB geometric delay and SSB station positions.

  Since our calculation starts with station locations given in the
geocentric frame, it is convenient to obtain an expression for
$ {\bf{r}}_2 (t_1) - {\bf{r}}_1 (t_1) $ in terms
of quantities defined in that same geocentric frame.  To obtain such an
expression consider two events $ \lbrack {\bf{r}}_1 ^{\prime} (t_1 ^{\prime}),~
{\bf{r}}_2 ^{\prime} (t_1 ^{\prime}) \rbrack $ that are geometrically separate,
but simultaneous, in the geocentric frame, and occurring at time $t_1
^{\prime} $.  These two events appear in the SSB frame as
\begin{equation}
{\bf{r}}_1 (t_1) = {\bf{r}} _1 ^{\prime} (t_1 ^{\prime}) + (\gamma - 1)
[{\bf{r}}_1 ^{\prime} (t_1 ^{\prime}) \cdot {\bbox{\beta}}] {\bbox{\beta}}
/ \beta ^2 + \gamma {\bbox{\beta}} t_1 ^{\prime}
\label{eqr1t1}
\end{equation}
and as
\begin{equation}
{\bf{r}}_2 (t_2) = {\bf{r}} _2 ^{\prime} ( t_1 ^{\prime} ) + (\gamma - 1)
[{\bf{r}}_2 ^{\prime} ( t_1 ^{\prime} ) \cdot {\bbox{\beta}}]
{\bbox{\beta}} / \beta ^2 + \gamma {\bbox{\beta}}
t_1 ^{\prime} ~,
\end{equation}
where
\begin{equation}
t_2 - t_1 = \gamma \lbrack {\bf{r}}_2 ^{\prime} (t_1 ^{\prime}) -
{\bf{r}}_1 ^{\prime} (t_1 ^{\prime}) \rbrack \cdot {\bbox{\beta}} ~.
\end{equation}
With these three equations and the expression
\begin{equation}
{\bf{r}}_2 (t_2) = {\bf{r}}_2 (t_1) + {\bbox{\beta}} _2 \lbrack
t_2 - t_1 \rbrack
\end{equation}
we may obtain the vector $ {\bf{r}}_2 (t_1)$:
\begin{eqnarray}
{\bf{r}}_2 (t_1) ~=~ &~&{\bf{r}} _2 ^{\prime} (t_1 ^{\prime}) + (\gamma - 1)
[{\bf{r}}_2 ^{\prime} (t_1 ^{\prime}) \cdot {\bbox{\beta}}] {\bbox{\beta}}
/ \beta ^2 \nonumber \\
&~&+ \gamma {\bbox{\beta}} t_1 ^{\prime}
- \gamma {\bbox{\beta}} _2 \lbrack {\bf{r}}_2 ^{\prime} (t_1 ^{\prime}) -
{\bf{r}}_1 ^{\prime} (t_1 ^{\prime}) \rbrack \cdot {\bbox{\beta}} ~.
\label{eqr2t1}
\end{eqnarray}
This is the position of station 2 at the time $t_1$ as
observed in ${\displaystyle \Sigma}$.  From this we obtain:
\begin{eqnarray}
{\bf{r}}_2 (t_1) - {\bf{r}}_1 (t_1) ~=~ {\bf{r}}_2 ^{\prime} (t_1 ^{\prime})
- {\bf{r}}_1 ^{\prime} (t_1 ^{\prime}) &+& (\gamma - 1) \biggl(
\lbrack {\bf{r}}_2
^{\prime} (t_1 ^{\prime}) - {\bf{r}}_1 ^{\prime} (t_1 ^{\prime}) \rbrack \cdot
{\bbox{\beta}}\biggr) {\bbox{\beta}} / \beta ^2  \nonumber \\
&-& \gamma {\bbox{\beta}} _2 \lbrack {\bf{r}}_2 ^{\prime} (t_1 ^{\prime}) -
{\bf{r}}_1 ^{\prime} (t_1 ^{\prime}) \rbrack \cdot {\bbox{\beta}} ~.
\label{eqdr}
\end{eqnarray}
As shown in Sec.~\ref{arrtim}, the vectors
$\lbrack {\bf{r}}_2 (t_1) - {\bf{r}}_1 (t_1) \rbrack$ and $ {\bbox{\beta}}_2 $
are all that is needed to obtain $t_2 - t_1 $ for the case of plane waves.
For curved wave fronts we will need to know the
individual station locations in the barycentric frame as well.
These we obtain from Eqs.~(\ref{eqr1t1}) and (\ref{eqr2t1})
with $t_1 ^{\prime} $ set equal
to zero.  Setting $t_1 ^{\prime} = 0$ is justified since the origin
of time is arbitrary when calculating time differences.

  In implementing these transformations, the relationship
for the transformation of velocities is also needed.
Taking differentials of Eqs.~(\ref{eqlorr}) and (\ref{eqlort}) we have:
\begin{equation}
d {\bf{r}} = d {\bf{r}} ^{\prime} + (\gamma - 1) (d {\bf{r}} ^{\prime}
\cdot {\bbox{\beta}}) {\bbox{\beta}} / \beta ^2 +
\gamma {\bbox{\beta}} dt ^{\prime} ~,
\end{equation}
\begin{equation}
dt = \gamma ( dt ^{\prime} + d {\bf{r}} ^{\prime} \cdot {\bbox{\beta}} ) ~.
\end{equation}
Dividing to obtain $ { d {\bf{r}} } / dt$ we obtain for station 2
in the SSB ${\displaystyle \Sigma}$ frame:
\begin{equation}
{\bbox{\beta}} _2 = \Bigl[ { {\bbox{\beta}} _2 ^{\prime} + ( \gamma -1 )
({\bbox{\beta}} _2 ^{\prime} \cdot {\bbox{\beta}})
{\bbox{\beta}} / \beta ^2 +
\gamma {\bbox{\beta}} } \Bigr] \bigg/ \Bigl[ { \gamma ( 1 + {\bbox{\beta}}
_2 ^{\prime} \cdot {\bbox{\beta}} ) } \Bigr] ~.
\label{eqbeta2}
\end{equation}
All barycentric positions and velocities required for the calculations
in this section are obtained from planetary ephemerides in the J2000.0
frame \cite{Standish82,StandNew85}.

   For station 2 relative to the geocentric origin, we have from
Eqs.~(\ref{eqq}) and (\ref{eqrcrt}):
\begin{equation}
{\bbox{\beta}} _2 ^{\prime} \approx {\bf \Omega} \thinspace {\bf P}
\thinspace {\bf N} \thinspace { d{\bf U} \over dH } \thinspace {\bf X}
\thinspace {\bf Y} \thinspace {\bf{r}} _2 ^{\prime} \thinspace
\omega_{\rm E} ~,
\end{equation}
where the inertial rotation rate of the Earth is
\begin{equation}
\omega_{\rm E} = 7.292115\times10^{-5} ~~~ {\rm radians~per~second}.
\label{erot}
\end{equation}
This is not a critical number since it is used only
for station velocities, or to extrapolate Earth rotation forward for very
small fractions of a day ($i.e.$, typically less than 1000 seconds).

   The assumption of rectilinear motion can be shown to result in negligible
errors.  Using the plane wave front approximation of Eq.~(\ref{eqgdel}),
an estimate of the error $ \delta \tau $ in the calculated delay
due to an error $ \Delta {\bbox{\beta}} _2 $ in the above value of
${\bbox{\beta}}_2$ is
\begin{equation}
\delta \tau = {\bf{\widehat k}} \cdot \lbrack {\bf{r}}_2 (t_1)
- {\bf{r}}_1 (t_1) \rbrack
{ \Biggl( { { 1 } \over { 1 - {\widehat {\bf{k}}} \cdot ( {\bbox{\beta}}_2 +
\Delta {\bbox{\beta}}_2 ) } } - { { 1 } \over { 1 - {\widehat {\bf{k}}}
\cdot {\bbox{\beta}}_2 } } \Biggr) } \approx \tau \Delta {\bbox{\beta}}_2 ~.
\end{equation}
Further, from Eq.~(\ref{eqbeta2}) above (since $\gamma \approx 1 + 10^{-8}$),
\begin{equation}
\Delta {\bbox{\beta}}_2 \approx \Delta {\bbox{\beta}}_2 ^{\prime} ~.
\end{equation}
For the vector $ {\bbox{\beta}}_2 ^{\prime} $ in a frame rotating with angular
velocity $\omega$, the error $\Delta {\bbox{\beta}}_2 ^{\prime} $ that
accumulates in the time interval $\tau$
due to neglecting the rotation of that frame is
\begin{equation}
\Delta {\bbox{\beta}}_2 ^{\prime} \approx
 {\bbox{\beta}}_2 ^{\prime} \omega \tau ~.
\end{equation}
Thus for typical Earth-fixed baselines, where $\tau \leq$ 0.02 s,
neglecting the curvilinear motion of station 2 due to the
rotation of the Earth causes an error of
$< 4 \times 10^{-14}$ s, or 0.012 mm, in the calculation of $\tau$.
Similarly, neglecting the orbital character of the Earth's motion causes
a maximum error on the order of 0.0024 mm.

  A related transformation from the ${\displaystyle \Sigma}$ to
${\displaystyle \Sigma ^{\prime}}$ frames
that is needed for models of antenna axis offsets and atmospheric effects
is an expression for the ``aberrated'' source direction unit vector
${\bf{\widehat s}}_0$.  In the SSB ${\displaystyle \Sigma}$ frame,
this vector is just
the negative of the propagation vector of Eq.~(\ref{celvec}):
${\bf{\widehat s}}_0 = -{\bf{\widehat k}}$.  The Earth's diurnal
rotational contribution to aberration is two orders of magnitude smaller
than that due to the Earth's orbital velocity and can be neglected for 
axis offset and tropospheric models.  The source unit vector
${\bf{\widehat s}}$ in the geocentric celestial frame is then given
by the Lorentz transformation
\begin{equation}
{\bf{\widehat s}} = a_s {\bf{\widehat s}}_0 + a_\beta {\bbox{\beta}}
\label{aberr}
\end{equation}
with the coefficients
\begin{equation}
a_s = { { 1 } \over {\gamma (1 + \bbox{\beta} \cdot {\widehat {\bf{s}}_0})}}
~~~~~ {\rm{and}} ~~~~~
a_\beta = a_s \bigl[ (1-\gamma) {{ \bbox{\beta} \cdot {\widehat {\bf{s}}_0} }
\over {\bbox{\beta}^2} } + \gamma \bigr] ~.
\end{equation}

  Working in a frame at rest with respect to the center of mass of the
Solar System causes relativistic effects due to the motion of the Solar
System in the ``fixed frame'' of the extragalactic radio sources to be
absorbed into the mean positions of the sources and their proper motions.
With observational data extending
over a sufficiently long time span, this motion in inertial space
should be separable from terrestrial and intra-Solar-System dynamics.
The known motions of the Solar System barycenter with respect to four
standards are listed in Table~\ref{mottab}.  Both the magnitudes and
directions are tabulated, with the latter given in terms of Galactic
longitude $l$ and latitude $b$.  In
order of increasing velocity, they are given with respect to the
local standard of rest (LSR), Galactic center (GC), Local Group (LG),
and cosmic microwave background (CMB).  Note that the motion of the
SSB relative to the CMB is on the order of 10$^{-3} c$,
indicating that relativity should play a significant
role in modeling its consequences for VLBI.  Since the extragalactic
radio sources are expected to be between the LG and CMB,
the motion of the Solar System relative to the ``fixed'' extragalactic
reference frame (EGRF) is also expected to be of similar magnitude.
Detection of this motion could contribute significantly to understanding
of the large-scale structure of the universe.

  We note at this point that the extragalactic radio sources have a median
redshift $z~\approx~1.2$, whereas the more distant CMB is characterized by
redshift $z=10^3$ \cite{Melchi90}.
If the extragalactic radio sources are assumed to be at rest relative
to the CMB, one then has an estimate of the SSB velocity in the extragalactic
radio frame.  Motion of the SSB relative to the EGRF potentially contributes
to the observed delays in VLBI experiments in three ways: geometric,
gravitational, and aberrational effects.

{\it Geometric effects:} The geometric effects of galactic rotation
can be easily estimated.  In the vicinity of the Sun, the period for
galactic rotation is approximately 240 million years.  Thus our angular
velocity about the galactic center is $\approx 2 \pi / 2.4 \times 10^8 =
2.6 \times 10^{-8}$ radians/year.  For sources within the Galaxy,
at distances approximately equal to our distance from the galactic center,
the apparent positions could change by $\approx 26$
nrad/yr.  An intercontinental baseline (10,000 km) could thus be in
error by as much as 26 cm/yr (1 nrad $\approx$ 1 cm) if measurements
were based on sources within the Galaxy.  Since our distance from the
galactic center is $8.5 \pm 1$ kpc $~\approx 2.7 \times 10^4$ light
years \cite{Binney},
and most extragalactic radio sources are believed to be $\approx 10^9$
light years distant, the potential baseline error is scaled by the ratio
of these two distances, $\approx 3 \times 10^{-5}$, and becomes only
$\approx$ 0.01 mm/year.  Even with the present near-20-year history
of VLBI data, the purely geometric systematic error due to galactic
rotation is negligible, and only exceeds the millimeter level
for sources closer than 10 million light years.

{\it Gravitational delay effects:}
Gravitational delay effects depend on the relative positions of the radio
source, the intervening masses, and the receiving station.  Thus one must
consider how the motion of the SSB changes these effects.  The central
bulge of the Milky Way galaxy contains $\approx10^{11}$ solar masses
and it is about $8.5$ kpc ($\approx2\times 10^9$ AU) from the Solar System.
The Galactic gravitational delay is thus estimated to be approximately 40 times
larger than the solar delay.  While the solar effect varies on the time scale
of a year, the galactic effect varies on a time scale of 240 million years.
As with the galactic aberration effects, we choose to absorb the static
portion (maximum of 4 arc seconds at $10^{\circ}$ from the GC) into the
reported radio source positions. We are left to model the variation of this
effect with time. Given that the SSB's angular motion about the GC is 5
mas/yr and that scintillations from the interstellar medium limit VLBI
observations at 2 to 8 GHz frequencies to no closer than $10^{\circ}$ from
the GC, the largest expected change in deflection would be $< 0.5~\mu$as
per year.

{\it Aberration effects:}
The 0.0012$c$ SSB--CMB velocity will cause large ($\approx 4$
arc min) changes in apparent source positions due to aberration.
To the extent that the SSB velocity does not change on time scales
of decades, however, one is free to absorb these aberration
effects into reported source positions.  With such a convention, the problem
is reduced to considering changes in the SSB velocity.

  Based on the results of studies in galactic dynamics
\citeaffixed{Binney}{$e.g.$}, one may construct a model of the
SSB's acceleration both toward the Galactic Center (GC) and toward the
Local Standard of Rest (LSR).  At present we have not modeled the motion
of the GC relative to the Local Group (LG) nor have we modeled the motion
of the Local Group relative to the cosmic microwave background.  This is
equivalent to assuming that the GC-LG and LG-CMB velocities are constant
on time scales of decades ($i.e.$, acceleration
$\ll 2 \times 10^{-11} c \rm /yr$).

SSB-LSR velocity: We consider a circular motion of radius $\approx 1$ kpc
with a period of $\approx 200$ million years.  This gives a change in velocity
\begin{equation}
 \dot V_{_{\rm SSB-LSR}} ~=~ 2 \times 10^{-12} ~ c / \rm yr
\end{equation}
directed toward the local standard of rest.
This acceleration will cause a maximum aberration of $\approx 0.5~\mu$as/yr
which can be ignored because of its small size.

LSR-GC velocity: We consider a circular motion of radius 8.5 kpc
with a period of 240 million years.  This gives a change in velocity
\begin{equation}
 \dot V_{_{\rm LSR-GC}} ~=~ 2 \times 10^{-11} ~ c / \rm yr
\end{equation}
directed toward the Galactic center ($\alpha = 17^{\rm h}45^{\rm m}.5,~\delta =
-28^{\circ}56^\prime$).
This velocity will cause a maximum aberration of $\approx 5~\mu$as/yr.
This aberration effect is at present not included in $a~priori$ VLBI
models.  It is, however, possible to estimate the SSB's velocity
{\it change} from the data.

\subsection{ Source structure }
\label{sstruc}

  By analogy with the time dependence of station coordinates caused by
tectonic and tidal motion, non-stationarity of the source coordinates
must also be considered.  In VLBI this complication arises because any
structure or motion of the radio sources is manifested in slight
variations of the observed delays.  Likewise, in generalization of
the model to ranging experiments (interplanetary spacecraft,
GPS, SLR, LLR) the ``structure''
of the ``source'' includes positions of the transmitter relative to
the center of mass of the satellite, and its orbital path relative
to the Earth ($i.e.$, the GPS, Lageos, or Moon ephemerides).  For VLBI,
such effects can be eliminated to a large extent by judicious choice of
objects in planning the experiments.  With the exception of special purpose
experiments (such as those of \citeasnoun{Lestrade95} on radio stars),
the sources are well outside our Galaxy, ensuring minimal proper motion.
Numerous astrophysical studies during the past two decades have shown
that compact extragalactic radio sources exhibit structure on a
milliarcsecond scale \citeaffixed{KellT81}{$e.g.$}.
Such studies are important for developing models of the origin of radio
emission of these objects.  Many radio source structures are found to be
quite variable with frequency and time
\cite{Zensus87,Taylor94a,Taylor94b,Polat95,Thkkr95,Henst94,Fey96}.
Survey maps of 187 radio
sources in the \citeasnoun{Taylor94a} reference showed that the structures
of only 8 sources did not exceed a scale of 1 mas at an observing
frequency of 5 GHz.  If extragalactic sources are to serve as reference
points in a reference frame that is stable at a level below 1 mas, it is
important to correct for the effects of their structures in astrometric
VLBI observations.

  Corrections for the effects of source internal structures are based
on work by \citeasnoun{Thomas80}, \citeasnoun{Ulvst88}, and
\citename{Charlot89}
~\citeyear{Charlot89,Charlot90a}.
A varying non-point-like distribution of the intensity of a source yields
time dependent corrections to the group delay and phase delay rate
observables, $\Delta\tau_s$ and $\Delta\dot\tau_s$, that may be written
in terms of the source's intensity distribution $I({\bf{s}}, \omega, t)$ as
\begin{equation}
\Delta \tau_s = {\partial \phi_s}/{\partial\omega},~~~~~~~~~~~
\Delta \dot\tau_s = {\partial \phi_s}/{\partial t} ~,
\end{equation}
with
\begin{equation}
\phi_s = \arctan(-Z_s/Z_c)
\label{eqphis}
\end{equation}
and
\begin{equation}
Z_{\{{s\atop c}\}} = \int\!\!\int d\Omega ~I({\bf{s}}, \omega, t)
{\Bigl\{{\sin \atop \cos} \Bigr\}} (2 \pi {\bf{B}}\cdot{\bf{s}}
/\lambda) ~.
\end{equation}
Here $\phi_s$ is the correction to the phase of the incoming signal,
$\bf{s}$ is a vector from the adopted reference point to a point
within the source intensity distribution in the plane of the sky,
$\omega$ and $\lambda$ are the observing frequency and wavelength,
$\bf{B}$ the baseline vector, and the integration is over solid angle
$\Omega$.  Source intensity distribution
maps are most conveniently parametrized in terms of one of two
models: superpositions of Dirac delta functions or Gaussians.  At a
given frequency, the corresponding intensity distributions
are written as
\begin{equation}
I({\bf{s}}) = \sum \limits_k S_k \delta (x-x_k, y-y_k)
\label{eqis}
\end{equation}
or
\begin{eqnarray}
I({\bf{s}}) = {\displaystyle \sum \limits_k {{S_k}\over{2 \pi a_k b_k}}}
 \thinspace {\exp} \Bigl(&-[(x-x_k)\cos\theta_k+(y-y_k)\sin\theta_k]^2
 /{2{a_k}^2} ~~~\nonumber  \\
~~~~~ &-[(x-x_k)\sin\theta_k-(y-y_k)\cos\theta_k]^2 /{2{b_k}^2} ~\Bigr),
\end{eqnarray}
where $S_k$ is the flux of component $k$, and $\bf{s}_k$
(with components $x_k, y_k$ in the plane of the sky)
is its position relative to the reference point.
For Gaussian distributions, $\theta_k$ is the
angle between the major axis of component $k$ and the $u$ axis
(to be defined below), and $(a_k, b_k)$ are the standard
deviations: full widths at
half maximum of the (major, minor) axes of component $k$
normalized by $2\sqrt{2\log2}$.  The quantities $Z_{\{{s\atop c}\}}$
entering the structure phase $\phi_s$ (Eq.~\ref{eqphis}) are
\begin{equation}
Z_{\{{s\atop c}\}} = \sum \limits_k S_k {\Bigl\{{\sin \atop \cos} \Bigr\}}
(2 \pi {\bf{B}}\cdot{\bf{s_k}} / \lambda)
\label{eqzsd}
\end{equation}
for delta functions, and
\begin{equation}
Z_{\{{s\atop c}\}} = \sum \limits_k S_k \exp[-2\pi^2(a_k^2U_k^2+b_k^2V_k^2)]
{\Bigl\{{\sin \atop \cos} \Bigr\}} (2 \pi {\bf{B}}\cdot{\bf{s}}_k/\lambda)
\label{eqzsg}
\end{equation}
for Gaussians.  Here
\begin{eqnarray}
U_k &=& ~~u \cos \theta_k + v \sin \theta_k \nonumber  \\
V_k &=& -u \sin \theta_k + v \cos \theta_k ~,
\end{eqnarray}
with $u$ and $v$ being the projections of the baseline vector $\bf{B}$
on the plane of the sky in the E-W and N-S directions, respectively.

  Maps may be specified in terms of an arbitrary number of either
Gaussian or delta function components.  At most, six
parameters characterize each component: its polar
coordinates and flux, and, for a Gaussian, its major and minor
axes and the position angle of the major axis.  The structural
correction for phase is computed via Eqs.~(\ref{eqphis}) and either
(\ref{eqzsd}) or (\ref{eqzsg}).  For the bandwidth synthesis
delay observable, the structure correction is the slope of a straight
line fitted to the individual structure phases calculated for
each frequency channel used during the observation.  For example,
for Mark III data there are typically 8 channels spanning
$\approx$8.2 to 8.6 GHz at X band, and 6 channels spanning
$\approx$2.2 to 2.3 GHz at S band.  Delay rate structure
corrections are calculated by differencing the structure phases
at the two times (Eq.~\ref{eqpdr}) used to form the theoretical rate
observable.  In the case of dual-band (S-X) experiments, a linear
combination (Eq.~\ref{eqiontau}) of the structure corrections calculated
independently for each band is applied to the dual-frequency observables.

  The practical question to be resolved is whether structural corrections
based on source intensity distribution maps yield significant and
detectable corrections to the observables at the present levels of
experimental and modeling uncertainty.  Maps are becoming available for
a considerable number of the sources currently observed by VLBI
(\citename{Fey96},~1996; available at
{\tt{http://maia.usno.navy.mil/rorf/rrfid.html}}).
Some of the extended sources show time
variability on a scale of months.  Since the corrections
$\Delta \tau_s$ and $\Delta {\dot{\tau}}_s$ are quite sensitive
to fine details of the structure, in such cases new maps
may be required on short time scales.  Depending on the relative
orientation of the source and VLBI baseline, the delay correction
can be as large as $\approx$1 ns for extended sources, which is
equivalent to tens of cm.  Despite the potentially large corrections
and difficulties in producing and applying maps, the prognosis appears
to be good.  \citename{Charlot90b}~\citeyear{Charlot90b} found that data
from a multiple baseline
geodynamics experiment are adequate to map source structures with
high angular resolution.  The use of such maps for the source 3C 273
yields structure corrections that substantially improve modeling of
observable delays \cite{Charlot94}.  Recent work has investigated
the problem of time dependent structures by interpolating the
positions and strengths of the components ejected from the core of
3C 273, from maps made many months apart, and using such interpolated
maps to correct observations made in the time period between the
maps \cite{CharSov98}.  Since maps of many sources are
becoming available from imaging studies using the VLBA \cite{Fey96},
it can be expected that structure corrections will become fairly
routine for numerous extended radio sources used in astrometric
work.

   Empirical evaluation of the effects of unknown source structure
on VLBI measurements could be made via the time rates of change of
the source right ascension $\alpha$ and declination $\delta$,
based on a time-linear model of the source coordinates
\begin{eqnarray}
\alpha &=& \alpha_0  + {\dot{\alpha}} (t - t_0) \nonumber  \\
\delta &=& \thinspace \delta_0 + {\dot{\delta}} (t - t_0) ~.
\end{eqnarray}
Non-zero estimates of the rate parameters
${\dot{\alpha}}$ and ${\dot{\delta}}$ could arise either from
genuine proper motion or from motion of the effective source centroid
sampled by VLBI measurements.  Unambiguous interpretation of such results
is problematic, but non-zero rates can be used as crude diagnostics for
the presence of structure effects.  Apparent source position rates have
been reported by \citeasnoun{Ma91}, \citeasnoun{Jacobs93}, and
\citeasnoun{Eubanks96}.  Thus far, statistically significant
rates are not believed to represent true proper motion, but rather
to be the consequence of a change in the interference pattern caused
by spectral redistribution or ejection of components from the central
engine of the source.  Galactic rotation and gravitational radiation
\cite{Pyne96,Gwinn97} may also contribute to apparent motions.
Apparent position shifts and proper motions may likewise result
from gravitational lensing within the Galaxy \cite{Hosok97}.

  In the absence of maps, using parameter estimation from astrometric
VLBI data, source structure may also be modeled as a superposition of
two $\delta$ functions centered at points $P_1(x_1,y_1)$ and
$P_2(x_2,y_2)$ respectively, as in Eq.~(\ref{eqis}) above. The parameters
describing the two components are: 1) flux ratio $K = S_2 / S_1$, where
$S_k$ is the flux of the $k$th component, 2)~\nobreak component separation
$s = {|{\bf{s}}| = |\overrightarrow{P_1 P_2}}|$, and 3) position angle
$\theta$.  The position angle is $\theta = 0^{\circ}$ when
$\overrightarrow {P_1 P_2}$ is in the direction of increasing declination
${\bbox{\widehat\delta}}$, and $\theta = 90^{\circ}$ when
$ \overrightarrow {P_1 P_2}$ is in the direction of increasing right
ascension ${\bbox{\widehat\alpha}}$.  From \citeasnoun{Charlot90b}, the group
delay has the following dependence on the structural parameters:
\begin{equation}
\tau ~=~
         {    { 2 \pi K ~ ( 1 - K )  }
              \over
              { \omega  ~ ( 1 + K )  }         }
         {    { R \lbrack { 1 - \cos( 2 \pi R )} \rbrack }
              \over
              { \lbrack { K^2 + 2K \cos(2 \pi R) + 1 } \rbrack } } ~,
\end{equation}
where
\begin{equation}
 R = {\bf{B}} \cdot {\bf{s}}  / \lambda ~.
\end{equation}
Note that this model is not linear in its three parameters.  This may
complicate parameter estimation in the absence of accurate $a~priori$
values.  For evaluating partial derivatives of $\tau$ that are needed
for parameter estimation, the component separation ${\bf s}$ and baseline
${\bf B}$ are most conveniently written in terms of their components in
the celestial system, as
\begin{eqnarray}
{\bf s} &=& {\bbox{\widehat\alpha}} ~s~ \sin\theta ~+~ {\bbox{\widehat\delta}}
 ~s~ \cos\theta    \nonumber \\
{\bf B} &=& {\bbox{\widehat \alpha}} ~u~ \lambda ~~~~~+~
 {\bbox{\widehat \delta}} ~v~ \lambda ~.
\end{eqnarray}
Then $R$ becomes
\begin{equation}
R = s \thinspace (u~\sin\theta ~+~ v~\cos\theta ) ~.
\end{equation}

\subsection{ Antenna structure }
\label{ant}

  The development in Secs.~\ref{arrtim} to \ref{emot} outlines how the
time delay model would be calculated for two points fixed with respect
to the Earth's crust.  Just as the sources are not point-like
(Sec.~\ref{sstruc}), the antenna system likewise does not necessarily
behave as an Earth-fixed point.  Not only are there instrumental delays
in the system, but portions of the antenna move relative to the Earth.
To the extent that instrumental delays are independent of the antenna
orientation, they are indistinguishable to the interferometer from clock
offsets (which are treated in Sec.~\ref{clk}).  If necessary, such
instrumental delays can be separated from clock properties by careful
calibration of each antenna system \citeaffixed{Rogers75}{$e.g.$}.
Physical motions of each antenna relative to the Earth's surface must
be considered, however, since they are part of the geometric model.
Fig.~\ref{antpic} shows the Deep Space Network's Deep Space Station 43,
which is a 70-m diameter antenna located in Tidbinbilla, Australia
(near Canberra).
It can easily be imagined that establishing and measuring the stability of
such a large structure is not a trivial matter.  For comparison, the majority
of contemporary steerable antennas are 20 to 35 meters in diameter.
\vbox{
\begin{figure}
\vskip 15pt
\hskip 2.0in
\epsfysize=3.0in
\epsffile{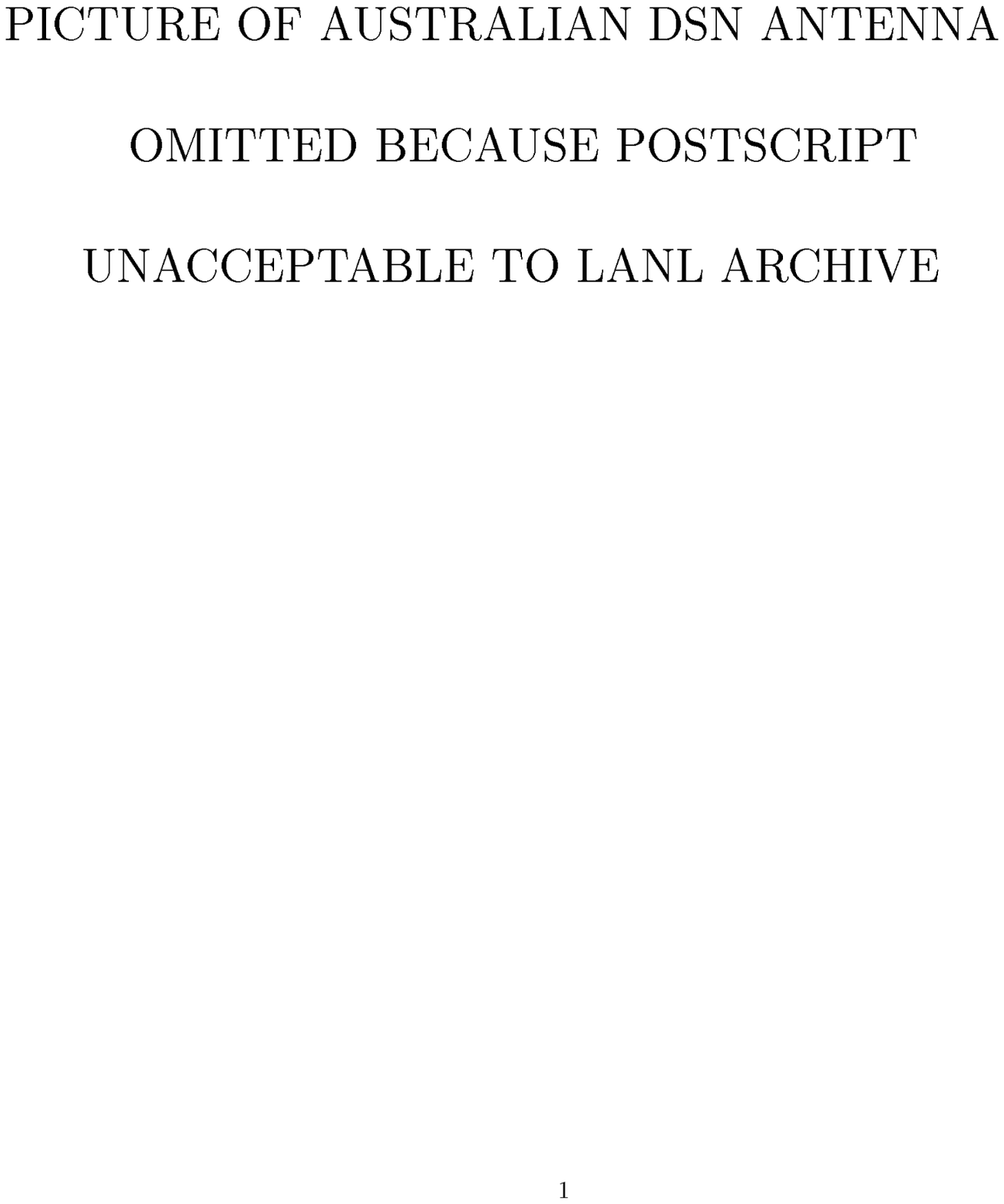}
\vskip -1.0in
\caption{ The 70-m diameter DSN antenna near Tidbinbilla, Australia. }
\label{antpic}
\end{figure}
}

  The model of the antenna structure is divided into several components.
First, we consider the geometric delay which arises when the two axes used
to steer the antenna do not intersect.  This effect is then modified to
account for the slight distortion by tropospheric refraction of the
direction in which the antenna is pointed.  Variation in ambient temperature
causes thermal expansion of the structure, and the force of gravity
causes deformations that vary as the antenna changes orientation.
For most types of antennas there is also a subtle effect caused
by the rotation of the feed horn -- relative to a fixed direction on the
celestial sphere -- as the antenna tracks the sidereal motion of
a source. A small correction to the model of the zenith troposphere
delay is required for antennas with non-intersecting axes.  Lastly,
we discuss the modeling of `site vectors' which account for
experiment-to-experiment variations in the position of a mobile antenna
relative to a fixed benchmark.

\vbox{
\begin{figure}
\hskip 2.0in
\epsfysize=2.0in
\epsffile{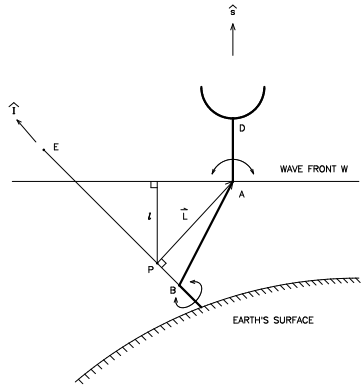}
\vskip 15pt
\caption{ Generalized schematic representation of the geometry of a
steerable antenna.  Two rotation axes and the ``reference point'' P
are indicated. }
\label{antgeom}
\end{figure}
}
  Before giving the details of our model, let us describe a general antenna
pointing system (shown schematically in Fig.~\ref{antgeom}) which
applies to all antennas that are steerable along two
coordinates.  The unit vector ${\bf{\widehat s}}$ to the aberrated
source position is shown.  Usually, a symmetry axis AD points parallel
to ${\bf{\widehat s}}$.  The point A on the figure also represents the
end view of an axis which allows rotation in the plane perpendicular to
that axis.  This axis is offset by some distance $L$ from a second
rotation axis BE.  For many antennas this offset is zero or a few meters,
but it can be as large as 15 m (for the 43-m diameter antenna at Green
Bank, West Virginia).  All points on this second rotation axis are fixed
relative to the Earth.  Consequently, any point along that axis is a
candidate for the fiducial point which terminates this end of the baseline.
The point we actually use is the point P.  A plane containing the rotation
axis A and perpendicular to BE intersects BE at the point P.  This is a
somewhat arbitrary choice, one of conceptual convenience.

  Consider the plane W which is perpendicular to the dish symmetry
axis, AD, and contains the antenna rotation axis A (perpendicular to the
plane of Fig.~\ref{antgeom}).  For plane wave fronts
this is an isophase plane (it coincides with the wave front).  For curved
wave fronts, however, it deviates from an isophase surface by $\approx L^2
/ (2R)$, where $R$ is the distance to the source, and $L$ is taken as a
typical antenna offset AP.  For $L~\approx$~10~meters, $R$~=~$R_{\rm moon}$
~=~60$R_{\rm E} \approx 3.6 \times 10^8 $ m, and the curvature correction
$L^2 / (2R) \approx 1.4 \times 10^{-4}$ mm is totally negligible.  The
source has to be as close as $R = 50\thinspace$km, or $0.01 R_{\rm E}$, before
this deviation approaches a 1 mm contribution to the delay.  Consequently,
for all anticipated applications of radio interferometry using high-gain
radio antennas, the curvature of the wave front may be neglected in
modeling the influence of antenna orientation on the time delay.
Likewise, gravitational effects are sufficiently constant over a dimension
$L$ to permit the use of a single Cartesian frame over the antenna
structure, to a very good approximation.  Provided the instrumental
delay of the antenna system is independent of the antenna orientation,
the recorded signal is at a constant phase delay, independent of antenna
orientation, at any point on the W plane.  Since this delay is
indistinguishable from a clock offset, it will be totally absorbed
by that portion of our model (see Sec.~\ref{clk}).  Any delay along
the symmetry axis AD up to the position of the feed (inside the dish,
along AD in the direction ${\bf{\widehat s}}$) will likewise be absorbed
into the clock model.

  However, for millimeter accuracy one must consider the Earth's orbital
velocity of $\approx 10^{-4}\thinspace c$ which causes Lorentz effects on
the order of $10^{-4} L$.  For a 10-m axis offset this amounts to
$1\thinspace$mm.   The Lorentz effects from the Earth's orbital velocity
are accounted for by aberrating the source direction
${\bf{\widehat{s}}}_0$ from the SSB frame into the geocentric
celestial frame, yielding
${\bf{\widehat{s}}}$ according to Eq.~(\ref{aberr}).  Since the Earth's
diurnal rotational velocity is 100 times smaller than the orbital velocity,
the diurnal Lorentz effects for a $10\thinspace$m offset would be negligible
($\leq 0.01\thinspace$mm).

  The advantage of choosing the W plane, rather than some other plane
parallel to it, is that the axis A is contained in this plane, and axis A
is fixed relative to the BE axis by the antenna structure.  If $l$ is the
length of a line from P perpendicular to the W plane, the wave front
will reach the Earth-fixed point P at a time $\tau_{ax} = l/c$ after
passing through axis A.  If $\tau_0$ is the model delay for a wave front
to pass between the reference points P of antennas 1 and 2, then the
model for the observed delay $\tau$ should be amended as:
\begin{equation}
\tau = \tau _0 - ( \tau_{{ax}_2} - \tau_{{ax}_1} ) = \tau _0 +
  { (l_1 - l_2 )/c } ~,
\end{equation}
where the subscripts refer to antennas 1 and 2.

  To calculate this ``axis offset'', we follow the treatment given
by \citeasnoun{Wade70}.  First define a unit vector ${\bf{\widehat I}}$
along BE, in the sense of positive away from the Earth.
For antennas with their steering axes in altazimuth, equatorial,
or X-Y mounts, the direction ${\bf{\widehat{I}}}$ points toward
the local geodetic zenith, the celestial pole, or the local geodetic
horizon, respectively. Next we define
a vector ${\bf{L}}$ from P to A, $L = ~\mid{\bf{L}}\mid$.  Without much
loss of generality in this antenna system, we assume that ${\bf{\widehat s}},
{\bf{L}}$, and ${\bf{\widehat I}}$ are coplanar.  Then:
\begin{equation}
{\bf{L}} = \pm L~ { { {\bf{\widehat I \times}} ( {\bf{\widehat s
  \times}} {\bf{\widehat I}} ) } \over { \mid {\bf{\widehat I \times}}
  ( {\bf{\widehat s} \times {\widehat I}} ) \mid } } ~,
\end{equation}
where the plus or minus sign is chosen to give $ {\bf{L}} $ the
direction from P to A.  When ${\bf{\widehat s}}$ and ${\bf L}$
are parallel or antiparallel, if the antenna comes closer to the source
as $L$ increases, the plus sign is used.  Since
\begin{equation}
{\bf{\widehat I} \times} ( {\bf{\widehat s} \times {\widehat I}}
  ) = {\bf{\widehat s}} ~-~ {\bf{\widehat I}} ~( {\bf{\widehat I}}
  \cdot {\bf{\widehat s}} ) ~,
\end{equation}
\begin{equation}
l = {\bf{\widehat s}} \cdot {\bf{L}} = \pm L \sqrt { 1 - { (
  {\bf{\widehat s}} \cdot {\bf{\widehat I}} ) } ^2 } ~,
\label{eqantl}
\end{equation}
where the sign choice above is carried through.

  The vector ${\bf{\widehat I}}$ is first defined in the terrestrial
frame for a station $s$ at geodetic latitude $\phi _{s ({\rm gd})}$
and longitude
$\lambda _s$.  We will consider four standard mount types \cite{Slzbrg67}:
altazimuth mounts, equatorial mounts, and X-Y mounts directed either
North-South or East-West.  Dropping the subscripts $s$ and ${\rm gd}$, for
an altazimuth mount ${\bf{\widehat I}}$ is in the direction of the
local geodetic zenith:
\begin{equation}
{\bf{\widehat I}} ~=~ ~\left( \matrix {
\cos \phi  \sin \lambda   \cr
\cos \phi  \cos \lambda   \cr
\sin \phi                 \cr } \right) ~.
\label{Ihat_azel}
\end{equation}
For an equatorial mount (also called HA-Dec), ${\bf{\widehat I}}$ is in
the direction of the celestial pole.  The $\pm 1$ is appropriate for a
northern/southern hemisphere station with its polar axis pointing toward
the North/South pole:
\begin{equation}
{\bf{\widehat I}} ~= ~\left( \matrix {
0  \cr
0  \cr
\pm 1\cr } \right) ~.
\label{Ihat_eq}
\end{equation}
For an X-Y mount antenna, ${\bf{\widehat I}}$ is in the direction of the
lower of the two axes (toward the local geodetic horizon), which can be
either North-South or East-West depending on the particular orientation
of the X-Y axes.  For an X-Y mount directed North-South:
\begin{equation}
{\bf{\widehat I}} ~= ~\left( \matrix {
-  \sin \phi  \sin \lambda  \cr
~~~\sin \phi  \cos \lambda  \cr
   \cos \phi                \cr } \right) ~,
\label{Ihat_XYNS}
\end{equation}
while for an X-Y East-West directed mount:
\begin{equation}
{\bf{\widehat I}} ~= ~\left( \matrix {
-  \sin \lambda  \cr
~~~\cos \lambda  \cr
0                \cr } \right) ~.
\label{Ihat_XYEW}
\end{equation}

  For completeness, we mention three unique antennas that have been
used in VLBI experiments, but do not fall into any of the standard
categories considered above.  One of these is an antenna that was
extensively used by the IRIS project of the
National Oceanic and Atmospheric Administration in experiments during
the 1980s.  It was unique because it was an equatorial mount designed
for the latitude of Washington, D.C. ($\phi_W$), but was deployed at
Richmond, Florida until it was destroyed in the hurricane of August
1992.  The considerable latitude difference, and the axis offset of
several meters, make it imperative that the antenna geometry be
properly modeled.  See \citeasnoun{Sovers96} for details.
Two other unique antennas, in Arecibo, Puerto Rico and Nancay,
France, are seldom used in astrometric and never in geodetic VLBI work.
The Arecibo antenna has hardware features which make it equivalent
to an altazimuth mount.  The Nancay array has been treated by
\citeasnoun{Ortega85}.

  Since ${\bf{\widehat I}}$ is given in the terrestrial frame $t$ and
${\bf{\widehat s}}$ is in the celestial geocentric frame $c$, we rotate
${\bf{\widehat I}}_t$ into the celestial frame using the matrix ${\bf Q}$
given in Sec.~\ref{tercel}:
\begin{equation}
{\bf{\widehat I}}_c = {\bf Q} {\bf{\widehat I}}_t ~,
\end{equation}
where the subscripts $t$ and $c$ indicate the terrestrial CIO 1903 frame
and the geocentric celestial frame, respectively.  With this done, one
may now obtain ${\bf{\widehat{s}}} \cdot {\bf{\widehat I}}_c$
and subsequently the axis offset delay from Eq.~(\ref{eqantl}).

  Note that for ``nearby'' sources parallax must also be included ($i.e.$,
geographically separate antennas are not pointing in the same direction).
If ${\bf{R}}_0$ is the position of the source as seen from the center of
the Earth, and ${\bf{r}}$ is the position of a station in the same frame,
then the position of the source relative to that station is
\begin{equation}
{\bf{R}} = {\bf{R}}_0 - {\bf{r}}
\end{equation}
and in Eq.~(\ref{eqantl}) we make the substitution
\begin{equation}
{ ( {\bf{\widehat s}} \cdot {\bf{\widehat I}} ) } ^2 =
  \Biggl[ { { ( {\bf{R}}_0 - {\bf{r}} ) \cdot {\bf{\widehat I}} }
  \over { | {\bf{R}}_0 - {\bf{r}} | } } \Biggr] ^2 ~.
\end{equation}
For an extreme antenna offset of 10$\thinspace$m = $10^4\thinspace$mm
and $|{\bf{r}}| = R_{\rm E} = 6.4 \times 10^3\thinspace$km, the parallax
contribution exceeds 1$\thinspace$mm only if the source is nearer
than $\approx 6 \times 10^7\thinspace$km, or 40\% of the Earth-Sun
distance.

  Another amendment to the antenna geometry is actually due to the
atmosphere. The antenna tracks the apparent position of the source after
the ray path has been refracted by an angle $\Delta E$ in the Earth's
atmosphere, rather than along the vacuum source direction
vector ${\bf{\widehat s}}$ (the aberrated source direction geocentric
celestial frame).  While this atmospheric refraction effect is already
implicitly included in the tropospheric delay correction through the
so-called mapping function (Sec.~\ref{trp}),  it must be explicitly
accounted for in the antenna axis offset model by modifying
${\bf{\widehat{s}}}$.  For an extreme case of an elevation angle
of $6^{\circ}$ the deflection can be as large as $2 \times 10^{-3}$
radians.  Thus, for an antenna with dimension $L$ = 10 meters, the
component of the antenna model $\delta l \approx L \Delta E \approx$
20 mm.  The Earth-fixed source direction vector ${\bf{\widehat s}}$ is
modified to take atmospheric refraction into account on the basis
of the change from the vacuum elevation $E$ to an apparent value
$E + \Delta E$.  In the notation of Sec.~\ref{trp}, a single homogeneous
spherical layer approximation yields the bending correction in terms
of the dry and wet zenith troposphere delays $Z_{\rm d,w}$, the first moment
of the wet troposphere refractivity $f{\rm _w}$, $M_{\rm w} =
\int \limits_0^\infty
dq f_{\rm w}(q)$ \cite{Sovers96}, the dry troposphere scale height
$\Delta$, and the Earth radius $R_{\rm E}$:
\begin{equation}
\Delta E = \cos^{-1}[\cos(E+\alpha_0)/(1+\chi_0)] - \alpha_0 ~,
\end{equation}
where
\begin{eqnarray}
\chi_0 &=& ({Z{\rm _d}}+{Z{\rm _w}}/M{\rm _w})/\Delta ~, \nonumber  \\
\alpha_0 &=& \cos^{-1}[(1+\sigma^{\prime})/(1+\sigma)] ~, \nonumber  \\
\sigma &=& \Delta / R_e ~, \nonumber  \\
\sigma^{\prime} &=& \Bigl( [1 + \sigma (\sigma+2)]^{1/2} / \sin E
- 1 \Bigr) \sin^2 E ~.
\end{eqnarray}
This formula agrees with atmospheric ray-tracing results to within 1\%
at $6^{\circ}$ and $\approx$15\% at $1^{\circ}$ elevation.

  A further geometric effect on the antenna structure also has its
origin in the environment: variations of the temperature cause vertical
displacements of the antenna reference point, by analogy with the model
for atmospheric loading in Sec.~\ref{atmld}.  For large antennas,
these can amount to several mm for ordinary diurnal and seasonal
temperature variations of 10$-$20 K.  If VLBI data acquired under
diverse weather conditions are to be modeled simultaneously, it may
be important to account for the vertical motion of the reference point.
A rudimentary model of the temperature effect assumes that the vertical
displacement $\Delta r$ of the antenna reference point, a distance
$\Delta h$ above the ground, is
\begin{equation}
\Delta r = \alpha (T - T_0) \Delta h ~,
\end{equation}
where $\alpha$ is the coefficient of thermal expansion and $T_0$
is the reference temperature.  The reference temperature is taken to be
equal to the long-term average temperature at each station.  A linear
expansion coefficient of 12 ppm per kelvin is appropriate for both of
the usual constituents of antenna structures, steel and concrete.  The
vertical motion is thus $\approx$ 0.42 mm/K for a 70-m antenna
(assumed to have $\Delta h$ = 35 m).  Refinements of the
simple model would have to consider details of the antenna structure,
and allow for thermal lag relative to the ambient temperature
\citeaffixed{McGinnis77,Nothn95}{$e.g.$}, including thermal effects
on the axis offsets.

  Gravity loading of the flexible dish structure changes an antenna's
focal length.  Because the component of the gravity load along the
antenna's primary axis of symmetry is proportional to $\sin E$ ($E$
= elevation angle), the changes in focal length also have sinusoidal
elevation dependence \cite{ClarkTh88}.  In some antennas the
subreflector may be moved (``autofocused'') to compensate for such
gravity deformations and thereby maintain focus.  However, this
procedure does not maintain a constant signal path length through
the antenna optics and thus introduces systematic errors in the
antenna position derived from the measurements unless such changes
in path are modeled.  For example, the Deep Space Network antennas
used in weekly Earth orientation measurements \cite{Steppe94}
are designed to be in focus with no subreflector compensation at
$E=45^{\circ}$.  For these 70-m and 34-m high efficiency antennas
the delays relative to nominal focus at $E=45^{\circ}$ have been
empirically determined to be, in mm (\citename{Jacobs89},~1989;~1990):
\begin{eqnarray}
\tau_{\rm sr}(70{\text{-m}}) &=& ~~77 \sin E - 54  \nonumber  \\
\tau_{\rm sr}(34{\text{-m}}) &=& 13.5 \sin E - 9.8 ~,
\end{eqnarray}
where the coefficients are known to approximately $5\%$.  This
functional form clearly exhibits the relationship between subreflector
motion and other model parameters.  The elevation-dependent term biases
the station vertical coordinate, while the constant term is equivalent
to the clock offset.  Other antennas are expected to require similar
corrections with different coefficients.

  One physical effect that pertains to both instrumental delays and
antenna geometry is the differential feed rotation for circularly
polarized receivers.  This is caused by the changing orientation of
the antenna feed relative to a fixed direction on the celestial
sphere.  The phase shift $\theta$ is zero for equatorially mounted
antennas.  For altazimuth mounts,
\begin{equation}
\tan \theta = \cos \phi \sin H / ( \sin \phi \cos \delta -
  \cos \phi \sin \delta \cos H ) ~,
\end{equation}
with $\phi$ = station latitude, $H$ = hour angle, and $\delta$ =
declination of the source.  For X-Y mounts, two cases are distinguished:
orientation N$-$S or E$-$W.  The respective rotation angles are
\begin{eqnarray}
\tan \theta &=& - \sin \phi \sin H / ( \cos \phi \cos \delta +
  \sin \phi \sin \delta \cos H ) ~~~({\rm N}-{\rm S})  \nonumber  \\
\tan \theta &=& \cos H / ( \sin \delta \sin H )
   ~~~~~~~~~~~~~~~~~~~~~~~~~~~~~~~~~~~({\rm E}-{\rm W}).
\end{eqnarray}
The effect cancels for group delay data, but can be significant for
phase delay and delay rate data (up to 100 fs/s for the latter).
The effect on phase delay is
\begin{equation}
\tau = (\theta_2 - \theta_1)/f ~,
\end{equation}
where $f$ is the observing frequency and $\theta_i$ the phase rotation
at station $i$.

  Another small correction, which couples atmospheric delay and antenna
geometry, accounts for the effect of orientation of hour angle-declination
(HA-Dec) and X-Y antennas on the tropospheric path delay.  Antennas with
non-zero axis offsets, whose second rotation axis (A in Fig.~\ref{antgeom})
moves vertically with changing orientation, have zenith troposphere delays
that may vary by 1 to 2 mm over the range of available orientations.
Equatorial and X-Y mounts fall in this class.  At low elevation angles
this zenith variation is magnified by the mapping function to 1-2 cm.
These variations must be modeled in experiments whose accuracies are at
the millimeter level ($e.g.$, short-baseline phase delay measurements).
For the highest accuracy, tropospheric mapping functions that depend on
altitude also need to account for variation of the altitude of the
antenna reference point.  Reports by Jacobs~\citeyear{Jacobs88,Jacobs91}
derive the
corrections based on considering only the dry troposphere component,
and include all terms necessary to achieve an accuracy of a few millimeters
at the lowest elevations.  The correction to be added to the zenith dry
tropospheric delay is
\begin{equation}
\delta Z{\rm _d} ~=~ -{Z{\rm _d}} ( L / \Delta) ~\psi ~,
\end{equation}
where $L$ is the antenna axis offset, $\Delta$ the dry troposphere
scale height ($\approx$ 8.6 km), and $\psi$ is an angular factor
that varies with the type of mount.  For equatorial mounts,
\begin{equation}
\psi ~=~ \cos \phi \cos H ~,
\end{equation}
where $\phi$ is the geodetic latitude and $H$ the local hour angle
east of the meridian.  The Richmond antenna correction has this
form with the Richmond $\phi$ replaced by $\phi_W$ for Washington,
and $H$ by a pseudo-hour angle $H_R$, where
\begin{equation}
H_R ~=~ \arctan \Bigl( \cos E \sin (\theta - \epsilon) \Big/
  \bigl[ \cos \phi _W \sin E - \sin \phi _W \cos E
  \cos (\theta + \epsilon ) \bigr] \Bigr) ~.
\end{equation}
For North-South oriented X-Y mounts,
\begin{equation}
\psi ~=~ \sin E / (1-\cos^2\theta \cos^2 E)^{1/2} ~,
\end{equation}
where $\theta$ is the azimuth (E of N), and for East-West oriented X-Y mounts,
\begin{equation}
\psi ~=~ \sin E / (1-\sin^2\theta \cos^2 E)^{1/2} ~.
\end{equation}

  Finally, we need to consider experiments involving transportable antennas
which are placed at slightly different locations each time a site is occupied.
The most important part of the antenna ``motion'' between successive
site occupations is expressed as an offset (``site vector'') between
the current antenna location and some Earth-fixed benchmark.  If it is
desired to describe a series of such experiments in terms of a single
set of site coordinates (those of the benchmark), then this offset
vector must be determined as part of each observing session.  It is
usually expressed in local geodetic coordinates (vertical, East, and
North).  Models of this offset vector normally assume that the local
geodetic vertical direction for the antenna is parallel to that for
the benchmark (flat Earth).  This implies that the changes derived
for the benchmark coordinates are identical to those for the antenna
coordinates.  The error introduced by this assumption in a baseline
adjustment is approximately $d \Delta B / R_{\rm E}$, where $\Delta B$
is the baseline adjustment from its $a~priori$ value, $d$ is the
separation of the antenna from the benchmark, and $R_{\rm E}$ is the radius
of the Earth.  To keep this error smaller than 1 mm for baselines
that differ from $a~priori$ values by $\approx$1 meter, it is sufficient
for $d$ to be $<$ 6000 meters.

  More troublesome is that an angular error $\delta \Theta$ in determining
the local vertical, when using an antenna whose reference point is a
distance $\Delta h$ above the ground, can cause an error of
$\Delta h~{\rm{sin}} \delta \Theta \approx \Delta h~\delta \Theta$
in measuring the baseline to the benchmark \cite{Allen82}.  Unless this
error is already absorbed into the measurement of the offset vector,
care must be taken in setting up the portable antenna so as to minimize
$\delta \Theta$.  To keep the baseline error $<\thinspace 1\thinspace$mm
for an antenna height of 10 meters, $\delta \Theta$ is required to be
$<\thinspace$20 arcseconds.  Often plumb bobs are used to locate the antenna
position relative to a mark on the ground.  This mark is, in turn,
surveyed to the benchmark.  Even the difference in geodetic vertical from
the vertical defined by the plumb bob may be as large as 1 arc minute,
thus causing a potential error of 3 mm for antennas of 10 meter height.
Consequently, great care must be taken in these measurements, particularly
if the site is to be repeatedly occupied by portable antennas.

\section{ INSTRUMENTAL DELAY MODELS }
\label{clk}

  The frequency standards (``clocks'') at each of the two antennas are
normally independent of each other.  Attempts are made to synchronize
them before an experiment by conventional synchronization techniques
such as GPS, but these techniques may be accurate to only
$\approx~1~\mu$s in epoch and $\approx~10^{-13}$ in rate.
More importantly, clocks often exhibit ``jumps'' and instabilities at a
level that would greatly degrade interferometer accuracy if not modeled.
To account for these clock effects, an additional ``delay'' $\tau_c$
is included in
the model delay, a delay that models the behavior of a station clock as
a piecewise quadratic function of time throughout an observing session.
Usually, however, only the linear portion of this model is needed.
For each station this clock model is given by
\begin{equation}
\tau _c = \tau _{c1} + \tau _{c2} ( t - t_0 ) + \tau _{c3}
( t - t _0 ) ^2 /2 ~.
\label{quadclk}
\end{equation}

  In addition to the effects of the lack of synchronization of clocks between
stations, there are other differential instrumental effects which may
contribute to the observed delay.  In general, it is adequate to model
these effects as if they were ``clock-like''.  Note that the instrumental
effects on delays measured using the multifrequency bandwidth synthesis
technique \cite{Rogers70} may be different from the instrumental effects
on delays obtained from phase measurements at a single frequency.
This is because the bandwidth synthesis process obtains group
delay from the slope of phase versus frequency
$\bigl[\tau_{\rm gd}=({\partial\phi} / {\partial\omega})\bigr]$
across multiple frequency segments spanning the receiver passband.
Thus, any fre\-quen\-cy-independent instrumental contribution to the
measured interferometer phase has no effect on the group delay determined
by the bandwidth synthesis technique. However, if the delay is obtained
directly from the phase measurement $ \phi $ at a given frequency
$\omega$ then the phase delay $\bigl(\tau_{\rm pd} = \phi / \omega \bigr)$
does have that instrumental contribution.

    Because of this difference, it is necessary to augment the ``clock''
model for phase delay measurements:
\begin{equation}
\tau_{c_{\rm pd}} = \tau_c + \tau_{c4} (t - t_0 ) + \tau_{c5}
(t - t_0 ) ^2 /2 ~,
\end{equation}
where $ \tau _c $ is the clock model for bandwidth synthesis observations
and is defined in Eq.~(\ref{quadclk}).  Since present systems measure
both bandwidth synthesis group delay and phase delay rate, all of the
clock parameters described above must be used.  However, in a perfectly
calibrated interferometer, $\tau_{c4}$ = $\tau_{c5}$ = 0.
This particular model implementation allows simultaneous use of delay
rate data derived from phase delay, with group delay data derived by
means of the bandwidth synthesis technique.

  A refinement of the clock model may be required for dual-frequency (S/X)
delays.  It originates from the differential instrumental delay for S-
and X-band data, which may be sizeable.  For dual-frequency observables,
the clock model depends on this differential instrumental delay and on
the frequencies $\omega_{\rm S}, \omega_{\rm X}$ in the individual bands as
\begin{equation}
\tau_{c6}~\omega_{\rm X}^2 / ( \omega_{\rm X}^2 - \omega_{\rm S}^2) ~.
\end{equation}
The differential instrumental delay $\tau_{c6}$ is normally highly
correlated with the usual clock offset $\tau_{c1}$, but under some
circumstances may convey additional information.  For example, if
the frequencies $\omega_{\rm S}, \omega_{\rm X}$ are not exactly the
same for all measurements, the $\tau_{c6}$ term will not be perfectly
absorbed into the clock offset $\tau_{c1}$.

  To model the interferometer delay on a given baseline, a difference of
station clock terms is formed:
\begin{equation}
\tau _c = \tau_c (2) - \tau_c (1) ~.
\end{equation}
Specification of a reference clock can be postponed until least-squares
parameter adjustment, and is of no concern in the model description.

\section{ ATMOSPHERIC DELAY MODELS }
\label{atmo}

  During its journey from the radio source to the two Earth-based
receivers, the radio wave front must pass through intergalactic,
interstellar, interplanetary, and terrestrial atmospheric media.
Both the neutral and charged components of these media (neutral
molecules and ionized plasma, respectively) modify the propagation
speed.  This produces an overall delay relative to propagation in
vacuum.  Differential effects among the divergent paths of the two
signals traversing different regions of the anisotropic medium
distort the wave front, and thus contribute to differential delay
in arrival times at the two Earth-based receivers.  Only in the
immediate vicinity of the Earth, however, do the two ray paths
diverge sufficiently to cause measurable differences in arrival
times.  This section is concerned with models which correct the
VLBI observables for such additional delays due to propagation
effects.  It is divided into two parts that consider contributions from
charged particles and neutral molecules, respectively.  In recognition
of the dominant influence of the near-Earth environment, they are
named ``Ionosphere'' and ``Troposphere''.

\subsection{ Ionosphere }
\label{ion}

  For a medium composed of charged particles (plasma), \citeasnoun{Spitzer}
gives the refractive index at frequency $\nu$ in the quasi-longitudinal
approximation:
\begin{equation}
n =\Bigl[ 1 - \Bigl( { { \nu _p } \over { \nu } } \Bigr) ^2
\Bigl( 1 \pm { { \nu _g } \over { \nu } } \cos \Theta \Bigr) ^{-1}
\Bigr] ^{1/2} ~,
\end{equation}
where $\Theta$ is the angle between the magnetic field $B$ and the
direction of propagation of the wave front.  The plasma frequency
$ \nu _p $ is
\begin{equation}
\nu _p = \Bigl( { { \rho c^2 r_0 } / { \pi } } \Bigr) ^{1/2}
\approx 8.98 \rho ^{1/2} ~~({\rm{Hz}})
\end{equation}
and the electron gyrofrequency $ \nu _g $ is
\begin{equation}
\nu _g = { { e B } \over { 2 \pi m c } } \approx 2.80 \times 10^6 B
~~({\rm{Hz}}) ~.
\end{equation}
Here $\rho$ is the number density of the electrons (m$^{-3}$), $c$
the speed of light, $r_0$ the classical electron radius, $e$ and $m$
the electron charge and mass, and $B$ is measured in gauss.

  Tables~\ref{plasma} and \ref{gyrof} give approximate values of the
plasma frequency $\nu_p$ and gyrofrequency $\nu_g$ for the three
regimes along a radio signal's ray path: Earth, interplanetary, and
interstellar space.  The plasma parameters $\rho$ and $B$ represent
typical conditions in these three regions, with the upper limit
adopted in cases of substantial variability \cite{Lang,Zombeck90}.
For example, the ``Earth'' values are appropriate to the daytime
ionosphere.  For ray paths close to the Sun, however, the tabulated
``interplanetary'' values are several orders of magnitude too small.
Intergalactic electron densities and magnetic fields are at least
two orders of magnitude smaller than the tabulated interstellar values.
Given the S-band ($\nu_{\rm S} = 2.3$ GHz) and X-band ($\nu_{\rm X} = 8.4$ GHz)
frequencies typically used in geodetic and astrometric VLBI experiments,
the corrections for plasma and gyrofrequency effects can be parametrized
by the ratios of $\nu_p$ and $\nu_g$ to $\nu_{\rm S,X}$ respectively.
It can be seen from Table~\ref{plasma} that there is an order-of-magnitude
falloff in the plasma frequency as we proceed outward from the Earth to
interplanetary and interstellar regions.  Table~\ref{gyrof} shows that
electron gyrofrequency effects are a factor of 10 smaller than plasma
effects near the Earth, and negligible in interplanetary and interstellar
environments.

  Relative to a perfect vacuum as a reference, the contribution
$\Delta _{\rm pd}$ to the VLBI phase delay $\tau _{\rm pd}$ for a
monochromatic signal traversing a medium of refractive index $n$ is
\begin{equation}
\def\nupnu{{\Bigl({{\nu_p^{\prime}}\over{\nu}}\Bigr)}}
\Delta _{\rm pd} = { 1 \over c } \int { ( n - 1 )dl }
\approx - { 1 \over {2c } } \int \nupnu ^2 \Bigl[
1 + { 1 \over 4 } \nupnu ^2 + { 1 \over 8 } \nupnu ^3 +
... \Bigr] dl ~,
\end{equation}
where
\begin{equation}
{ { \nu _p^{\prime} } \over { \nu } } = { \Bigl( { {\nu _p} \over \nu } \Bigr) }
\Bigl[ 1 \pm \Bigl( { { \nu _g } \over { \nu } } \Bigr) \cos \Theta \Bigr]
^{-1/2} ~.
\end{equation}
For 8.4 GHz, we may approximate this effect to parts in $10^6 - 10^7$ by:
\begin{equation}
\Delta _{\rm pd} \approx { { -q } \over { \nu ^2 } }
\Bigl[ 1 \pm { \Bigl\langle { { \Bigl( { { \nu _g } \over { \nu } } \Bigr) }
\cos \Theta } \Bigr\rangle } \Bigr] ^{-1} \approx { { -q } \over { \nu ^2 } }
\Bigl[ 1 - { \Bigl\langle { { \Bigl( { { \pm \nu _g } \over { \nu } } \Bigr) }
\cos \Theta } \Bigr\rangle } \Bigr] ~,
\end{equation}
where
\begin{equation}
q = { { c r_0 } \over { 2 \pi } } \int { \rho dl }
= { { c r_0 I_e } \over { 2 \pi } }
\end{equation}
and where $I_e$ is the total number of electrons per unit area along the
integrated line of sight.  The angular brackets symbolize a geometrical
average.  If we also neglect the term $\langle \nu_g \cos\Theta \rangle
/ \nu $, then the expression for $\Delta _{\rm pd}$ becomes simple and
independent of the direction of travel of the wave front through the plasma:
\begin{equation}
\Delta _{\rm pd} = { { - q } / { \nu ^2 } } ~.
\label{eqionpd}
\end{equation}
This additional delay is negative.  Thus, a phase {\it advance} occurs
for a monochromatic signal.  Since phase delay is obtained at a single
frequency, observables derived from the phase delay ($e.g.$, phase delay
rates) experience an increment which is negative (the observable with
the medium present is smaller than it would be without the medium).
In contrast, group delays measured by a technique such as bandwidth
synthesis $\bigl[ \tau _{\rm gd} = ({\partial \phi} /
{\partial \omega }) \bigr]$
experience an additive delay which can be derived from Eq.~(\ref{eqionpd})
by differentiating $\phi = \nu \Delta _{\rm pd}$ with respect to frequency:
\begin{equation}
\Delta _{\rm gd} = { { q } / { \nu ^2 } } ~.
\end{equation}
The group delay is of
the same magnitude as the phase delay advance.  For group delay
measurements, the measured delay is larger with the medium present
than without the medium.

  The uppermost component of the Earth's atmosphere, the ionosphere,
is a layer of plasma whose density peaks at about 350 km altitude,
but is widely distributed between 80 and 1000 km above the Earth's
surface.  It is created primarily by the ultraviolet portion of the
sunlight and the solar wind, and
is a superposition of three Chapman layers \cite{BassH93}:
E, F$_1$, and F$_2$, with height parameters of 110, 210, and 350 km,
respectively.  During daytime in a year near the maximum of the
11-year solar activity cycle the electron density
peaks in the F$_2$ layer at about $3.7 \times 10^{12}$ electrons/m$^3$.
For a typical ionosphere, $\Delta _{\rm gd} \approx$ 0.1 to 2 ns at local
zenith for $\nu$ = 8.4 GHz.  This effect has a maximum at approximately
1400 hours local time and a broad minimum during local night.  For long
baselines, the effects at each station can be quite different.  Thus, the
differential effect may be of the same order as the maximum.

  For the interplanetary medium at an observing frequency of 8.4 GHz
and assuming that the ionized region extends over 10$^7$ km, a single
ray path experiences a delay of approximately $6 \times 10^{-7}$ s in
traversing the Solar System.  However, the differential delay
between the ray paths to the two stations on the Earth is considerably
smaller, since the gradient between the two ray paths should also be
inversely proportional to the dimensions of the plasma region ($i.e.$,
millions as opposed to a few thousand kilometers).  An
interstellar signal from a source at a distance of 1 megaparsec
($3 \times 10^{19}$ km) would experience an integrated plasma delay of
approximately 600 seconds at a frequency of 8.4 GHz if the plasma density
were that of Table~\ref{plasma} everywhere along the ray path.
In the near vacuum of intergalactic space, the
average plasma density is $\approx$~2 orders of magnitude lower than
this, reducing the delay to $\approx$10 s.  In this case, however, the
typical dimension is also that much greater, and so the differential
effect on two ray paths separated by one Earth radius is still not as
large as the differential delay caused by the Earth's ionosphere.

  Plasma effects can best be calibrated by the technique of observing
the sources at two frequencies, $\nu _1$ and $\nu _2$, where
$\nu _{1,2} \gg \nu _p$ and where $ { { | \nu _2 ~-~ \nu _1 | }
/ { (\nu _2 ~+~ \nu _1 }) } \approx 1$.
For the VLBI delays $\tau$ at the two frequencies $\nu _1$ and $\nu _2$
we obtain
\begin{equation}
\tau _{\nu 1} = \tau + q / { \nu _1^2 } ~~~~~~ {\rm{and}} ~~~~~~
\tau _{\nu 2} = \tau + q / { \nu _2^2 } ~.
\end{equation}
Multiplying each expression by the square of the frequency involved and
subtracting,
\begin{equation}
\tau = a~ \tau _{\nu 2} + b~ \tau _{\nu 1} ~,
\label{eqiontau}
\end{equation}
where $a = {\nu _2^2} / (\nu _2^2 - \nu _1^2)$ and $b = {- \nu _1^2} /
(\nu _2^2 - \nu _1^2)$.  This linear combination of the observables at
two frequencies thus removes the charged particle contribution to the delay.
For uncorrelated errors in the frequency windows, the overall error in the
derived delay can be modeled as
\begin{equation}
\sigma _{\tau}^2 = a^2 \sigma _{\tau _{\nu 2}} ^2
+ b^2 \sigma _{\tau _{\nu 1}} ^2 ~.
\end{equation}
Modeling of other error types is more difficult and will not be treated
here.  Since the values of $a$ and $b$ are independent of $q$, these same
coefficients apply both to group delay and to phase delay.

   If we had not neglected the effect of the electron gyrofrequency in the
ionosphere, then instead of Eq.~(\ref{eqiontau}) above, the combined S/X
delay would have been
\begin{equation}
\tau = a \tau _{\nu 2} + b \tau _{\nu 1} +
{ q~ { \langle { \nu _g \cos \Theta } \rangle } } / [ \nu _2 \nu _1
( \nu _2 - \nu _1 ) ] ~,
\end{equation}
where $a $ and $b$ are defined as in Eq.~(\ref{eqiontau}).  If the third
term on the right-hand side is expressed in units of the contribution
of the ionosphere at frequency $\nu _2$, we obtain
\begin{equation}
\tau = a \tau _{\nu 2} + b \tau _{\nu 1} +
{ \Delta _{\rm pd} \nu _2 ~ { \langle { \nu _g \cos \Theta } \rangle } } /
[ \nu _1 ( \nu _2 + \nu _1 ) ] ~.
\end{equation}
For X band $\Delta _{\rm pd} \approx$ 0.1 to 2 ns at
the zenith.  When using S band as the other frequency in the pair, this
third term is $ \approx 2 \times 10^{-4} \Delta _{\rm pd} \cos \Theta
\approx$ 0.02 to 0.4 ps at zenith.  In the worst case of high ionospheric
electron content, and at low elevation angles, this effect could reach
1 mm of total error in determining the total delay using the simple
formula of Eq.~(\ref{eqiontau}) above.  Notice that the effect becomes
much more significant at lower frequencies.  \citeasnoun{BassH93}
present a more detailed discussion of higher order ionospheric effects.

  To avoid the complications of modeling the ionosphere, most
contemporary VLBI experiments are performed at two frequencies.
The dual-frequency calibration outlined above then permits us to
consider the ionosphere problem to be solved at currently required
accuracy levels, and to ignore ionospheric propagation effects in
modeling.  For completeness, we present a model of ionospheric
delay that can be used for single-frequency experiments.  In these
cases, the interferometer model must use whatever measurements of
the total electron content are available.  The model is very simple:
the ionosphere is modeled as a spherical shell whose lower boundary
is at the height $h_1$ above the geodetic surface of the Earth, and
the upper boundary is at the height $h_2$ above that same surface (see
Fig.~\ref{ionshell}).  For each station the ionospheric delay assumed to be
\begin{equation}
\tau_{_{\rm I}} = { { k g I_e S(E) } / {\nu ^2} } ~,
\end{equation}
where
\begin{equation}
k = 0.1 c r_0 / 2 \pi ~.
\end{equation}
$I_e$ is the total electron content at zenith (in electrons per meter
squared $\times 10^{-17}$), and $g = 1 (-1)$ for group (phase) delay.
$E$ is the apparent geodetic elevation angle of the source, $S(E)$ is
a slant range factor discussed below, and $\nu$ is the observing
frequency in gigahertz.

  The slant range factor (see Fig.~\ref{ionshell}) is
\begin{equation}
S(E) = \Bigg( { { \sqrt { R^2 \sin ^2 E + 2 R h_2 + h_2^2 } } -
{ \sqrt { R^2 \sin ^2 E + 2 R h_1 + h_1^2 } } } \Bigg) \bigg/ (h_2 - h_1) ~,
\end{equation}
which is strictly correct for a spherical Earth of radius $R$, and a
source at apparent elevation angle $E$.  The model uses this expression,
with the local radius of curvature of the geoid surface at the receiving
station $R$ taken to be equal to the distance from the station to the
center of the Earth.  The model also assumes this same value of $R$ can
be used at the ionospheric penetration points, $i.e.$,
\begin{equation}
R_i = R + h_i ~.
\end{equation}
\vbox{
\begin{figure}
\hskip 2.0in
\epsfysize=3.0in
\epsffile{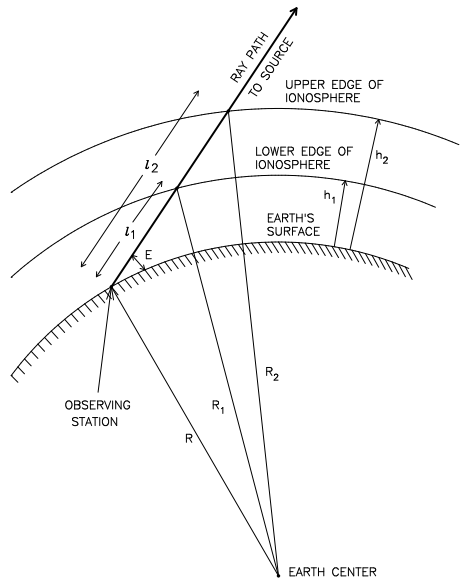}
\vskip 15pt
\caption{ Geometry of the spherical ionospheric shell used for ionospheric
corrections.  The slant range through the shell is $l_2 - l_1$ at elevation
angle $E$. }
\label{ionshell}
\end{figure}
}
This is not strictly true, but is a very close approximation, especially
in view of the relatively crude nature of the total electron content
determinations on which the model also depends.  The total ionospheric
contribution on a given baseline is
\begin{equation}
\tau_{_{\rm I}} = \tau_{_{\rm I}}(2) - \tau_{_{\rm I}}(1) ~,
\end{equation}
where the arguments 1 and 2 identify the stations.
The ionospheric total electron content $I_e$ is obtained by some external
set of measurements such as Faraday rotation or GPS techniques.  Such
external measurements, in general, are not along directions in the
ionosphere coincident with the ray paths to the interferometer.  Thus, for
each antenna, it is necessary to map a measurement made along one ray path
to the ray paths used by the interferometer.  Many different techniques to
do this mapping have been suggested and tried; \citeasnoun{Wilson95}
discuss recent progress.

  The deficiencies of these ionosphere models for single-frequency observations
are compounded by the lens effect of the solar plasma.  In effect, the Solar
System is a spherical plasma lens which causes the apparent positions of the
radio sources to be shifted from their actual positions by an amount which
depends on the solar weather and on the Sun-Earth-source angle.  Since both
the solar weather and the Sun-Earth-source angle change throughout the year,
very accurate single-frequency observations over the time scale of a year are
virtually impossible unless simultaneous auxiliary experiments are performed.

\subsection{ Troposphere }
\label{trp}

  The lower few tens of kilometers of the Earth's atmosphere are
known as the troposphere.  In contrast to the ionosphere, this
layer is to a good approximation electrically neutral
\cite{Hought86}.  Radio signals passing through the troposphere
experience delay, bending, and attenuation relative to an
equivalent path through a vacuum, because the index of refraction is not
equal to the vacuum value.  The additional delay is $\approx 2$~m at
zenith and increases to $\approx 20$~m at $6^{\circ}$ above the horizon;
bending amounts to $\approx 0.1^{\circ}$ at elevations $6^{\circ}$ above
the horizon.  This makes it imperative for accurate VLBI models
to account for tropospheric delay.  We review the refractivity of the
moist air composing the troposphere, discuss mapping functions which
model the integrated path length through the troposphere, and finally
consider the limitations to the mapping function models due to azimuthal
asymmetry and turbulence in the water vapor distribution.

  Permanent and induced dipole moments of the molecular species present
in the atmosphere modify its index of refraction and thus delay the
passage of radiation at microwave frequencies.  The excess path delay
caused by the troposphere relative to a vacuum is
\begin{equation}
\tau_{\rm tr} ~=~ \int_S~ ( n - 1 )~ dS ~,
\label{eqraytrace}
\end{equation}
where $n$ is the index of refraction and $S$ represents the signal's path
through the  troposphere.  Since departures of $n$ from unity are small,
normally the ``refractivity'' $N = 10^6 ( n - 1 )$ is used instead of
the index of refraction.  Detailed discussions of the refractivity of
moist air are found in \citeasnoun[Chap.~13]{Thmpsn86},
\citeasnoun{BeanDut}, or \citeasnoun{Thayer}.  To summarize these reviews,
the refractivity of moist air at microwave frequencies depends on the
permanent and induced dipole moments of the molecular species present
in the atmosphere.  The principal species, nitrogen and oxygen, have no
permanent dipole moments, and contribute only via their induced moments.
On the other hand, water vapor does have a substantial dipole moment.
Induced and permanent dipole moments contribute to the refractivity as
$\propto p/T$ and $\propto p/T^2$, respectively \cite{Debye}, where $p$
and $T$ are the pressure and temperature of the species under consideration.
This is the basis of the Smith-Weintraub equation \cite{SmWntr53}:
\begin{equation}
   N ~=~ 77.6 ~{p_{_D} \over T}
     ~+~ 64.8 ~{p_{_V} \over T}
     ~+~ 3.776 \times 10^5~{p_{_V} \over T^2} ~,
\end{equation}
where $p_{_D}$ and $p_{_V}$ are the partial pressures of ``dry'' air and water
vapor in units of millibars.  The coefficients are taken from
\citeasnoun{Thayer}.
The first term represents the aggregate induced dipole refractivity from
all the dry constituents, and the second and third terms the induced dipole
and permanent dipole refractivity from water vapor, respectively.
Thus the troposphere problem becomes a matter of modeling pressure and
temperature ($p_{_D}$, $p_{_V}$, $T$) along the ray path.  Typically this
problem is simplified by assuming the troposphere to be a mixture of
perfect gases with unit compressibilities which are nearly in hydrostatic
equilibrium.  Parenthetically, as noted by \citeasnoun{Thmpsn86},
at visible wavelengths water vapor has a smaller influence on refractivity,
so that the third term is not needed to model optical refractivity.
This simplifies troposphere modeling for laser techniques.

\vbox{
\begin{figure}
\hskip 2.5in
\epsfysize=2.0in
\epsffile{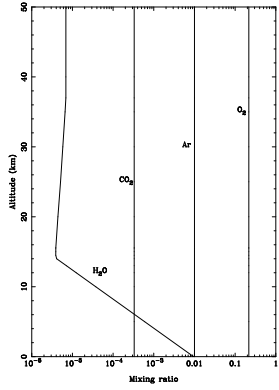}
\vskip 15pt
\caption{ Mixing ratios of four major constituents of the Earth's
atmosphere $vs$. altitude (based on Goody and Yung, 1989). }
\label{troph2o}
\end{figure}
}

  One of the biggest obstacles to modeling the tropospheric delay is
the non-uniform and highly variable distribution of water vapor.  In
contrast to the smooth exponential decrease of N$_2$ and O$_2$
concentrations with altitude, the H$_2$O concentration generally decreases
more rapidly within 1-2 km of the surface, and normally becomes negligible
above $\approx$5 km altitude \cite{Arya,AGU95}.
Figure~\ref{troph2o} shows the altitude
dependence of the mixing ratios of four major atmospheric constituents.
The fractions of O$_2$, Ar, and CO$_2$ are seen to remain nearly
perfectly constant from the surface up to 50 km, while the water vapor
fraction falls by over 3 orders of magnitude up to the tropopause
altitude ($\approx$15 km) before leveling off at 7 ppm.  Water vapor
content can also easily vary by 50\% during the few hours around sunrise
and sunset.  Weather changes have more pronounced effects on the wet
than on the dry troposphere.  The nitrogen and oxygen concentrations
essentially scale with pressure while water vapor undergoes much larger
concentration changes and redistribution with changing weather.

  The next step in generating the troposphere model is to obtain the
integrated refractivity for a signal propagating from the zenith.
The dry portion, primarily oxygen and nitrogen, is very nearly in hydrostatic
equilibrium. As a result the zenith delay does not depend on the details
of $p_{_{\rm D}}$ and $T$ along the signal path. The delay can be quite
accurately estimated simply by measuring the barometric pressure at the surface.
The dry zenith delay ${Z_{\rm d}}$ (m) is related to the total surface pressure
$p$ (mbar) as
\begin{equation}
{Z_{\rm d}} = 2.2768\times 10^{-3} p /
\bigl(1 - 0.00266 \thinspace \cos 2\phi - 0.00028 h \bigr) ~,
\label{eqtrpzen}
\end{equation}
where the factor in the denominator corrects for a non-spherical Earth
\cite{Saast72}, $\phi$ is the station latitude, and $h$ is the
station altitude in km (see Eq.~\ref{flatht}).  \citeasnoun{Davis85}
discuss the reason for using the
total surface pressure $p$ instead of the dry partial pressure $p_{_{\rm D}}$.
They also explain the related and subtle distinction between `dry' and
`hydrostatic' delay, which has been glossed over in the present discussion.
Typically, at sea level in the local zenith direction, the additional dry
(or hydrostatic) delay that the incoming signal experiences due to the
troposphere is $\approx 7.7$ nanoseconds or 2.3 meters.

   At each station $i$ the delay experienced by the incoming signal due to the
troposphere can most simply be modeled using a spherical-shell troposphere
consisting of wet and dry components $\tau_{{\rm w}i}$ and $\tau_{{\rm d}i}$:
\begin{equation}
\tau _{\rm tr} (i) = \tau _{{\rm w}i } + \tau _{{\rm  d}i } ~.
\end{equation}
The total troposphere delay model for a given baseline is then:
\begin{equation}
\tau _{\rm tr} = \tau _{\rm tr} (2) - \tau _{\rm tr} (1) ~.
\end{equation}
If $ E_i $ is the unrefracted geodetic elevation angle of the observed
source at station $i$, we have (dropping the subscript~$i$):
\begin{equation}
\tau_{\rm tr}(E) = {\cal M}{\rm _d} (E) {Z{\rm _d}} + {\cal M}{\rm _w} (E)
{Z{\rm _w}} ~,
\label{eqtrpdel}
\end{equation}
where $Z_{\rm d,w}$ is the additional (dry, wet) delay at local zenith due to
the presence of the troposphere, and $\cal M$ are the so-called ``mapping
functions'' which relate delays at an arbitrary elevation angle $E$ to
the zenith delays.  Note that since $E$ is the unrefracted elevation angle,
refraction effects are contained within the mapping functions
${\cal M}_{\rm d,w}$.  These will be considered in some detail below.

  For some geodetic experiments, the observed delay can be accurately
calibrated for the total tropospheric delays at the two stations, which
are in turn calculated on the basis of surface pressure measurements
for the dry component, and water-vapor radiometer (WVR) measurements
for the wet component.  At a fixed location, the dry zenith delay (m)
is related to the surface pressure $p$ (mbar) via Eq.~(\ref{eqtrpzen}).
The wet zenith delay ${Z{\rm _w}}$ can be inferred from WVR measurements
performed in the vicinity of the VLBI stations at the time of the experiment.
These corrections can be removed and replaced by an alternate model if desired.
In the absence of such external calibrations, it was found that estimating
the zenith delay as a linear function of time can improve troposphere
modeling considerably.  The dry and wet zenith parameters are written as
\begin{equation}
{Z_{\rm d,w}} = {Z_{\rm d,w}^0} + {\dot{Z}}_{\rm d,w} (t - t_0) ~,
\end{equation}
where $t_0$ is a reference time.  The time rates of change
${\dot{Z}}_{\rm d,w}$ may then be estimated from fits to the data.

\subsubsection{ Mapping functions }
\label{map}

  The term ``mapping function'' is used to describe the relation
between the tropospheric delay at zenith and an arbitrary angle $E$
above the horizon.  Throughout the history of VLBI, extensive attention
has been paid to tropospheric mapping functions, in view of the dominance
of tropospheric delay mismodeling in the error budget.  The simplest
way to relate the tropospheric delay for an oblique path to the
delay for a signal received from directly overhead is to assume
a flat Earth covered by an azimuthally symmetric troposphere
layer.  Since the delay is proportional to the path length through
this layer, the mapping function is then simply equal to the
cosecant of the elevation angle: ${\cal M}(E) = 1 / \sin E$.
Even in the early days of VLBI it was found that this approximation
was inadequate, and \citeasnoun{Marini} considered corrections accounting
for the Earth's curvature.  This led to a mapping function in
the form of a continued fraction,
\begin{equation}
{\cal M} = { 1 \over { \sin E + { { \displaystyle{\strut a} } \over
{ \displaystyle{\sin E + { \displaystyle{\strut b \over {\sin E + c }
} } } } } } } ~.
\end{equation}

  The simplest function, which was widely used in early VLBI modeling
in the 1970s and early 1980s, was obtained by C.~C. \citeasnoun{Chao74}.
It truncates Marini's continued fraction and replaces the second
$\sin E$ by $\tan E$ in order to force ${\cal M} = 1$ at zenith:
\begin{equation}
{\cal M}_{\rm d,w} = { 1 \over { \sin E + { {
 \displaystyle{\strut {A_{\rm d,w}} } }
 \over { \displaystyle{{\strut{\tan E + B_{\rm d,w}}}} } } } } ~.
\end{equation}
The dry and wet coefficients were determined from empirical fits to
ray tracing results through measured average atmospheres:
$A{\rm _d} = 0.00143$, $B{\rm _d} = 0.0445$, $A{\rm _w} = 0.00035$,
and $B{\rm _w} = 0.017$.
It was claimed to be accurate at the level of 1\% at $E = 6^{\circ}$,
and to become rapidly more accurate as zenith is approached.

  As more accurate measurements in the 1980s demanded more accurate
troposphere modeling, numerous improved mapping functions were developed.
Many of these have mathematical forms that are slight variants of
Marini's continued fraction, and contain constants derived from analytic
fits to ray-tracing results either for standard atmospheres or for
observed atmospheric profiles based on radiosonde measurements.
The functions of \citeasnoun{Davis85}, \citeasnoun{Ifadis},
\citeasnoun{Herr92}, and \citeasnoun{Niell96} (NMF,
Niell Mapping Function) fall into this category.
Most of them contain parameters that are to be determined from surface
meteorological measurements.  The NMF function is unique in that it
attempts to represent global weather variations analytically as a function
of location and time of year, and contains no adjustable parameters.
The NMF mapping functions provide an excellent model for situations
in which data are not available to supply optimum parameters for some
of the alternative mappings.  Their seasonal, latitude, and altitude
variations are based on interpolations of the U.~S.
standard atmospheres of \citeasnoun{Cole65}.  Niell presents evidence that
short term (hours to days) surface temperature variations are
unimportant compared to the seasonal, latitude, and altitude variations
of the temperature in the region from a few km up to the tropopause.
Another tropospheric mapping function, due to \citeasnoun{Lanyi} is unique
in that it does not fully separate the dry and wet components and thus
attempts to give a more faithful representation of the physical effects.
It also contains the most complete set of atmospheric parameters, and we
will therefore present some details for this function.  Several reviews
have recently evaluated the multitude of tropospheric mapping functions
that are now available: \citeasnoun{Gallini94}, \citeasnoun{Mendes94},
and \citeasnoun{Estef94}.

  Because of its potential to provide the most complete description of
the tropospheric delay, including use of $in~situ$ temperature profile
data, we will give some details of the Lanyi mapping function.
Motivation for and details of its development were given by
\citeasnoun{Lanyi}.  It is based on an ideal model atmosphere whose temperature
is constant from the surface to the top of the inversion layer $h_1$,
then decreases linearly with height at a rate $W$ (lapse rate) from $h_1$
to the tropopause height $h_2$, and is constant again above $h_2$.
Lanyi's approach was to develop a semi-analytic approximation to the
ray trace integral (\ref{eqraytrace}) which retains explicit temperature
$vs.$ altitude profile parameters that can be supplied from meteorological
measurements.  The goal was to provide an accurate approximation
to ray tracing in a form that was less computationally intensive
than performing a full numerical ray trace for every observation.
The mapping function is expanded as a second order polynomial
in $Z_{\rm d}$ and $Z_{\rm w}$ (plus the largest third order term).
Unlike all the other functions mentioned above, the
Lanyi mapping function is nonlinear in $Z_{\rm d}$, $Z_{\rm w}$.
It differs from other mapping functions in the sense that it does
not conform to the linear expression (\ref{eqtrpdel}).
In particular, we note that there is a $Z_{\rm d} Z_{\rm w}$
cross term which couples the dry and wet delays.  This nonlinear term
is present because for a given dry (wet) zenith delay the geometric
path through the atmosphere is not independent of the amount of wet
(dry) zenith delay.  Hence the coefficients of the nonlinear terms
have a subscript ``$b$'' to indicate that these terms arise from
the {\it bending} of the signal path through the atmosphere.

  Here we give only a brief summary of the functional form.
The tropospheric delay is written as:
\begin{equation}
\tau _{\rm tr} = F ( E ) / \sin E ~,
\end{equation}
in order to factor out the $1/\sin E$ ``flat Earth'' model, where
\begin{eqnarray}
F ( E ) &=& F{\rm _d} ( E ) {Z{\rm _d}} + F{\rm _w} ( E ) {Z{\rm _w}}
 \nonumber \\
&+& \Bigl[ F_{b1} ( E ) {Z{\rm _d}}^2 +
2 F_{b2} ( E ) {Z{\rm _d}} {Z{\rm _w}} +
F_{b3} ( E ) {Z{\rm _w}}^2 \Bigr] \Big/ \Delta
+ F_{b4} ( E ) {Z{\rm _d}}^3 / \Delta ^2 ~.
\label{eqtrpfe}
\end{eqnarray}
The quantities $ {Z{\rm _d}} $ and $ {Z{\rm _w}} $ are the zenith dry and wet
tropospheric delays, while $ \Delta$ is the dry atmospheric scale
height, $\Delta = k T_0 / mg_{\rm c} $, $ k $ = Boltzmann's constant,
$ T_0 $ = daily average surface temperature, $ m $ = mean molecular
mass of dry air, and $ g_{\rm c} $ = gravitational acceleration at the
center of gravity of the air column.  With the standard values
of $k$, $m$, $g_{\rm c} = $ 978.37 cm/s$^2$, and an average mid-latitude
temperature $ T_0 $ = 292 K, the scale height $ \Delta$ = 8.6 km.
The dry, wet, and bending contributions to the delay,
$ F{\rm _d} (E) $, $ F{\rm _w} (E) $, and $ F_{b1,b2,b3,b4} (E) $,
are expressed in terms of moments of the refractivity.
The latter are evaluated for the ideal model atmosphere and thus
give the dependence of the tropospheric delay on the four model
parameters $T_0$, $W$, $h_1$, and $h_2$.  Note that Lanyi's
formulation (Eq.~\ref{eqtrpfe}) differs from the simple model
(Eq.~\ref{eqtrpdel}) in the presence of the ``bending'' terms
$F_{b1-4}$.  These account for the influence of the dry and wet
constituents in bending the incoming ray path.

   Four meteorological parameters describe the temperature $vs.$
altitude profile in the Lanyi model.  These have already been
mentioned above: the surface temperature $T_0$, temperature lapse
rate $W$, inversion and tropopause altitudes $h_1$ and $h_2$.
Table~\ref{lanyisurf} summarizes their standard values, and also
gives the approximate sensitivities of the tropospheric delay to the
meteorological parameters.  These values are calculated at $ 6^{\circ}$
elevation, which is the approximate lower limit of validity of the
Lanyi model.  At this elevation, the ray path traverses the equivalent
of approximately 10 ``air masses''.  A fifth parameter, the surface
pressure $ p_0 $, may be used to calibrate the dry zenith delay via
Eq.~(\ref{eqtrpzen}).  The full potential of the Lanyi function can
only be realized if complete meteorological information is available
for the time and place of a VLBI experiment.  When interpolated
standard global values \citeaffixed{Cole65}{$e.g.$} are used for the four
parameters, it is essentially equivalent to the NMF mapping.

  While an exceptional amount of research has been devoted to improvement
of tropospheric mapping functions during the past two decades,
it is nevertheless becoming obvious that present measurement accuracy
also demands characterization of azimuthal asymmetry and short-term
weather variations.  This may be achievable via real-time auxiliary
measurements (such as WVRs), or statistical models of temporal and
spatial correlations.

\subsubsection{ Limitations of mapping }
\label{maplim}

  In recent years the limitations of the mapping function have become
apparent in VLBI observations.  Winds at high altitudes, unusually strong
lee waves behind mountains ($e.g.$, Owens Valley, California), and very
high pressure gradients may all limit the accuracy of an azimuthally
symmetric dry troposphere model based on measurements of surface barometric
pressure.  Rough estimates indicate that, except in such unusual cases,
errors in the simple model cause sub-centimeter errors in the baseline.
The accuracy in more typical situations is expected to be limited by
horizontal refractivity gradients caused by equator-to-pole (North-South)
gradients in temperature, pressure, and humidity.  These gradients were
first accounted for in analyses of satellite laser ranging experiments
by \citename{Gardner76}~\citeyear{Gardner76,Gardner77}.
East-West gradients caused by motion of weather
systems passing over a site may also contribute to the breakdown of the
simplified mapping function model.  The limits of validity of the
azimuthal symmetry assumption in VLBI analysis are starting to be
investigated \cite{MacMil95,MacMil97,Chen97}.

  While the limitations of dry mapping are small and in the case of
North-South gradients may be modelable, the wet component of the atmosphere
(both water vapor and condensed water in the form of clouds) is not so
easily modeled.  It is known to be highly variable in time and space
\cite{Fleag80}.  The experimental evidence \cite{Resch84}
is that it is ``clumpy'', and not
azimuthally symmetric about the local vertical at a level which can cause
many centimeters of error in a baseline measurement.  Furthermore, because
of incomplete mixing, surface measurements are inadequate in estimating
this contribution which even at zenith can reach 20 or 30 cm.
Ideally, this portion of the tropospheric delay should be determined
experimentally at each site at the time of the VLBI measurements.
Often the interferometer data themselves are used to quantify the effect
of the water vapor as part of the parameter estimation process.

  Independent quantitative estimates of the amount of water vapor along
a path through the atmosphere can be made by employing ``water vapor
radiometers'' (WVR).  These instruments measure the intensity of thermal
emission due to transitions between rotational energy levels of the water
molecule at microwave frequencies \cite{Elgered93}.  They are steerable
in both elevation and azimuth, have half-power beam widths of
6$^\circ-$ \thinspace 9$^\circ$, and measure sky brightness temperatures
at several
frequencies near 22$\thinspace$ and 31$\thinspace$GHz.  While there is
not a unique correspondence between the sky brightness temperature and
water vapor content, current data retrieval procedures appear to be
capable of accuracies as good as several mm of path delay \cite{Keihm96}.
Recent measurements along the lines of sight of VLBI observations
\cite{Linf96,Ttlbm96} have yielded wet troposphere delays that agree
with VLBI parameter estimates on the level of a few mm, and give a
threefold reduction in residuals.  Similar results have been obtained
by \citeasnoun{Elgered91}.  Despite optimism in the early years of WVR
development, such calibrations have not been routinely available for the
bulk of VLBI data collected during the past two decades.  Because
state-of-the-art WVR measurements have not been routinely available,
VLBI analyses should at the minimum model the neutral atmosphere at each
station as a two-component effect, with each component being an azimuthally
symmetric function of the local geodetic elevation angle.

  In an effort to properly account for unmodeled wet troposphere errors,
a model of the spatial and temporal spectrum of wet troposphere refractivity
fluctuations was developed by \citeasnoun{Trhft87}.  This
Treuhaft-Lanyi model
assumes that tropospheric delay errors are dominated by fluctuations in the
distribution of water vapor.  Furthermore, the model assumes that these
fluctuations are well described by a Kolmogorov spatial distribution
\citeaffixed{Ttrski61}{$e.g.$} that occurs in the bottom 1-2 km of the
troposphere and is carried over the site by a constant wind on the order
of 10 m/s.   It introduces realistic variations of the troposphere with
both time and geometry through this ``frozen flow'' assumption.  The model
is parametrized in terms of a ``rockiness'' coefficient, wind direction
and speed, and the height of the turbulent layer.

  The Treuhaft-Lanyi model is used to generate an $a~priori$ observable
covariance matrix which is then included in the least squares parameter
estimation procedure.  It provides estimates of correlations of the
tropospheric delays observed in different parts of the sky at different
times.  Correlations between phase rate observables are ignored.
\citeasnoun{Linf95} demonstrates that they typically do not exceed $10\%$,
and decay to much smaller values on time scales of a few minutes.  By
accounting for spatial variations and not just temporal variations, the
Treuhaft-Lanyi technique differs from filters which parametrize the
troposphere as a stochastic time-varying parameter \cite{Herr90}.
Lack of knowledge of the parameters required by the covariance model
presently limits its potential for improving VLBI parameter estimates.
Some of these parameters, which characterize the strength, extent, and
direction of the turbulent flow, are starting to be quantified
\cite{Naudet95}.  An important benefit of this technique is that data
strength is not used to estimate frequent troposphere parameters in cases
where the latter are not of primary interest.  Thus the variances of the
estimates of some important parameters ($e.g.$, baselines, source positions)
are smaller than would otherwise be obtained.  This benefit is in addition
to the reduction of systematic parameter biases relative to estimates which
use an inferior observation covariance (data weights).

\section{ APPLICATIONS AND RESULTS }
\label{curres}

  The model of VLBI observables which was developed in
Secs.~\ref{gdel}$-$\ref{atmo} has been used, with slight variations,
by a number of research groups to analyze geodetic and astrometric
experiments performed with various networks during the past two
decades \citeaffixed{Kondo92,Ma92,SJG93,NEOS93,Johnst95}{$e.g.$}.
The differences among groups generally involve minor variations
in tropospheric and tidal modeling.  Such analyses have presently
reached an approximate accuracy level of 1 cm on intercontinental
baselines (1 ppb).  In this section we highlight some of the new
insights into a wide variety of geophysics that have resulted from
VLBI measurements.  We intentionally exclude discussion
of literature concerning internal structures of extragalactic radio
sources and the associated astrophysical processes, since it is
sufficiently extensive to merit a separate review.  Thus, we again
focus on the accomplishments of purely astrometric and geodetic VLBI.
Some of these have already been introduced in the sections on model
description.  We will divide the applications of VLBI into three major
categories: reference frames (both celestial and terrestrial),
structure and dynamics of the Earth, and the orientation of the Earth
in the quasi-inertial celestial frame of reference.  Additional
information can be extracted that relates to the Earth's atmosphere,
relativistic bending, and rudimentary details of source structure.
In the near future, data that have accumulated over nearly
two decades are also expected to permit quantification of the Earth's
long-term motion in inertial space (galactic rotation).  Earlier in
this decade, \citeasnoun{Robrtsn91} presented a survey of VLBI applications that
we attempt to update here with our brief overview.  More detailed reviews,
including presentation of recent experiments, analyses, and results
can be found in Volumes 24 and 25 of \citeasnoun{Smith93}, which
report the achievements of NASA's Crustal Dynamics Project.  A general
introduction of the applications of many of the new space techniques,
including radio interferometry, is provided by the book of
\citeasnoun{Lambeck}.
Much of the astrometric/geodetic data that have been accumulated since
the 1970s are publicly available from the Crustal Dynamics Data
Information System at Goddard Space Flight Center.  This repository
includes both raw data and selected results of analyses, and is
accessible by computer at {\tt{http://cddisa.gsfc.nasa.gov/cddis.html}}.

\subsection{ Reference frames }
\label{apfram}

  The accomplishment of VLBI that is of greatest
significance is the establishment of reference frames which allow
quantitative treatment of the dynamics of the Earth and Solar System
at unprecedented levels of accuracy.  There are two such reference
systems: celestial and terrestrial.  The remainder of this section
summarizes determination and applications of these two fundamental
reference frames, their hundred-fold accuracy improvement since the
inception of VLBI and related methods, and problems of stability
related to these new accuracy levels.

  Astronomical objects have been used for millenia to construct celestial
reference systems for measuring the passage of time, for navigation, and
for investigating Solar System dynamics.  To review some of the important
milestones in the development of celestial reference frames: early
astronomers measured the
motions of the planets ($lit.$ ``wandering stars'') against the background
of ``fixed'' stars.  With improved observational precision, motions of
these fixed stars became evident.  Hipparchus is credited with
recognizing precession circa 129 B.C.  Proper motions of
individual stars were observed in 1718 by Halley.  Nutation of the
Earth was discovered by Bradley in 1748.  As observing precision continued
to improve, Herschel and Laplace suggested using extremely distant objects
to define astrometric reference frames.  Such objects reduce or eliminate
the effects of the proper motions on reference frame definition.  The
1781 catalog of Messier \cite{RobMuir}, and the New
General Catalog of \citeasnoun{Dreyer} were important steps in
identifying these more distant objects.  The work of \citeasnoun{Leavitt}
and \citeasnoun{Hubble} helped to establish the extreme distance of what
are now classified as extragalactic objects.  Present-day Earth-based
optical measurements have culminated in the latest ``fundamental
catalog'' FK5 \cite{Fricke88}, and the recent Hipparcos measurements
from Earth orbit (\citename{ESA97}, 1997) have inaugurated a new era for
optical catalogs.

  The historical celestial reference frame, based on measurements at
optical wavelengths, has been the unifying backdrop for centuries-long
astronomical observations.  It is a coordinate system whose origin is
located at the barycenter of the Solar System, and which is used to
establish locations of objects within the Solar System and beyond.
Its present accuracy level in the FK5 realization is on the order
of 0.1 arcsecond \cite{Fricke88}.  Extension of observations
into the microwave spectral region by early VLBI measurements already
improved this accuracy by an order of magnitude.  By the late 1990s,
yet another order-of-magnitude improvement has brought it to 1 mas or
better.  Unfortunately there are very few objects emitting sufficient
flux density to be observable at both optical and radio wavelengths.
Thus special techniques must be employed in order to connect the VLBI
celestial reference frame to the historical optical frame
\cite{Lestrade95}.

  Early determinations of the positions of extragalactic radio sources
by \citeasnoun{Wade70} and \citeasnoun{CohShaf71} yielded angular coordinates
accurate at approximately the 1 arcsecond level.  Within five years,
this accuracy was improved by more than an order of magnitude
\cite{WadJhn77}, and started to rival the best optical
determinations.  Improvement by approximately another two orders
of magnitude has been achieved during the past two decades for several
hundred radio sources.  Launch of the astrometric satellite Hipparcos
in 1989 \cite{Koval95} has likewise extended the accuracy
of optical positions of $\approx10^5$ sources to the milliarcsecond
level.  Since the median optical apparent magnitude of the radio sources
is $\approx$ 18 (well beyond Hipparcos sensitivity), a direct tie to the
radio reference frame is not possible.  A few optical sources (radio
stars) also emit weakly at radio frequencies.  These can be observed
in both the optical and radio regions in order to relate the two
reference frames \cite{Lestrade95}.

\vbox{
\begin{figure}
\hskip 0.5in
\epsfysize=3.0in
\epsffile{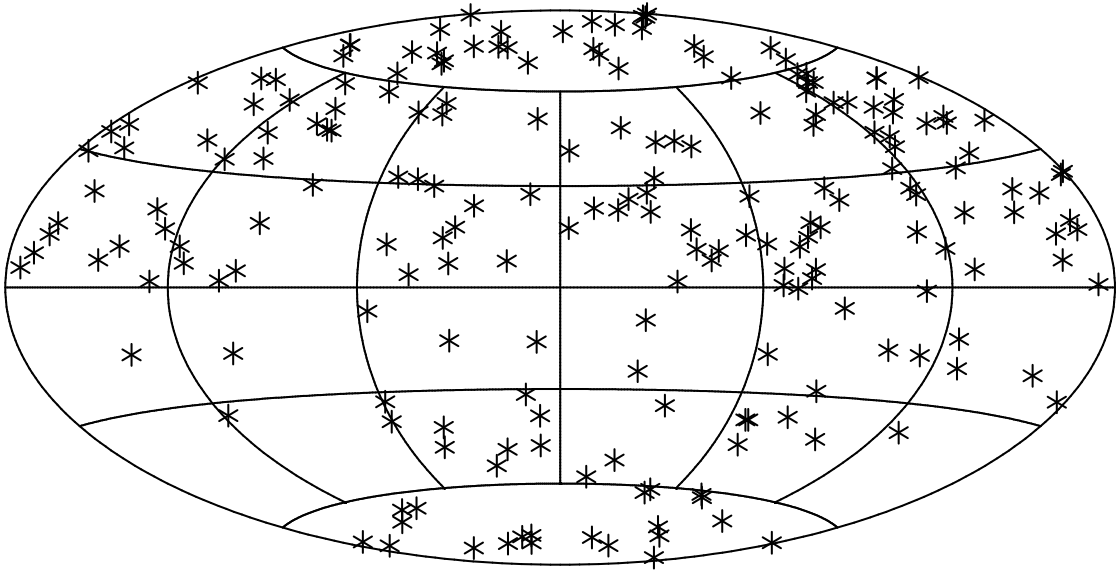}
\vskip 15pt
\caption{ Distribution of the 212 best-observed extragalactic sources
comprising the new IAU celestial reference frame (ICRF).
A conventional equal-area projection is used: declination increases from
$-90^\circ$ to $+90^\circ$ (bottom to top), and right ascension from
$-$12 to 12 hr (left to right).  Note the relatively sparse population at
negative declinations. }
\label{icrf}
\end{figure}
}

  During the last two decades, several research groups have used the
VLBI technique to catalog the positions of extragalactic radio sources
\cite{RobrtsnF86,Sovers88,Ma90,Johnst95}.
The number of objects whose positions are known at the mas level now
exceeds 400.  A considerable fraction of these continue to be monitored
in periodic experiments.  Survey campaigns
\citeaffixed{Patnaik92}{$e.g.$} with the MERLIN and VLA arrays), meanwhile,
have accumulated somewhat less precise coordinates of several thousand
additional radio sources, which form a valuable pool of candidates for
inclusion in the
higher-precision reference frame.  Some applications of extragalactic
radio reference frames have been to deep space navigation
\citeaffixed{Border93}{$e.g.$}, Earth orientation measurements
\citeaffixed{IERS94}{$e.g.$}, geodesy \citeaffixed{Fallon92}{$e.g.$},
and astrometry \citeaffixed{Bartel85,TrhftL91,Lebach95}{$e.g.$}.
In addition to the intrinsic scientific
interest in the stability of dynamical systems, these diverse applications
require accurate and stable positions of the objects composing the
reference frames.  The International Astronomical Union has now adopted
a new fundamental celestial reference frame (the International Celestial
Reference Frame, ICRF) that is based on the angular coordinates
of 212 radio sources \cite{IAU96,Ma97}.  Figure~\ref{icrf}
illustrates the distribution of the defining sources over the sky.
This is the first time that the fundamental celestial coordinate
system is no longer based on observations at visible wavelengths.
The ICRF is approximately 100 times more accurate than the FK5 catalog,
the present realization of the fundamental optical celestial frame.

  Limitations of reference frames that are based on extragalactic radio
sources have been recognized for some time.  While estimates of the proper
motion of quasars presently give null results at the approximate level of
50 prad/yr \cite{Eubanks96}, the extended structure of the
emitting objects at the nrad level is ubiquitous.  Structure
that is constant in time can be handled by careful definition of a fiducial
point for each source.  Unfortunately, many sources show considerable
time variability, and thus require constant monitoring to ensure
stability of the reference frame.  Such variability can be considered
to belong in the category of systematic errors.  Specialized imaging
experiments are being carried out in order to correct for source structure
and its variation \cite{Fey96}.  Imaging is also feasible, however,
via reanalysis of older data collected primarily for geodetic purposes
\citeaffixed{Charlot94,Piner97}{$e.g.$}.
Figure~\ref{strucplt} shows such results
for the source 3C 273 in 1986.  It can be seen that the central core
and neighboring components (``jet'') extend over several mas, making
it impossible to establish a fiducial mark at the sub-milliarcsecond
level without detailed study.
Similar images at other epochs show that the components of the jet
move rapidly, and new components are ejected sporadically.  After
this structure and its time variation was characterized in the 1980s,
it was dropped from most observing schedules.  The vast majority of
the defining sources of the ICRF, however, are much more point-like.

\vbox{
\begin{figure}
\epsfysize=2.2in
\epsfbox[-20 300 280 520]{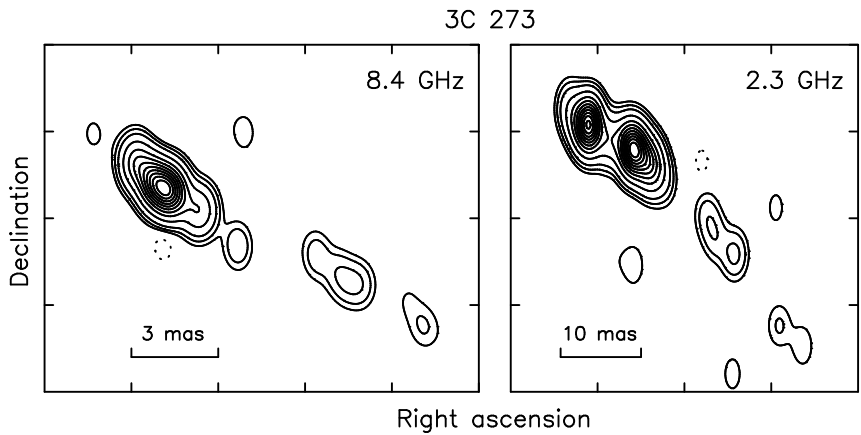}
\vskip 15pt
\caption{ Typical maps of the S- and X-band intensity distribution of
the source 3C 273, generated from geodetic VLBI data for the epoch
1986.27. }
\label{strucplt}
\end{figure}
}
  Another source of instability of the extragalactic reference frame
originates in the motion of the Solar System due to the rotation of our
Galaxy relative to the extragalactic radio sources.  Detection of
the acceleration of the SSB is now on the borderline of possibility.
As mentioned in Sec.~\ref{emot}, the motion
of the Solar System in inertial space makes its presence felt if
observational data extend over a sufficiently long time span.
To the extent that the Earth moves along with the uniform
rotation of our Galaxy and the distance to the Galactic Center
is known, this would amount to an indirect measurement of the period
of galactic rotation.  Such results could also make a contribution
to our understanding of cosmology from comparisons to motion of the
SSB with respect to the cosmic background radiation.  A related
topic is the proposed use of VLBI measurements to determine the
distance to the Galactic center \cite{Reid93} which would be of
importance in establishing a more reliable cosmic distance scale.

  The terrestrial reference frame has its origin at the Earth's center
of mass and establishes fiducial points on the surface of the Earth.
Prior to the advent of space geodetic experiments, there was no global
connection of national geodetic grids at a level better than $\approx$1
meter.  Doppler tracking of artificial Earth satellites improved this
situation, and the advent of laser ranging, VLBI, and global positioning
satellite tracking has further increased the accuracy of global reference
frames to approach 1 cm in the late 1990s.  In the arena of terrestrial
reference frames, VLBI measurements have contributed significantly to
the establishment of more than 100 fiducial points fairly uniformly
distributed over the Earth's surface, which collectively comprise the
ITRF: International Terrestrial Reference Frame \cite{Boucher96}.
Figure~\ref{itrf} shows the global distribution of VLBI stations that
contribute to the ITRF, while Fig.~\ref{pltmot} of Sec.~\ref{aprot}
gives an example of such results.  The VLBI technique has not been the
only contributor to an improved terrestrial reference frame.  Laser
ranging to satellites and the recent proliferation of GPS measurements
have begun to play a dominant role in forming a unified terrestrial
coordinate system that is accurate at the centimeter level.  This
provides a global reference for the national geodetic networks which have
been traditionally established by conventional triangulation methods;
these networks are being densified by GPS measurements.  The
result is a globally coherent reference system which provides a
backdrop for numerous studies of the Earth's behavior.  Some limitations
to its stability are discussed in the next section.

\vbox{
\begin{figure}
\hskip 0.5in
\epsfysize=3.0in
\epsffile{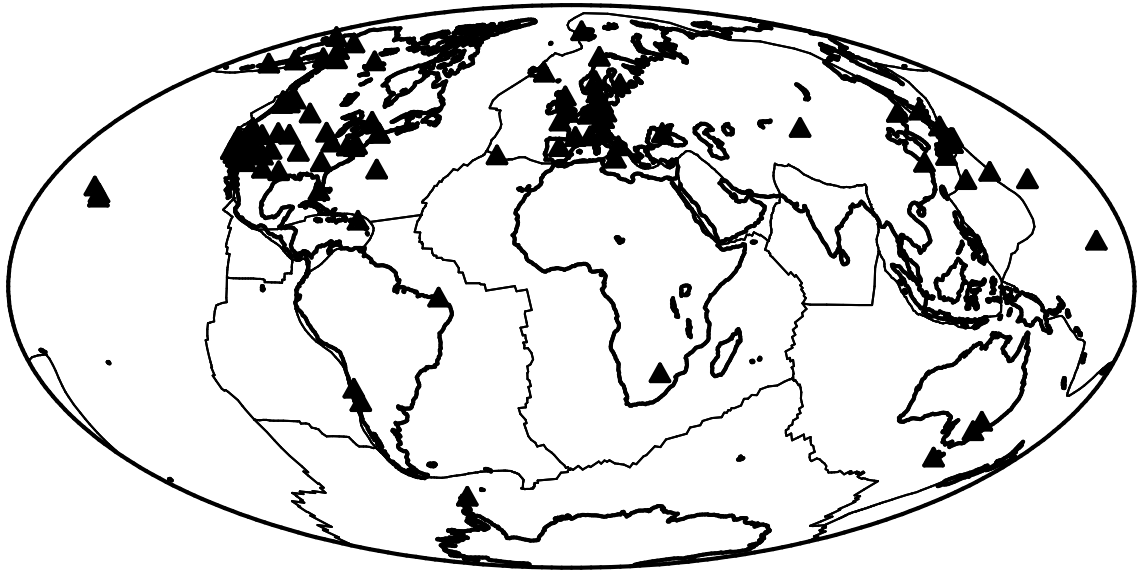}
\vskip 15pt
\caption{ Distribution of 104 VLBI observing stations that contribute to
the International Terrestrial Reference Frame (ITRF). An equal-area
projection of the Earth's surface is used.  Note the heavy concentration
in the northern hemisphere. }
\label{itrf}
\end{figure}
}

  It must be noted that both current celestial and terrestrial reference
frames exhibit a pronounced northern bias.  Due to the dominance of
observing stations in the northern hemisphere, there has been nonuniform
coverage of both the Earth's surface and the celestial sphere.  For
example, fewer than one fifth of the 50 best-determined ITRF sites are
in the southern hemisphere, and fewer than one third of the 212 sources
that are included in the new ICRF are at negative declinations
(Figs.~\ref{icrf} and \ref{itrf}).  It is expected that future
development of observatories in the southern hemisphere will remedy
this situation to the extent possible with the limited land area
available south of the Earth's equator.

\subsection{ Earth orientation and structure }
\label{aprot}

  The motion of the Earth as a whole, as well as its surface and bulk
composition and structure, are also areas in which VLBI measurements
have made significant contributions.  There are basically two classes
of Earth-related applications.  They pertain either to orientation of
the structure as a whole, or to motions of its crust.  Both categories
have varied root causes: Solar System dynamics and internal, oceanic,
and atmospheric processes on the Earth.  Manifestations of most of these
complicated motions are for the most part far removed from ordinary
human experience.  One exception may be the irregular slowing of the
Earth's rotation, which necessitates the introduction of occasional
``leap seconds'' in order to keep synchrony between the Earth's rotation
and atomic time.  Many of the remaining effects have amplitudes that
are only on the order of a meter or less at the Earth's surface, but
their characterization is of crucial importance in building a coherent
picture of our planet and its environment.

  Precise specification of the orientation of the Earth is of
practical importance, for example, in navigating spacecraft to
other planets.  The seemingly minuscule errors of parts per billion
are magnified into errors of kilometers near Jupiter.  Predictions of
future Earth rotation values (particularly UT1) lose accuracy very
rapidly for times past a few weeks after the date of measurement and
extrapolation.  Continuous monitoring of Earth rotation is therefore
imperative for applications that require high-quality real-time
UTPM.  Substantial help in this regard comes from a perhaps
unexpected source: the global distribution of atmospheric winds
or more precisely total atmospheric angular momentum (AAM),
which is widely monitored in real time for weather forecasting.

  Measurements of the rotation of the Earth and the wandering of its spin
axis relative to the crust contain a great deal of information related
to diverse physical processes.  While to a good approximation the
rotation rate is constant, it has slowed considerably over geological
time scales.  The length of a day was only 18 hours 900 million years
ago as determined from recent analyses of tidal sediments
\cite{Sonett96}.  A number of phenomena come into play in
determining the rotation rate,
ranging from the lengthening of the Earth-Moon distance, to frictional
and electromagnetic forces between the Earth's core and mantle, to
frictional forces between winds and ocean water and the surface.
Presently the rate
of deceleration is much larger than the billion-year average, on the
order of 2 ms per day, but it is known to have reversed sign within
the last hundred years \cite{Arch92b}.  The spin
axis has a quasi-periodic
motion that describes an approximate circle of $\approx$ 20 m diameter
(see Fig.~\ref{pmplt}) in $\approx$~300 days whose center also wanders
by tens of meters over decades; the size of the circle fluctuates with
a period of approximately 7 years.  In
addition to motions with respect to the Earth's crust, the spin axis
also exhibits motion in inertial space (nutation).  The two nutation
angles are usually included along with the three quantities describing
Earth rotation to form a set of five ``Earth orientation'' parameters.
The radio interferometric technique, along with other space geodetic
experiments, has brought about orders-of-magnitude improvements in
measurements of small irregularities in these motions, thereby opening
prospects for detecting numerous periodic and aperiodic processes.
We summarize the present accuracy of Earth orientation determinations,
and present examples of contributions of the results to two diverse
fields of geophysics.

  With a network of several stations that are reasonably uniformly
distributed over the surface of the Earth, and separated by baselines
on the order of an Earth radius, it is possible to infer the Earth
orientation parameters to an accuracy level that is better than 1 mas
(a few nanoradians, or parts per billion).  To illustrate VLBI
measurements of Earth orientation, Figs.~\ref{pmplt} and \ref{rotplt}
show the variation of the position of the spin axis relative to the
crust, and the rate of rotation, during a part of the 1980s and 1990s.
As seen in Fig.~\ref{pmplt} the spin axis follows a nearly circular path,
\vbox{
\begin{figure}
\hskip 2.0in
\epsfysize=3.0in
\epsffile{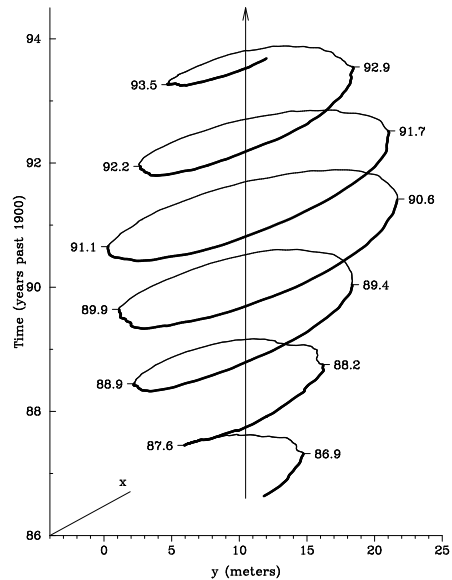}
\vskip 15pt
\caption{ Observed crust-fixed position of the Earth's spin axis during
the years 1986-1993.  The origin is at the conventional pole of 1903,
and the upward arrow indicates the approximate present-day average pole
position. }
\label{pmplt}
\end{figure}
}
with a diameter that oscillates between a few meters and 20 m with
a period of approximately seven years.  The deceleration of the Earth's
rotation rate is seen in the bottom part of Fig.~\ref{rotplt} as a
lag behind atomic time (TAI).  The resulting excess length of day
is seen in the top part of the figure: it exhibits a rich spectrum
which is just now beginning to be understood in detail.  The
dominant oscillations are caused by the annual interchange of
atmospheric angular momentum with the solid Earth.

  Experimental determinations of the Earth's rotation rate show that
it is not constant at the level of a few parts per billion.  There are
annual and semiannual cycles in the length of day, both with amplitudes
of approximately 0.3 ms and minima in January and July \cite{Eubanks93}.
Since the early 1980s it has been known that most of this variability is
well correlated with the total AAM \cite{Morgan85}.  Refinement
of this correlation was made possible by the superior precision of VLBI
relative to classical Earth orientation measurements using optical
techniques.  Variations in
the atmospheric angular momentum are transferred (via friction and
forces on mountain ranges) to variations of the angular momentum of
the solid Earth, conserving total angular momentum.  This discovery
has made substantial improvements in our ability to perform short-range
forecasts of Earth rotation.  Global weather measurements play a
significant role in this process, and their rapid availability is
crucial in improving forecast quality \cite{Freedm94}.
As with most global geophysical measurements, the spatial and
temporal density is inadequate at some level.  Relatively few
weather stations operate in the southern hemisphere, and the strong
winds at high altitudes (pressures of 100 to 1 mbar) are significant
\cite{Rosen85}.  In addition to the dominant 0.6-ms effect
of atmospheric angular momentum, a number of smaller meteorological
and tidal effects contribute to irregularities in Earth rotation.
Before considering some selected examples of these, we discuss the
remaining components of Earth orientation: the two nutation angles
that specify Earth orientation in an inertial frame.

\vbox{
\begin{figure}
\hskip 2.0in
\epsfysize=3.0in
\epsffile{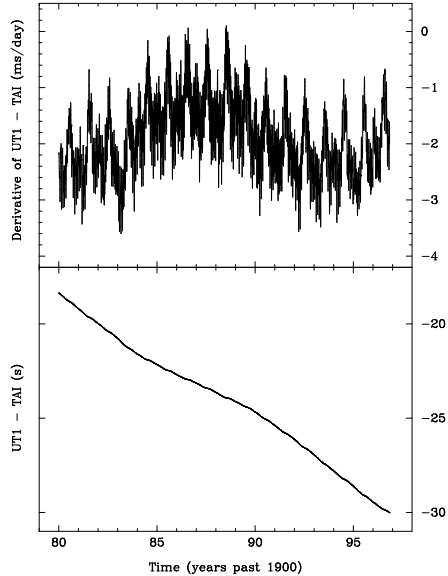}
\vskip 15pt
\caption{ Observed rotation rate of the Earth during the years 1980-1996.
The bottom plot shows the discrepancy of time measured by atomic clocks
and by the Earth's rotation, while the top plot shows the increase
of a few ms per day in the length of the day.  The annual variations
are caused by wind-surface friction. }
\label{rotplt}
\end{figure}
}
  In response to torques exerted by the Sun and Moon, the Earth's spin
axis oscillates in fundamental modes distributed over a wide range
of frequencies.
The amplitudes of these motions can be calculated fairly precisely
based on classical dynamics.  At the discrimination level of modern
measurement techniques, however, the theoretical models are not adequate.
The relatively poorly understood internal structure of the Earth plays
a large role in some of these motions, and empirical measurements
have just begun to shed light on the underlying mechanisms.
In the 1970s it became increasingly clear that the then-standard
Woolard nutation model was insufficiently accurate for analyses of modern
data.  Based on the theoretical work of \citeasnoun{Gilb75}
and \citeasnoun{Kinosh77}, Wahr developed a new nutation series
\citeyear{Wahr79,Wahr81},
which was adopted by the International Astronomical Union
in 1980 \cite{Sdlmn82}.  These two
series describe the time variation of the nutations in longitude
and obliquity by superposing oscillations at a total of 106
frequencies, with amplitudes of 0.1 mas precision.  Only a few
years after the adoption of the 1980 IAU model, VLBI measurements
revealed significant errors, predominantly at
semiannual, annual, and 18.6-yr periods.  \citeasnoun{Herr86}
initially pointed out these discrepancies, and gave improved estimates
of a number of amplitudes from VLBI analyses.  \citeasnoun{Gwinn86}
subsequently interpreted them as manifestations of irregularities
in the figure of the Earth's inner core.

  As an example of the ability and evolution of VLBI techniques in
the measurement of the Earth's spin axis in inertial space,
Fig.~\ref{nutplt} shows plots of the two nutation angles $\Delta \psi$
and $\Delta \varepsilon$ during the history of regular astrometric
VLBI measurements.  The plots show corrections to the standard
(1980 IAU) theoretical model of nutation.  It is apparent that
there are highly regular deficiencies at the 10-mas level.  Yearly
and semiannual frequencies are clearly visible, along with long-term
(linear and 18.6-y period) discrepancies of considerable size.

  The dominant nutation in longitude, with a period of 18.6 years,
is difficult to separate from precession (26,000-yr period) with
short data spans.  Only recently has the history of VLBI observations
reached a time span of one 18.6-yr period, permitting this decoupling.
Determinations of the nutation amplitudes and precession at the
levels of a milliarcsecond and a few tenths of mas/yr, respectively,
have been made by \citeasnoun{Herr88}, \citeasnoun{Charprc95},
\citeasnoun{WltrMa94}, and
\citeasnoun{WltrSov96}.  The value of the precession constant
which was determined from optical measurements in the 1970s
\cite{Fricke77} has been found to be in error by 3 mas/yr.
Most recently, it has been found that torques from bodies other
than the Sun and Moon can contribute to nutation of the Earth at
a level of 0.1 mas \cite{Kinosh90}.  These have
not yet been directly detected, but evidence is accumulating that
accounting for their cumulative effect improves the fit of
the VLBI model to experiment.  Another revision of the official
IAU nutation series is expected to include several hundred
additional terms, and to correct many amplitudes on the basis of
empirical VLBI determinations.

\vbox{
\begin{figure}
\hskip 2.0in
\epsfysize=3.0in
\epsffile{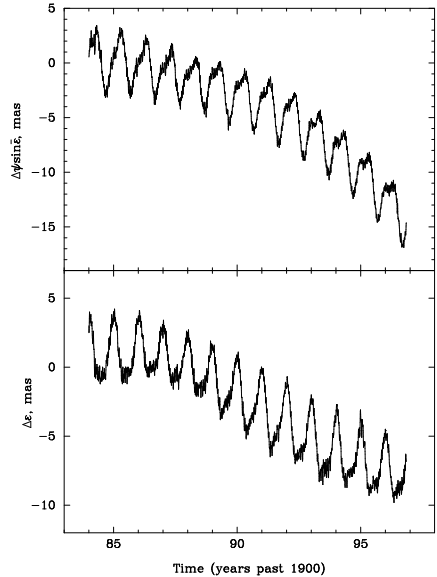}
\vskip 15pt
\caption{ Discrepancies of measured nutation angles and the 1980 IAU
models during the years 1984-1996.  Both the obliquity $\Delta \varepsilon$
and longitude $\Delta \psi \sin \bar \varepsilon$ are shown.  Short-term
(annual and semiannual) and long-term (linear and 18.6-year) discrepancies
are evident. }
\label{nutplt}
\end{figure}
}
  Gravitational forces from the Solar System, predominantly the
Sun and Moon, cause a rich spectrum in the tidal response of the
solid Earth and its oceans and atmosphere.  Direct motions of the
Earth's crust (with amplitudes on the order of a meter) are quite
obvious in VLBI analyses, but smaller secondary effects on station
positions and Earth orientation also cause measurable effects, and
need to be considered.  One of the first estimates of the solid Earth's
tidal response was the determination of Love numbers from VLBI by
\citeasnoun{Herr83}: $h = 0.62\pm0.01$, $l = 0.11\pm0.03$, and
a phase lag $\phi = 1^\circ\pm1^\circ$.  More recent work
\cite{Haas96,Gipson96} shows that the response factors
at numerous tidal frequencies can be reliably obtained from analyses
of VLBI measurements.

  Perhaps the most important of the secondary tidal effects is the
modification of Earth orientation by mass redistribution in the
oceans by global tides.  This affects both the rotation rate (UT1)
and orientation of the spin axis (PM).  \citeasnoun{Brosche89} made the
initial attempts to model these ``fast UTPM'' variation amplitudes.
In the early 1990s several independent determinations by VLBI and SLR
of the Earth orientation amplitudes induced by 8 ocean tidal components
gave results in good agreement with each other
\cite{SJG93,Herr94,Watkins94}.  Initially their
agreement with purely theoretical values based on global tide models
was not good, but it is improving \cite{Gross93,Ray94,ChaoR96,Gipson96},
as the global tide model
is refined by TOPEX/Poseidon results.  Agreement is now at the level of
10$\thinspace \mu$as.

  In addition to affecting the orientation of the whole Earth,
global redistribution of mass in the oceans by tidal action also
causes local motions of stations on the Earth's crust.  These
``ocean loading'' amplitudes have been incorporated in VLBI models
since the 1980s \cite{Schrnk83}, and can amount to motions of
several cm.  Although their small size places them close to the
current resolution limit of VLBI, long data spans are
capable of yielding significant direct determination of their
amplitudes \cite{Sovers94}.  By analogy with oscillations in the
distribution of ocean water, variations in the atmospheric pressure
at an observing site can also cause long-term vertical motions with
amplitudes of many mm.  This ``atmospheric loading'' has been
detected in VLBI analyses \cite{Rabbel86,Manabe91,vanDam94}.

  Another minor systematic effect which has only very recently been
identified is post-glacial rebound: the slow relaxation of the Earth's
crust in response to melting of the thick ice sheets which covered it
during the last glaciation.  The dominant motion is vertical, and
theoretical models produce estimates of several mm/year \cite{Peltier95}.
Unfortunately, determination of the vertical component of
station motion is notoriously weak relative to the two horizontal
components, because of the strong correlation of the vertical with
the zenith tropospheric delay.
Nevertheless, recent empirical estimates for a number of VLBI observing
stations with sufficiently long observation histories
\cite{Argus96,Ryan97} have produced results that are in fair agreement
with theoretical models \cite{Mitrov93}.

  A limitation to the stability of the terrestrial reference frame,
which is also of great interest for its own sake, is the motion of
the tectonic plates that constitute the Earth's solid surface.  Since
this possibility was proposed by Wegener in the early part of the
20th century, the existence of tectonic motion has been confirmed for
time scales of millions of years.  This was achieved by detailed
cataloging of magnetization properties of crust near active plate
boundaries \cite{MnstJ78,DeMets90}.  As
the time base for space geodetic station motions became sufficiently
long in the 1980s, the astonishing result emerged that present-day
(decade scale) plate motions are very similar to those inferred from
paleomagnetic data (Myr scale).  This attests to the relatively smooth
character of the forces driving plate tectonics.  At present, the
rates of motion of more than a dozen tectonic plates that have been
occupied by VLBI stations are known to within fractions of mm/year
\cite{Fallon92,ArgG96,Boucher96}.

\vbox{
\begin{figure}
\hskip 2.0in
\epsfysize=3.0in
\epsffile{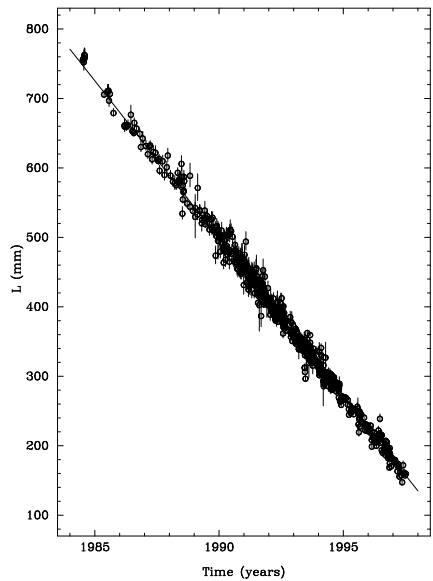}
\vskip 15pt
\caption{ Measured baseline between Alaska and Hawaii during 514
VLBI observing sessions, 1984-1997.  L is the baseline length in mm,
minus a constant offset of 4728114 meters.  The weighted RMS scatter
about the linear trend of $-$46 mm/yr is 7 mm. }
\label{pltmot}
\end{figure}
}

  As an example of the manifestation of tectonic station motion,
Fig.~\ref{pltmot} shows results of VLBI measurements of the length of the
baseline connecting stations at Gilmore Creek, Alaska and Kokee Park,
Hawaii (island of Kauai) during the years 1984-1997.  This figure is
derived from results of Goddard Space Flight Center 1997 data analyses,
archived at the World Wide Web site
{\tt{http://lupus.gsfc.nasa.gov/vlbi.html}}.  Several
points of interest are illustrated by the time history of this
particular baseline.  First, the North American (Alaska) and Pacific
(Hawaii) plates are one of the fastest moving pairs of tectonic plates,
converging at nearly 5 cm/yr.  Second, there is nearly perfect agreement
of the measured rate of change ($-$46 mm/yr) with the Nuvel model rate
of $-$45 mm/yr.  Third, the weighted root-mean-square scatter about a
perfectly linear trend is approximately 7 mm, which is a typical result
for a baseline of this length ($\approx$4700 km), and amounts to 1.5 ppb.
Fourth, the evolution of the VLBI technique is illustrated by the increased
frequency of experiments and the decrease in baseline length uncertainties
during the 1980s and 1990s.  Fifth, even the current theoretical models
have some systematic errors, which may be manifested in some of the
departures from linearity, $e.g.$, in 1994-96.  Finally, the general
stability of the experimental techniques is shown by the smooth transitions
at the epochs of both a large natural and a man-made perturbation at Kauai:
the hurricane damage in late 1992 and the transition to a new antenna,
40 meters distant, during the summer of 1993 are hardly noticeable in
Fig.~\ref{pltmot}.

  Proximity of VLBI stations to the epicenters of strong earthquakes,
and to active faults (especially the San Andreas fault in California)
is starting to contribute to improved understanding of the forces that
are involved in these motions.  Several pairs of
measurements of station coordinates before and after substantial
earthquakes now exist \cite{ClarkAK90}.  They confirm
permanent, essentially instantaneous displacements of several cm, but
the resolution is not sufficient to detect any relaxation mechanisms in
the subsequent hours and days.  The relatively dense network of stations
on both sides of the San Andreas fault show that the magnitude of
horizontal motion associated with the fault movement decreases with
increasing distance from the fault \cite{Ward88}.  Two
reviews \cite{DeMets95,Larson} contain more detailed discussion
of recent results related to global tectonics and crustal deformation.
A recently developed technique of active interferometry, synthetic
aperture radar (SAR), shows promise of accurate and timely
determination of regional deformations associated with tectonic and
earthquake activity \cite{Zebker94}.

\subsection{ Troposphere and ionosphere }
\label{apatm}

  Atmospheric effects have long been the bane of Earth-based astronomy.
They continue to play that role for interferometry at radio wavelengths.
In contrast to many of the more deterministic parts of the VLBI model,
the effects of the atmosphere are not easily predicted or externally
quantified.  The experiments themselves must therefore serve to
characterize atmospheric behavior.  In addition to the global effects
of the atmosphere (AAM) that were already mentioned in connection with
the above discussion of Earth rotation, we consider two additional aspects
here.  At best, empirical estimation can yield time series of accurate
atmospheric parameters, whose applicability is unfortunately limited
to the times and locations of the experiments.

  Determination of ionospheric electron content is possible whenever
the VLBI measurements are performed at two widely separated
frequencies.  The details of this dual-frequency calibration process
can be used to determine the integrated electron density along the
incoming ray path.  Limited sampling in both time and space, however,
limits the utility of VLBI as a method for studying the global ionosphere.
Satellite techniques give a much denser data set, and routine generation
of ionospheric maps is currently being implemented using GPS measurements
both with ground-based receivers and other satellites in low Earth orbit
\cite{Wilson95,Ho96}.

  Estimating the zenith tropospheric delay at frequent time intervals
for each VLBI station during data analysis permits monitoring the time
variation of water vapor content above that station.  These intervals
can even be reduced to essentially the intervals between observations
if a stochastic filtering scheme is used \citeaffixed{Herr90}{$e.g.$}.
If the observing schedule has ensured a good spatial distribution of
measurements (including low elevation angles), then the uncertainties
of the derived zenith wet delays are typically on the order of a few mm.
Comparisons of water vapor radiometer and VLBI results have demonstrated
the ability to track the atmospheric water vapor content quite faithfully
\cite{Elgered91,Ttlbm96}.  For the highest
accuracy, it is important to account for both the temporal and spatial
correlations of fluctuations of the tropospheric delay.  Stochastic
estimation schemes can account for temporal correlations.  To describe
atmospheric spatial asymmetry, horizontal gradient parameters may
be estimated.  An analysis scheme utilizing a theoretical model of
atmospheric turbulence which also contains parameters to quantify the
``clumpiness'' of the atmosphere \cite{Trhft87,TrhftL91}
accounts for correlations between VLBI delays
as a function of both time and space.  It also gives a better description
of the influence of both the water vapor content and wind distribution
surrounding a site \cite{Naudet95}.

\subsection{ Relativity }
\label{aprel}

  At the precision level of intercontinental VLBI measurements, the
incoming signals are affected by their passage through the varying
gravitational potential within the Solar System.  Since the positions
and velocities (ephemerides) of all the major gravitating bodies are
known to a high degree of accuracy, analyses of the experiments provide
a means of testing alternative relativistic theories.
\citeasnoun{Shapiro64}
first suggested that timing experiments on radio waves propagating
through the Solar System can be used to measure the parameter
$\gamma _{_{\rm PPN}}$
of parametrized post-Newtonian relativity theory.  These original
measurements of radar reflections from the inner planets \cite{Shapiro67}
gave the result $\gamma _{_{\rm PPN}}$ = 0.9 $\pm$ 0.2.
Determinations of $\gamma _{_{\rm PPN}}$ from
geodetic VLBI data spanning more than a decade gives
$\gamma _{_{\rm PPN}}$ = 1.000$\pm$0.002 \cite{Robrtsn84,RobrtsnC91}.
This uncertainty, $\sigma_\gamma$ =
0.002, is competitive with the results of specifically designed experiments
of much shorter duration.  An example of the latter is a VLBI experiment
which also observes at a third frequency (K band, $>$12 GHz) in addition
to the standard S-band and X-band wavelengths \cite{Lebach95}.
This yields $\gamma _{_{\rm PPN}}$ = 0.9996$\pm$0.0017, a somewhat improved
$\sigma_\gamma$ despite an order of magnitude fewer data.  Experiments
scheduling numerous observations in the vicinity of the Sun can achieve
equivalent results with just a fraction of the data, because of the
extremely steep variation of gravitational delay with closest approach
to the Sun.  The higher K-band frequency is less sensitive to the
corrupting influence of plasma-induced phase scintillation.
The estimated value of unity for $\gamma _{_{\rm PPN}}$
(within one standard deviation
in all cases) indicates agreement with Einstein's general relativity.

  Unfortunately none of the other parameters of post-Newtonian relativity
theory are accessible to measurement with VLBI experiments.  Nonlinearity
in the superposition of gravity $\beta$ and the time dependence of the
universal gravitational constant $\dot G/G$ cannot be determined from
VLBI, but limits are being placed on their magnitudes by analysis of lunar
laser ranging data \cite{Wllms96}.  Initial searches for VLBI
evidence of the passage of gravitational waves have placed loose bounds
on their existence \cite{Pyne96,Gwinn97}.  We also note that VLBI is
weakly sensitive to the difference between Newtonian and Lorentzian
aberration [$\propto 1/4 (v/c)^2$], which is a 1-cm effect for long
baselines and is implicitly included in the VLBI model.

\section{ PROGNOSIS FOR FUTURE MODEL IMPROVEMENTS }
\label{impr}

  Data analysis has provided continuous feedback to experiments
by characterizing previously unknown or poorly
quantified aspects of the model ($e.g.$, nutation, tidal variations,
troposphere).  Such interplay is certain to continue in the future,
as both experimental and theoretical techniques are refined to
eliminate remaining systematic errors.  This section gives condensed
summaries of areas which will require close attention in the future
if improvements of the current VLBI model are to permit achieving
true part-per-billion accuracy.

  Both special and general relativistic aspects of the current model
could be improved.  Second-order general relativistic effects have not
yet been thoroughly investigated, but probably do not contribute at the
picosecond level.  Theoretical studies such as that of \citeasnoun{Tryshv96}
may make a valuable contribution here.  If variations of the
gravitational potential along the path of the baseline through the
Earth are taken into account in calculating the proper distance,
this correction was estimated by \citeasnoun{Thomas91} to amount
to 2 mm for a 10,000~km baseline.  Similarly, variations in the
gravitational potential at the station clocks are only approximately
accounted for by means of Eq.~(\ref{eqtdb}).  Concerning special
relativistic aspects, the \citeasnoun{Fairh90} extension of the work
of \citeasnoun{Moyer81} on the ``time ephemeris'' produces higher-order
terms that contribute to TDB$-$TDT at the $\mu$s level.

  Galactic effects may soon emerge above the detection threshold.
The rotational motion of the Galaxy produces aberrational effects
which change by 20 prad/yr.  This will need to be taken into account
for observations spanning more than two decades.

  The rich tidal spectrum of the solid Earth and its oceans will be a
source of model refinements for some time to come.  Direct contributions
of the planets to solid Earth tidal displacements can reach the millimeter
level.  In addition to the eight frequencies considered in the model
described in Sec.~\ref{tidutpm}, short-period variations of UTPM
have additional components \cite{Seiler95,Gipson96}.  Those which are
significant at the mm level will emerge as data analyses are refined.
Empirical estimates of ocean loading amplitudes for several IRIS stations
\cite{Sovers94} indicate that the best theoretically derived amplitudes
may be in error by several mm.  Future refinements in data analyses, and
improved global ocean models from the recent TOPEX/Poseidon mission
\cite{LeProv95,Fu94} may improve the accuracy of the
theoretical ocean loading model to the mm level.  Resonance with the Earth's
free core nutation may modify some of the amplitude corrections at nearly
diurnal frequencies by $\approx$ 1 mm.  Ocean tides cause motion of the
center of mass of the solid Earth due to motion of the center of mass of
the oceans \cite{BrosWun93}.  The amplitude of this
displacement can be as large as 1 cm at the usual diurnal and semidiurnal
tidal frequencies.  Non-conservation of the total angular momentum of
the Earth via interchange with the Sun and Moon is also not included
in the tidal models \cite{BrosSeil96}.  The effects of both
of these refinements on VLBI observations must be assessed.  The retarded
tidal potential effect mentioned in Sec.~\ref{soltid} can be as large
as several tenths of mm.  Thus, for correct modeling at the mm level,
the light travel time should be accounted for.

  Non-point like flux distributions of the radio sources are being
efficiently mapped by experimenters using the VLBA \cite{Fey96}.
Their results are expected to provide improved models for astrometric
and geodetic VLBI.  Meanwhile, estimates of parameters for simplified
structural models may improve data analyses.

  There are short-period deficiencies in the present International
Astronomical Union models for
the orientation of the Earth in space that may be as large as 1 to 2
milliarcseconds, and longer-term deficiencies on the order of 1 mas
per year (3 cm at one Earth radius).  VLBI measurements made during
the past decade indicate the need for revisions of this order of the
annual nutation terms and the precession constant
\cite{Eubanks85,Herr86}.  The 18.6-year components of the
IAU nutation series are also in error, and present data spans are just
approaching durations long enough to separate them from precession.
Options to improve the nutation model were discussed in Sec.~\ref{nut}.
Any of these constitute a provisionally improved model,
especially for the annual and semiannual nutations, until the IAU
series is officially revised.  Future refinements of the equation of
the equinoxes (Eq.~\ref{eqeq}) will probably lead to changes on the
order of tens of $\mu$s in the hour angle.

  The large dish antennas that are used to collect signals from extragalactic
sources are susceptible to various instabilities at the level of many
mm.  Gravity loading may cause systematic variations in the position of
the reference point of a large antenna that are as large as 1 cm in the
local vertical direction.  Such systematic errors and their dependence on
antenna orientation and temperature may be modeled \cite{ClarkTh88,Jacobs89}.
The geometric structure of each
antenna, as well as its alignment with respect to local site features,
should be carefully checked against design specifications.  For example,
hour angle misalignment on the order of 1 arc minute can cause 1 mm
delay effects for HA-Dec antennas with 7-m axis offsets.  Thermal
expansion of the portion of an antenna above the reference point may
induce delay signatures that are several mm peak-to-peak for a typical
VLBI antenna \cite{Nothn95}.  It is thus imperative
to model this effect for achievement of the highest accuracy.

  The limits of validity of the dual-frequency calibration procedure
for ionospheric effects need to be carefully established, in
conjunction with consideration of plasma effects for ray paths near
the Sun.  In addition, corrections for the gyrofrequency effect may
reach a millimeter.

  New techniques for characterizing the atmosphere are expected to
allow more realistic modeling of the tropospheric delay than the
simple spherical-shell model underlying all the results of Sec.~\ref{trp}.
When comprehensive atmospheric data from a region surrounding
each observing site are available, present computer speeds should permit
estimating the tropospheric delay by means of a complete ray-tracing
solution for every observation.  Meanwhile, improvements in tropospheric
mapping can be sought by modeling variations of the temperature $vs.$
altitude profile as a function of season, latitude, altitude, and diurnal
cycle.  Efforts are also under way \citeaffixed{MacMil97,Chen97}{$e.g.$},
to model azimuthal gradients in the troposphere.  Persistent
equator-to-pole gradients
in pressure, temperature, humidity, and tropopause height suggest that
$a~priori$ modeling of North-South gradients may be beneficial.  East-West
gradients, which are probably dominated by weather systems passing over a
site, are likely to be more difficult to model without extensive weather data.

  By the late 1990s astrometric and geodetic radio interferometry has
largely fulfilled the promise seen for the technique at its inception
in the late 1960s.  VLBI has yielded a new nearly inertial celestial
reference frame that is accurate at a level of
nanoradians, achieved point positioning on the Earth at the centimeter
level, and produced the capability of determining the instantaneous
orientation of Earth in space at similar accuracy levels.  The VLBI
technique has intersected many fields of physics, and its evolution
has involved both routine and unexpected components.  In addition to
improved quantification of known physics, there was need
to take into account a number of unexpected contributions, which led
to enhanced understanding of the behavior of the Earth.  The field
is still developing vigorously both on the experimental and theoretical
fronts.  It can be hoped that progress will not be unduly impeded by
numerous processes effective at the millimeter level.  The future
holds the promise of exciting and unexpected results characterizing
the behavior of the Earth, the structure and emission mechanisms of
extragalactic radio sources, and the motion the Earth in the universe.

\acknowledgments

Numerous people at many organizations on several continents have
contributed to the evolution of the present VLBI model during the past
two decades, and we have personally benefitted from interacting with
many of them.  Our own initiation into VLBI studies at JPL was guided
by J.~B. Thomas and J.~G. Williams during the 1970s.  Critical internal
reviews of this manuscript were kindly done by J.~H. Lieske,
W.~M. Owen,~Jr., and J.~G. Williams, and externally by T.~M. Eubanks,
J.-F. Lestrade, L.~Yu. Petrov, and H.~G. Walter.  Suggestions of a
referee stimulated us to broaden its scope.
The work described in this paper was performed at the Jet Propulsion
Laboratory, California Institute of Technology, under contract with
the National Aeronautics and Space Administration.

\vfil\eject
\begin{table}
\caption{Maximum Magnitudes and Present Uncertainties of Portions of
 the VLBI Delay Model~(mm). }
\label{errbudg}
\begin{tabular}{lcr}
Model component & Maximum delay & Present model \\
                &               & uncertainty   \\
\tableline
BASELINE GEOMETRY      &               &    \\
~~Zero-order geometric delay     & $6\times10^9$ & ...\\
~~Earth orbital motion & $6\times10^5$ &  1 \\
~~Gravitational delay  & $2\times10^3$ &  2 \\
STATION POSITIONS      &                    \\
~~Tectonic motion      &    100        &  1 \\
~~Tidal motion         &    500        &  3 \\
~~Non-tidal motion     &     50        &  5 \\
EARTH ORIENTATION      &               &    \\
~~UTPM                 & $2\times10^4$ &  2 \\
~~Nutation/precession  & $3\times10^5$ &  3 \\
SOURCE STRUCTURE       &     50        & 10 \\
ANTENNA STRUCTURE      &  $10^4$       & 10 \\
INSTRUMENTATION        & $3\times10^5$ &  5 \\
ATMOSPHERE             &                    \\
~~Ionosphere           &  $10^3$       &  1 \\
~~Troposphere          & $2\times10^4$ & 20 \\
\end{tabular}
\end{table}

\begin{table}
\caption{Tectonic Plate Rotation Velocities: NNR-Nuvel-1A Model
 \protect \\ (Units are nanoradians/year). }
\label{nuv1a}
\begin{tabular}{lddd}
Plate & $\omega_{\rm x}$ & $\omega_{\rm y}$ & $\omega_{\rm z}$ \\
\tableline
Africa        &    0.891 & $-$3.099 &    3.922 \\
Antarctica    & $-$0.821 & $-$1.701 &    3.706 \\
Arabia        &    6.685 & $-$0.521 &    6.760 \\
Australia     &    7.839 &    5.124 &    6.282 \\
Caribbean     & $-$0.178 & $-$3.385 &    1.581 \\
Cocos         & $-$9.705 & $-$21.605&   10.925 \\
Eurasia       & $-$0.981 & $-$2.395 &    3.153 \\
India         &    6.670 &    0.040 &    6.790 \\
Juan de Fuca  &    5.200 &    8.610 & $-$5.820 \\
Nazca         & $-$1.532 & $-$8.577 &    9.609 \\
North America &    0.258 & $-$3.599 & $-$0.153 \\
Pacific       & $-$1.510 &    4.840 & $-$9.970 \\
Philippine    &   10.090 & $-$7.160 & $-$9.670 \\
Rivera        & $-$9.390 &$-$30.960 &   12.050 \\
Scotia        & $-$0.410 & $-$2.660 & $-$1.270 \\
South America & $-$1.038 & $-$1.515 & $-$0.870 \\
\end{tabular}
\end{table}

\begin{table}
\caption{Frequency Dependent Solid Earth Tide Parameters.}
\label{frqtidh2}
\begin{tabular}{rdr}
Component $k$ & $h_2^k$ & $H_k$ (mm) \\
\tableline
 $\psi _1$ (166554) &  0.937 &       3  \\
           (165565) &  0.514 &      50  \\
 $K_1    $ (165555) &  0.520 &     369  \\
           (165545) &  0.526 &    $-$7  \\
 $P_1    $ (163555) &  0.581 &  $-$122  \\
 $O_1    $ (145555) &  0.603 &  $-$262  \\
\end{tabular}
\end{table}

\begin{table}
\caption{Lunar Node Companions to Ocean Tides.}
\label{compan}
\begin{tabular}{ccc}
$i$       & $n_{ki}$  & $r_{ki}$: Relative \\
Component & Companion & Amplitude          \\
\tableline
 $K_2$ (275555)       & $-$1 & $-$0.0128  \\
                      &   +1 &   +0.2980  \\
                      &   +2 &   +0.0324  \\
 $S_2$ (273555)       & $-$1 &   +0.0022  \\
 $M_2$ (255555)       & $-$2 &   +0.0005  \\
                      & $-$1 & $-$0.0373  \\
 $N_2$ (245655)       & $-$1 & $-$0.0373  \\
 $K_1$ (165555)       & $-$1 & $-$0.0198  \\
                      &   +1 &   +0.1356  \\
                      &   +2 & $-$0.0029  \\
 $P_1$ (163555)       & $-$1 & $-$0.0112  \\
 $O_1$ (145555)       & $-$2 & $-$0.0058  \\
                      & $-$1 &   +0.1885  \\
 $Q_1$ (135655)       & $-$2 &   +0.0057  \\
                      & $-$1 &   +0.1884  \\
 $M_{\rm f}$ (075555) &   +1 &   +0.4143  \\
                      &   +2 &   +0.0387  \\
 $M_{\rm m}$ (065455) & $-$1 & $-$0.0657  \\
                      &   +1 & $-$0.0649  \\
$S_{\rm sa}$ (057555) & +1 & $-$0.0247  \\
\end{tabular}
\end{table}

\begin{table}
\caption{Ocean Tidally Induced Periodic Variations in Polar Motion
 (Sovers $et~al.$, 1993). }
\label{shptid}
\begin{tabular}{rrrrrrrrrrrrrr}
Index & Tide, period & \multicolumn{6}{c}{Argument coefficient} &
 $A_{i1}$ & $B_{i1}$ & $A_{i2}$ & $B_{i2}$ &  $A_{i3}$ & $B_{i3}$  \\
i   &      (hours) & $k_{i1}$ & $k_{i2}$ & $k_{i3}$ & $k_{i4}$ &
$k_{i5}$ & $n_j$ & \multicolumn{4}{c}{~~($\mu$as)} &
\multicolumn{2}{c}{$~~(0.1\mu s)$}   \\
\tableline
1 & $K_2$ \ \ \ 11.967 & 0& 0& 0&   0& 0& $-$2 &
  $-$2&     65&     44&$-$57&  $-$9&    26 \\
2 & $S_2$ \ \ \ 12.000 & 0& 0& 2&$-$2& 2& $-$2 &
   101&    166&    126&$-$89&  $-$4&    52 \\
3 & $M_2$ \ \ \ 12.421 & 0& 0& 2&   0& 2& $-$2 &
    26&    283&    247& $-$2&$-$104&   149 \\
4 & $N_2$ \ \ \ 12.658 & 1& 0& 2&   0& 2& $-$2 &
 $-$15&     56&     19&$-$11& $-$23&    20 \\
5 & $K_1$ \ \ \ 23.934 & 0& 0& 0&   0& 0& $-$1 &
$-$583&$-$2780&$-$2950&  376&    35&   151 \\
6 & $P_1$ \ \ \ 24.066 & 0& 0& 2&$-$2& 2& $-$1 &
   154&     46&     42&$-$17& $-$32& $-$64 \\
7 & $O_1$ \ \ \ 25.819 & 0& 0& 2&   0& 2& $-$1 &
   242& $-$152&      2&$-$30&$-$135&$-$166 \\
8 & $Q_1$ \ \ \ 26.868 & 1& 0& 2&   0& 2& $-$1 &
    72&  $-$32&     26&    7& $-$40& $-$53 \\
\end{tabular}
\end{table}

\begin{table}
\caption{ Solar System Velocity Relative to Various Standards of Rest. }
\begin{tabular}{lr@{}l@{${}\pm{}$}r@{}lrr}
Reference & \multicolumn{4}{c}{Velocity~(km/s)} & $l$ (deg) & $b$ (deg) \\
\tableline
Local standard of rest\tablenotemark[1]
     & ~~~~~~20 &&  1 &&  57 &   23 \\
Galactic center\tablenotemark[1]$^,$\tablenotemark[2]
          & 240 &&  5 &&  90 &    0 \\
Local Group\tablenotemark[3]
          & 308 && 23 && 105 & $-$7 \\
Cosmic microwave background\tablenotemark[4]
          & 370 &&  3 && 264 &   48 \\
\end{tabular}
\label{mottab}
\tablenotetext[1]{ \citeasnoun{Kerr86} }
\tablenotetext[2]{ \citeasnoun{Fich89} }
\tablenotetext[3]{ \citeasnoun{Yahil77} }
\tablenotetext[4]{ \citeasnoun{Kogut93} }
\end{table}

\begin{table}
\caption{Plasma Effects.}
\label{plasma}
\begin{tabular}{lcccc}
Plasma & $\rho (e/m^3)$ & $\nu_p$ (kHz) & $(\nu_p/\nu_{\rm S})$ &
 $(\nu_p / \nu_{\rm X})$  \\
\tableline
Earth           & $10^{12}$ & 8980 & $4 \times 10^{-3}$ &
     $10^{-3}$  \\
Interplanetary  & $10^{10}$ &  898 & $4 \times 10^{-4}$ &
     $10^{-4}$  \\
Interstellar    & $10^7$    &   28 & $1 \times 10^{-5}$ &
 $3 \times 10^{-6}$  \\
\end{tabular}
\end{table}

\begin{table}
\caption{Electron Gyrofrequency Effects.}
\label{gyrof}
\begin{tabular}{lcccc}
Magnetic field & $B$ (gauss) & $\nu_g$ (kHz) & $(\nu_g/\nu_{\rm S})$ &
   $(\nu_g / \nu_{\rm X})$  \\
\tableline
Earth           & $0.3$     & 840   & ~~$4 \times 10^{-4}$ &
 $10^{-4}$           \\
Interplanetary  & $10^{-5}$ & 0.028 & $1.2 \times 10^{-8}$ &
 $3 \times 10^{-9}$~ \\
Interstellar    & $10^{-6}$ & 0.003 & $1.2 \times 10^{-9}$ &
 $3 \times 10^{-10}$ \\
\end{tabular}
\end{table}

\begin{table}
\caption{Sensitivity of Tropospheric Delay to Lanyi Mapping Function
Parameters.}
\label{lanyisurf}
\begin{tabular}{ccc}
Parameter & Standard value & Sensitivity ($6^{\circ}$) \\
\tableline
$T_0$ & 292 K~~     &  $-$7 mm/K~~~~~~~ \\
$W$   & 6.8165 K/km &  20 mm/K/km       \\
$h_1$ & 1.25 km     & $-$20 mm/km~~~~~~ \\
$h_2$ & 12.2 km     &     5 mm/km~~     \\
\end{tabular}
\end{table}

\end{document}